\documentclass[prb,twocolumn,showpacs,amsmath,amssymb,superscriptaddress]{revtex4}

\usepackage{graphicx}
\usepackage{longtable}

\newcommand{\Id}{\boldsymbol{1}}

\newcommand{\La}{L^a}
\newcommand{\Lb}{L^b}
\newcommand{\Lc}{L^c}
\newcommand{\Lh}{L^h}

\newcommand{\myla}[1]{{\mathcal{L}}_1({#1})}
\newcommand{\mylc}[1]{{\mathcal{L}}_2({#1})}

\newcommand{\iL}{_{\text{L}}}
\newcommand{\iR}{_{\text{R}}}
\newcommand{\nd}{_{\text{nd}}}

\begin{document}

\title{Real-time renormalization group in frequency space:
A two-loop analysis of the nonequilibrium anisotropic Kondo model at finite magnetic field}
\author{Herbert Schoeller}
\author{Frank Reininghaus} 
\affiliation{Institut f\"ur Theoretische Physik, Lehrstuhl A, RWTH Aachen, 52056 Aachen, Germany}
\affiliation{JARA-Fundamentals of Future Information Technology}
\date{\today}
\begin{abstract}
We apply a recently developed nonequilibrium real-time renormalization
group (RG) method in 
frequency space to describe nonlinear quantum transport through
a small fermionic quantum system coupled weakly to several reservoirs 
via spin and/or orbital fluctuations. Within a weak-coupling two-loop analysis, 
we derive analytic formulas for the nonlinear conductance and the kernel determining the time 
evolution of the reduced density matrix. A consistent formalism is presented how the RG flow 
is cut off by relaxation and dephasing rates. We apply the general formalism 
to the nonequilibrium anisotropic Kondo model at finite magnetic field. We consider the 
weak-coupling regime, where the maximum of voltage and bare magnetic field is larger than 
the Kondo temperature. 
In this regime, we calculate the nonlinear conductance, the magnetic susceptibility, the 
renormalized spin relaxation and dephasing rates, and the renormalized $g$~factor. All 
quantities are considered up to the first logarithmic correction beyond leading order at 
resonance. Up to a redefinition of the Kondo temperature, we confirm previous results 
for the conductance and the magnetic susceptibility in the isotropic case. In addition, 
we present a consistent calculation of the resonant line shapes, including the determination 
whether the spin relaxation or dephasing rate cuts off the logarithmic divergence. Furthermore, we 
calculate quantities characterizing the exponential decay of the time evolution of 
the magnetization. In contrast to the conductance, we find that the derivative
of the spin relaxation (dephasing) rate with respect to the magnetic field is
logarithmically enhanced (suppressed) for voltages smaller (larger) than the renormalized magnetic field,
and that the logarithmic divergence is cut off by the opposite rate.
The renormalized $g$~factor is predicted to show a symmetric logarithmic suppression at resonance,
which is cut off by the spin relaxation rate. We propose a three-terminal setup
to measure the suppression at resonance. For all quantities, we analyze also the anisotropic 
case and find additional nonequilibrium effects at resonance.

\end{abstract}
\pacs{05.10.Cc, 72.10.Bg, 73.63.Nm}

\maketitle

\section{Introduction}
\label{sec:introduction}
\begin{figure}
  \includegraphics[scale=0.5]{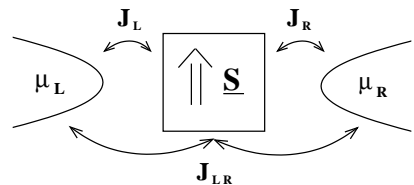}
  \caption{A spin-${1\over 2}$ quantum system coupled via exchange
to two reservoirs. $J_L = J_{LL}$ and $J_R =J_{RR}$ involve exchange between
the spins of the left/right reservoir with the local spin, and 
$J_{LR}=J_{RL}$ transfers a particle from one reservoir to the
other during the exchange process.} 
\label{fig:kondo}
\end{figure}

One of the basic unsolved problems of dissipative quantum mechanics and quantum transport
through mesoscopic systems is the nonequilibrium Kondo model. In its simplest version, it consists of a 
spin-${1\over 2}$ system coupled via exchange processes to the spins of two fermionic reservoirs,
which are kept at two different chemical potentials $\mu_{L/R}=\pm e{V\over 2}$, see Fig.~\ref{fig:kondo}.
Besides the importance of the Kondo model for many aspects of strongly correlated Fermion systems
(see Ref.~\onlinecite{hewson} for an overview), it was suggested to realize this model in transport 
experiments through quantum dots.\cite{kondo_theo} This has been achieved 
\cite{kondo_exp} with the particular advantage of full control over
all parameters such as
temperature, voltage, magnetic field, and exchange couplings. The central idea is to lower 
temperature and bias voltage such that only one
single-particle level of a quantum dot will contribute to transport. Adjusting the gate voltage 
such that charge fluctuations of this level are suppressed (Coulomb blockade regime), the dot can 
either be occupied by a spin up or a spin down electron, and the spin can fluctuate via second-order cotunneling 
processes, leading precisely to the model depicted in Fig.~\ref{fig:kondo}. In this realization, one
obtains an antiferromagnetic
exchange coupling of the order $J\sim \rho t^2/E_c$ (in dimensionless units), where $\rho$ is the 
density of states in the reservoirs, $t$ denotes the tunneling amplitude (hopping parameter) between
the leads and the dot, and $E_c$ is the energy of the virtual intermediate state (charging energy). 
In equilibrium, the Kondo model has been analyzed by various many-body methods (for an overview,
see Ref.~\onlinecite{hewson}) and can be solved exactly by Bethe ansatz \cite{bethe_ansatz} 
or conformal field theory.\cite{conformal_field_theory} Powerful numerical
techniques such as the numerical renormalization group have been developed \cite{wilson,costi} from which all
thermodynamic and spectral properties can be calculated. The basic physics can already be understood
from poor man scaling arguments.\cite{poor_man_scaling} If all exchange couplings are the same 
$J_L=J_R=J_{LR}=J_0$, one obtains the renormalization group (RG) equation 
\begin{equation}
\label{poor_man_scaling}
{dJ\over dl}\,=\,2J^2\quad,\quad l\,=\,\ln{D\over\Lambda}\quad,
\end{equation}
with the solution
\begin{equation}
\label{solution_poor_man_scaling}
J\,=\,{1\over 2\ln{\Lambda\over T_K}}\quad,\quad T_K\,=\,D\,e^{-{1\over 2J_0}}\quad.
\end{equation}
Here, $T_K$ is the Kondo temperature and $J$ denotes an effective exchange coupling corresponding 
to an effective band width $\Lambda$ of the reservoirs ($D\sim E_c$ is the original band width). 
For the antiferromagnetic case $J>0$, one obtains an enhancement of the exchange coupling by 
reducing the effective band width until $\Lambda$ reaches $T_K$, where $J$ diverges. This indicates a logarithmic
enhancement of the linear conductance for temperatures above the Kondo temperature (weak-coupling regime), 
and a complete screening of the dot spin by the reservoir spins below $T_K$ (strong-coupling regime). In 
the latter case, it can be shown that the conductance becomes unitary,
i.e., $G=2e^2/h$, at zero temperature $T=0$ and
zero bias voltage $V=0$.

In nonequilibrium, the Kondo model is not yet solved completely. Some numerical techniques have already
been developed to describe either the time evolution of an
out-of-equilibrium initial state by using time-dependent numerical
renormalization group (TD-NRG)
(Ref.~\onlinecite{TD_NRG}) or to describe the stationary current in the presence of a bias by using NRG in a scattering wave
basis \cite{NRG_scattering} or using an iterative real-time path integral approach.\cite{thorwart_egger} 
Scattering wave Bethe ansatz methods based on the Lippmann-Schwinger equation
are currently under way to solve the single-impurity Anderson model in nonequilibrium.\cite{palacios_andrei}
Concerning analytical RG methods, it has been emphasized that it is important
to understand how the RG equations are cut off by the bias voltage and by relaxation and dephasing rates.\cite{coleman,kaminski_etal,rosch_kroha_woelfle_PRL01} The problem can be stated as follows. Performing
perturbation theory in the bare exchange coupling $J_0$ (disregarding for the moment the differences
between $J_L$, $J_R$, and $J_{LR}$), logarithmic terms occur, which, at zero temperature and
depending on the physical observable under consideration, have the form [we indicate only the order
in the exchange coupling leaving out numbers of $O(1)$ and other physical energy scales from $V$ and
$h_0$ in the prefactor]
\begin{equation}
\label{log_pert}
\left\{
\begin{array}{cl}
I,\tilde{\Gamma}_1,\tilde{\Gamma}_2 \\ \tilde{g}-2 \\ M
\end{array}
\right\}\rightarrow
\left\{
\begin{array}{cl}
J_0^2 \\ J_0 \\ 1
\end{array}
\right\}\,
J_0^k\,\ln^l{D\over |nV-mh_0|}\quad,
\end{equation}
with $k\ge l\ge 0$, $n=0,\pm 1,\pm 2,\dots$ and $m=0,\pm 1$.
Here, $I$ is the current, $\tilde{\Gamma}_{1/2}$ are the spin relaxation/dephasing rates, $\tilde{g}$ is the 
renormalized $g$~factor, and $M$ denotes the magnetization. $D$ is the band width of the reservoirs and $h_0$ 
is the bare magnetic field. $n$ denotes
the number of particles transferred between the reservoirs and $m=0,\pm 1$ characterizes whether
spin flip processes occur or not. The points $nV=mh_0$ correspond to resonance positions,
where certain higher-order processes are allowed by energy conservation, e.g., the value $n=m=1$
corresponds to the onset of inelastic cotunneling processes.\cite{cotunneling} Higher-order terms
with $n\ge 2$ have so far not been discussed in the literature for the Kondo model, but are generically 
expected \cite{RTRG_FS} (for other models in the charge fluctuation regime, corresponding terms have been
calculated in Ref.~\onlinecite{von_oppen}). For given perturbation order $k$, the allowed
values for $l$ depend on the value of $n$. For $n=0,1$ it is known that $l=0,1,\dots,k$. Even in
the weak-coupling regime
\begin{equation}
\Lambda_c\,=\,\text{max}\{V,h_0\}\,\gg\,T_K \quad,
\end{equation}
the logarithmic terms can lead to a breakdown of perturbation theory if $J_0\ln{\Lambda_c\over T_K}$ 
is small enough such that $J_0\ln{D\over\Lambda_c}={1\over 2}-J_0\ln{\Lambda_c\over T_K}\sim O(1)$.
Furthermore, at resonance $nV=mh_0$, the logarithmic terms even diverge. Therefore, it is necessary
to resum the logarithmic terms in an appropriate way using RG and, at the same time, introduce the
physics of relaxation and dephasing rates to cut off the divergencies at resonance. This idea
has been proposed in Ref.~\onlinecite{glazman_pustilnik_05}. In a first step, within a standard poor 
man scaling approach, one resums all leading order logarithmic terms of the form
\begin{equation}
\label{log_leading}
J_0^r\,\left(J_0\,\ln{D\over\Lambda_c}\right)^k\quad,\quad k=0,1,2,\dots\quad,
\end{equation}
where $r=0,1,2$ depends on the physical observable under consideration [see Eq.~(\ref{log_pert})].
This leads to an effective exchange coupling $J_c$, given by Eq.~(\ref{solution_poor_man_scaling})
evaluated at $\Lambda=\Lambda_c$:
\begin{equation}
\label{J_c}
J_c\,=\,{1\over 2\ln{\Lambda_c\over T_K}}\,=\,
{1\over 2\ln{\text{max}\{V,h_0\}\over T_K}}\quad.
\end{equation}
In a second step, one tries to expand the physical observable systematically in powers of the
effective coupling constant $J_c$, leading to a series with terms similiar to 
Eq.~(\ref{log_pert}), but with the replacements $D\rightarrow\Lambda_c$ and $J_0\rightarrow J_c$.
If, in addition, one cuts off the resonances by an appropriate relaxation or dephasing rate $\tilde{\Gamma}$, 
a new series of the form
\begin{equation}
\label{log_RG}
J_c^{k+r}\,\ln^l{\Lambda_c\over |nV-m\tilde{h}+i\tilde{\Gamma}|}\quad,
\end{equation}
is expected, where $\tilde{h}$ is the renormalized magnetic field.
As pointed out in Ref.~\onlinecite{glazman_pustilnik_05} this perturbation series in $J_c$ is 
well-defined for $J_c\ll 1$, because the maximum
value of the logarithm at resonance is given by $\ln{\Lambda_c\over\tilde{\Gamma}}\sim\ln J_c$,
where we have used the rough estimate
\begin{equation}
\tilde{\Gamma}\,\sim\,J_c^2\,\Lambda_c
\end{equation}
based on a simple dimensional analysis. Therefore, at resonance, we expect terms of the form 
\begin{equation}
\label{resonance_RG}
J_c^{k+r}\,\ln^l J_c\quad,
\end{equation}
which are not dangerous since $J_c \ln J_c\ll 1$ if $J_c\ll 1$ (note that $k\ge l$). The leading
order result is the term $\sim J_c^r$ (denoted as one-loop in this paper), whereas the first logarithmic 
correction $\sim J_c^{r+1}\ln J_c$ is the first subleading term (denoted as two-loop in this paper). 

The purpose of the present paper is to present a well-defined two-loop RG approach to work out the
above-described procedure. Thereby we will apply a recently proposed real-time renormalization
group method in frequency space (RTRG-FS).\cite{RTRG_FS} This method has the advantage that
formally exact RG equations can be set up in nonequilibrium which include the relaxation and
dephasing rates in all resolvents appearing on the right hand side (r.h.s.) of the RG equations. Furthermore, in
leading order, precisely poor man scaling equation (\ref{poor_man_scaling}) is obtained.
This provides the possibility to proceed in two steps: First one expands the exact RG equations 
systematically around the poor man scaling solution for $D>\Lambda>\Lambda_c$, and, in the 
second step, one solves the RG equations perturbatively in $J_c$ for $\Lambda_c > \Lambda >0$. 
As we will show in this paper, in both steps two-loop terms are important to obtain the first
logarithmic corrections beyond leading order. In the first step, two-loop terms arising from
higher-order terms on the r.h.s. of the RG equation generate terms $\sim J_c^{r+1}\ln {J_c\over J_0}$ 
which depend only weakly on the voltage and are incorporated into a 
redefinition of the coupling constant $J_c$ (or, equivalently, the Kondo temperature). 
These terms lead to an overall increase (or decrease) in the physical observable under 
consideration but show no interesting dependence on voltage or magnetic field. In contrast,
in the second step, logarithmic contributions of form (\ref{log_RG}) are generated which give
rise to a logarithmic enhancement (suppression) at resonance. This means that the
various two-loop terms (leading all to terms of the same order of magnitude) can be systematically
divided into important and unimportant terms concerning their dependence on voltage and magnetic field. 
In accordance with results of 
Refs.~\onlinecite{rosch_kroha_woelfle_PRL01} and~\onlinecite{rosch_paaske_kroha_woelfle_PRL03}, we emphasize that
it is very important to include the frequency dependence of the vertices generated from
the first step $D>\Lambda>\Lambda_c$ since this influences the prefactor of the 
logarithmic contributions calculated in the second step $\Lambda_c>\Lambda>0$.

In one-loop order (but including certain two-loop terms from the
frequency dependence of the vertices), 
the conductance and the magnetization have been calculated previously
for the nonequilibrium Kondo model. Pioneering works are 
Refs.~\onlinecite{rosch_kroha_woelfle_PRL01} and~\onlinecite{rosch_paaske_kroha_woelfle_PRL03}, where 
the slave particle approach was used in combination with the Keldysh formalism and
quantum Boltzmann equations. The RG was formulated only on one part of the Keldysh contour and
a real-frequency cutoff was used. In these works, it was investigated 
how the voltage and the magnetic field cut off the RG flow and it was emphasized that
the frequency dependence of the vertices is important to obtain the first logarithmic 
contributions beyond leading order. In fact, the result of the present paper concerning
the conductance and the magnetization is precisely the same as that of 
Refs.~\onlinecite{rosch_kroha_woelfle_PRL01} and~\onlinecite{rosch_paaske_kroha_woelfle_PRL03}, up to the
redefinition of the Kondo temperature. This means that we will prove in this paper that
all two-loop contributions neglected in 
Refs.~\onlinecite{rosch_kroha_woelfle_PRL01} and~\onlinecite{rosch_paaske_kroha_woelfle_PRL03}
do not influence the prefactor of the first logarithmic correction beyond
leading order. Furthermore, in 
Refs.~\onlinecite{rosch_kroha_woelfle_PRL01} and~\onlinecite{rosch_paaske_kroha_woelfle_PRL03},
diagrams connecting the upper with the lower part of the Keldysh contour have been neglected.
Within these works, it was therefore not possible to describe the cutoff of the logarithmic 
terms by relaxation and dephasing rates on a full microscopic level. In this paper, we will 
show how this can be achieved within RTRG-FS, which provides a consistent formalism to calculate
the line shape at resonance and to see whether
the spin relaxation rate $\tilde{\Gamma}_1$ or the spin dephasing rate $\tilde{\Gamma}_2$ 
cuts off the logarithmic divergencies (in Ref.~\onlinecite{paaske_rosch_kroha_woelfle_PRB04}, the
latter question was adressed only in bare perturbation theory and zero magnetic field).
In addition, in this paper we will also discuss the anisotropic case and calculate the spin 
relaxation/dephasing rates together with the renormalized $g$ factor up to the first 
logarithmic contribution (corresponding to a three-loop calculation in conventional RG methods).
Besides the known reduction in the magnetic field in first order in
$J$,\cite{garst_etal_PRB05} 
we find that the renormalized magnetic field in second order in $J$ is proportional to 
logarithmic terms similiar to Eq.~(\ref{log_RG}) with a significant dependence on voltage and
magnetic field. We propose an experimental setup with a weakly coupled third lead to measure
the voltage dependence of the renormalized $g$ factor. Moreover, we find that the logarithmic 
terms of $\tilde{\Gamma}_2$ and $\tilde{g}$ are
controlled by $\tilde{\Gamma}_1$, whereas those of $\tilde{\Gamma}_1$ are controlled by
$\tilde{\Gamma}_2$. In the anisotropic case this leads to the effect that the logarithmic
resonances of ${d\tilde{\Gamma}_2\over h_0}$ become sharper with decreasing $J^\perp$ since 
$\tilde{\Gamma}_1$ does not contain any terms proportional to $J^z$ in
second order.
Furthermore, we will show that the susceptibility depends only weakly on the tranverse coupling
$J^\perp$ and the logarithmic resonances even survive in the limit $J^\perp\rightarrow 0$.
The anisotropic Kondo model has recently been proposed to be realizable in low-temperature
transport through single molecular magnets \cite{SMM_theory} and experiments are starting
to investigate such systems.\cite{SMM_experiment} Since the transverse coupling is induced
by small magnetic quantum tunneling terms, giving rise to rather small Kondo temperatures, the
susceptibility might be an interesting physical quantity to measure signatures of the Kondo 
effect even for very small values of $J^\perp$. 

Using flow equation methods,\cite{flow_eq} a consistent two-loop approach including
the cutoff by spin relaxation/dephasing rates has been presented in Ref.~\onlinecite{kehrein_PRL05}
for the isotropic Kondo model in the absence of a magnetic field. Within this
method, the cutoff from the rate $\tilde{\Gamma}$ occurs due to a competition of certain
one-loop and two-loop terms on the r.h.s. of the RG equation for the vertex. This is
a completely different picture compared to RTRG-FS, where the cutoff
parameter $\tilde{\Gamma}$ together with the voltage occurs already in the one-loop terms as an additional
term in the denominator of the resolvents. Thus, the RTRG-FS method is closer
to conventional poor man scaling and the physics of relaxation
and dephasing rates occurs naturally as a resummation of a geometric
series similiar to self-energy insertions in Green's function techniques. In this sense
the RTRG-FS method proves that conventional scaling equations (properly generalized to
the Keldysh contour) can account for the cutoff of the RG flow by rates and the voltage.
Furthermore, the RTRG-FS method provides a generic proof in all orders of perturbation
theory in the renormalized vertices that the various cutoff scales
$\tilde{\Gamma}_k$ are the physical relaxation and dephasing rates governing the time evolution
of the reduced density matrix of the quantum system. 

The RTRG-FS method used in this paper has been proposed in Ref.~\onlinecite{RTRG_FS} and
is a natural generalization of an earlier developed real-time RG 
method.\cite{hs_koenig_PRL00,hs_lecture_notes_00,korb_reininghaus_hs_koenig_PRB07} 
As described in detail in Ref.~\onlinecite{RTRG_FS}, many technical improvements have
been incorporated, the main ones being a formulation of the RG in pure frequency
space, integrating out the symmetric part of the Fermi distribution function
before starting the RG, and formulating the nonequilibrium RG on the imaginary 
frequency axis.
As a consequence, the rates determining the cutoff of the RG flow obtain the right scale,
and it is possible to show generically that relaxation and dephasing rates cut off the 
RG flow in all orders of perturbation theory and within all truncation schemes.
Furthermore, the dependence on the Keldysh indices can be completely avoided, 
and one can calculate the time evolution and the nonequilibrium stationary
state in pure Matsubara space without the need of any analytic continuation.
The latter idea has first been proposed in Ref.~\onlinecite{jakobs_meden_hs_PRL07} in 
the context of nonequilibrium functional renormalization group within 
the Keldysh formalism. A particular advantage of the RTRG-FS method is that
relaxation and dephasing rates occur naturally as the negative imaginary part
of the eigenvalue of the kernel determining the kinetic equation of the reduced
density matrix, and do not arise from more involved combinations of self-energies
and vertex corrections as in slave particle
formalism.\cite{paaske_rosch_kroha_woelfle_PRB04} Furthermore, it is straightforward
to calculate the time evolution from RTRG-FS since the RG gives directly the
result for the kernel in Laplace space. 

For generic problems with spin and/or orbital fluctuations, it was described in 
Ref.~\onlinecite{RTRG_FS} how to solve the RG equations analytically
in the weak coupling regime up to one-loop order. In this paper we will provide the
technically much more involved two-loop case, which is necessary to calculate
consistently the important logarithmic terms at resonance discussed above, see
Eq.~(\ref{log_RG}). The result will be applied to the calculation of the conductance
and the magnetic susceptibility of the anisotropic Kondo model in
the presence of a magnetic field. Furthermore, we will also analyze
quantities characterizing the time evolution of the Kondo
model. Thereby, we will concentrate on the calculation of
the dominant exponential decay of the magnetization in two-loop order, which is determined by the
spin relaxation and dephasing rates $\tilde{\Gamma}_{1,2}$ and the renormalized
magnetic field $\tilde{h}$. Finally, in addition to Ref.~\onlinecite{RTRG_FS},
we will generically show that precisely these physical quantities control all resonant 
line shapes (in Ref.~\onlinecite{RTRG_FS}, the question whether the
cutoff scales of the logarithmic terms are exactly identical to the physical 
relaxation/dephasing rates was still open). 

The paper is organized as follows. In Sec.~\ref{sec:generic_model}, we will set up
the generic model and the perturbative series. Section \ref{sec:rg_equations} is devoted
to the general RG formalism and the derivation of the two-loop RG equations for an 
arbitrary quantum dot coupled via 
spin and/or orbital fluctuations to reservoirs. The systematic way to analytically solve these
RG equations up to two-loop order is presented in Sec.~\ref{sec:generic_2_loop}.
The general formalism is applied to the nonequilibrium Kondo model in
Sec.~\ref{sec:kondo}. In Sec.~\ref{sec:kondo_model_algebra}, we will set up
the algebra in Liouville space needed to evaluate all expressions explicitly, and
Sec.~\ref{sec:kondo_2_loop} describes the evaluation of the general two-loop 
equations for the Kondo model. The final results for the conductance, the
magnetic susceptibility, the spin relaxation and dephasing rate, and the renormalized
$g$~factor are presented for the isotropic case in
Secs.~\ref{sec:relaxation_dephasing_rates}, \ref{sec:magnetization}
and \ref{sec:current}, whereas
the anisotropic case is discussed in section~\ref{sec:kondo_results_anisotropic}.
We close with a summary in Sec.~\ref{sec:conclusion}. A list of all symbols used in this 
papers is presented in Sec.~\ref{sec:list_of_symbols}. 

We emphasize that a reader, who is not interested in the formal derivation but only in
the final physical results of the paper, can skip the following formal
Secs.~\ref{sec:generic} and \ref{sec:kondo},  
and can directly move over to Sec.~\ref{sec:kondo_results}, where the notations needed to
understand the results are again repeated.

\section{Generic case}
\label{sec:generic}
In this section, we describe a generic quantum dot coupled via spin and/or orbital
fluctuations to several reservoirs. In Secs.~\ref{sec:generic_model} and 
\ref{sec:rg_equations}, we introduce the basic notations, the perturbative series and 
summarize shortly the setup of the RG equations as explained in
more detail in Ref.~\onlinecite{RTRG_FS}. In Sec.~\ref{sec:generic_2_loop}, we
present a systematic way how to solve the RG equations analytically
up to two-loop order in the weak-coupling regime (the one-loop case has been treated
in Ref.~\onlinecite{RTRG_FS}). Throughout this paper, we use units $e=\hbar=k_B=1$.

\subsection{Model and perturbative series}
\label{sec:generic_model}
{\it Model.} We consider a quantum dot with fixed charge in the Coulomb blockade regime where only
spin and/or orbital fluctuations are possible via the coupling to external reservoirs.
As shown in detail in Ref.~\onlinecite{korb_reininghaus_hs_koenig_PRB07}, a standard Schrieffer-Wolff
transformation leads to a Hamiltonian of the form
\begin{equation}
\label{H_total}
H\,=\,H_{res}\,+\,H_S\,+\,V\,=\,H_0\,+\,V \quad,
\end{equation}
where $H_{res}$ is the reservoir part, $H_S$ characterizes the isolated quantum dot, and
$V$ describes the coupling between reservoirs and quantum dot. They
are given explicitly by
\begin{eqnarray}
\label{H_res}
H_{res}\,&=&\,\sum_{\nu\equiv \alpha\sigma\dots}
\int d\omega\,(\omega+\mu_\alpha)\,a_{+\nu}(\omega)a_{-\nu}(\omega)\quad,
\label{H_S}\\
H_S\,&=&\,\sum_s\,E_s\,|s\rangle\langle s|\quad,\\
\label{coupling}
V\,&=&\,{1\over 2}\,\sum_{\eta\eta'}\sum_{\nu\nu'}\int d\omega \int d\omega'\\
\nonumber
&&\hspace{1cm}\,
g_{\eta\nu,\eta'\nu'}(\omega,\omega'):a_{\eta\nu}(\omega)\,a_{\eta'\nu'}(\omega'): \quad.
\end{eqnarray}
Here, $a_{\eta\nu}$ are the creation ($\eta=+$) and annihilation ($\eta=-$) operators of the
reservoirs, and $\nu$ is an index characterizing all quantum numbers of the
reservoir states. In the absence of further symmetries, $\nu$ contains the
reservoir index $\alpha$ and the spin quantum number $\sigma$ (for two 
reservoirs and spin-${1\over 2}$ particles, we use the notation
$\alpha\equiv L,R\equiv \pm$ and $\sigma\equiv \uparrow,\downarrow\equiv\pm$).
$\omega$ is the energy of the reservoir state relative to the chemical potential
$\mu_\alpha$ of reservoir $\alpha$. The eigenstates and eigenenergies of the isolated
quantum dot are denoted by $|s\rangle$ and $E_s$. The interaction $V$ is quadratic
in the reservoir field operators, which arises from second-order processes of one
electron hopping off and on the quantum dot coherently (for negative charging
energies, also two electrons can hop off or on the dot \cite{von_oppen_negative_U}). 
This keeps the charge
fixed and allows only spin and orbital fluctuations. The coupling vertex
$g_{\eta\nu,\eta'\nu'}(\omega,\omega')$ is an arbitrary operator acting on the dot states.
It is written in its most general form, depending on the quantum numbers and
energies of the reservoir states in an arbitrary way. However, as explained in
Ref.~\onlinecite{RTRG_FS}, the RG approach can be set up in its most convenient form
if one assumes that the frequency dependence of the initial vertices is rather
weak and varies on the scale of the band width $D$ of the reservoirs. Therefore,
we will assume this in the following and introduce below [see Eq.~(\ref{cutoff_contraction})]
a convenient cutoff function into the free reservoir Green's functions.

To achieve a more compact notation for all indices, we write $1\equiv \eta\nu\omega$
and sum (integrate) implicitly over all indices and frequencies. The interaction is
then written in the compact form
\begin{equation}
\label{spin_orbital}
V\,=\,{1\over 2}\,g_{11'}\,:a_1 a_{1'}:\quad.
\end{equation}
$:\dots:$ denotes normal ordering of the reservoir field operators, meaning that no
contraction is allowed between reservoir field operators within the normal ordering.
A contraction is defined with respect to a grand-canonical distribution of the
reservoirs, given by
\begin{equation}
\label{contraction}
{a_1\,a_{1'}
  \begin{picture}(-20,11) 
    \put(-22,8){\line(0,1){3}} 
    \put(-22,11){\line(1,0){12}} 
    \put(-10,8){\line(0,1){3}}
  \end{picture}
  \begin{picture}(20,11) 
  \end{picture}
}
\,\equiv\,
\langle a_1 a_{1'}\rangle_{\rho_{res}}
\,=\,\delta_{1\bar{1}'}\,f_\alpha(\eta\omega)\quad.
\end{equation}
$f_\alpha(\omega)=(e^{\omega/T_\alpha}+1)^{-1}=1-f_\alpha(-\omega)$ is the Fermi distribution
function corresponding to temperature $T_\alpha$ (note that the chemical 
potential does not enter this formula since $\omega$ is measured relative
to $\mu_\alpha$). Furthermore,
$\delta_{11'}\equiv\delta_{\eta\eta'}\delta_{\nu\nu'}\delta(\omega-\omega')$ is
the $\delta$ function in compact notation, and $\bar{1}\equiv -\eta,\nu,\omega$.
The cutoff by the band width $D$ can be introduced in many different ways into
the reservoir contraction. We use a Lorentzian cutoff and replace the contraction by
\begin{equation}
\label{cutoff_contraction}
{a_1\,a_{1'}
  \begin{picture}(-20,11) 
    \put(-22,8){\line(0,1){3}} 
    \put(-22,11){\line(1,0){12}} 
    \put(-10,8){\line(0,1){3}}
  \end{picture}
  \begin{picture}(20,11) 
  \end{picture}
}
\,\rightarrow\,
\delta_{1\bar{1}'}\,\rho(\omega)\,f_\alpha(\eta\omega)\quad,
\end{equation}
with
\begin{equation}
\label{band_cutoff}
\rho(\omega)\,=\,{D^2\over D^2 + \omega^2}\quad.
\end{equation}

Within the normal ordering of Eq.~(\ref{spin_orbital}), the field operators can be 
arranged in an arbitrary way (up to a fermionic sign), therefore the coupling vertex can
always be chosen such that antisymmetry holds,
\begin{equation}
\label{antisymmetry}
g_{11'}\,=\,-\,g_{1'1}\quad.
\end{equation}
Furthermore, due to the hermiticity of $V$, the vertex has the property
\begin{equation}
\label{hermiticity}
g_{11'}^\dagger\,=\,g_{\bar{1}'\bar{1}}\quad.
\end{equation}

The particle current operator flowing from reservoir $\gamma$ to the quantum dot 
is defined by $I^\gamma=-{d\over dt}N^\gamma_{res}=-i[H,N^\gamma_{res}]$, 
where $N^\gamma_{res}$ is the particle number in reservoir $\gamma$. Using 
Eqs.~(\ref{H_total}) and (\ref{coupling}), a straightforward calculation leads to
\begin{equation}
\label{current_operator}
I^\gamma\,=\,{1\over 2}\,i^\gamma_{11'}\,:a_1 a_{1'}:\quad,
\end{equation}
with
\begin{eqnarray}
\label{current_vertex}
i^\gamma_{11'}\,&=&\,-2i\,c^\gamma_{11'}\,g_{11'}\quad,\\
\label{c_symbol}
c^\gamma_{11'}\,&=&\,-{1\over 2}(\eta\delta_{\alpha\gamma}
\,+\,\eta'\delta_{\alpha'\gamma})\quad.
\end{eqnarray}

We are interested in the time evolution of the reduced density matrix $\rho_S(t)$ of the
quantum dot and in the average $\langle I^\gamma \rangle(t)$ of the current operator.
Formally, they follow from the solution of the von Neumann equation
\begin{eqnarray}
\label{rho_S_formal}
\rho_S(t)\,&=&\,
\mbox{Tr}_{res}\,e^{-iL(t-t_0)}\,\rho_S(t_0)\,\rho_{res}\quad,\\
\label{current_formal}
\langle I^\gamma \rangle(t) \,&=&\,
\mbox{Tr}_S\,\mbox{Tr}_{res}\,(-iL_{I^\gamma})e^{-iL(t-t_0)}\,\rho_S(t_0)\rho_{res}\quad,
\end{eqnarray}
where
\begin{equation}
\label{L_LI}
L\,=\,[H,\cdot]_- \quad,\quad
L_{I^\gamma}\,=\,{i\over 2}\,[I^\gamma,\cdot]_+ 
\end{equation}
are operators in Liouville space acting on usual operators in Hilbert space via
the (anti)commutator $[A,B]_\pm=AB\pm BA$. Initially, we have assumed that the
density matrix is a product of an arbitrary dot part $\rho_S(t_0)$ and a 
grandcanonical distribution $\rho_{res}$ for the reservoirs. It is convenient
to introduce the Laplace transform
\begin{eqnarray}
\nonumber
\tilde{\rho}_S(E)\,&=&\,\int_{t_0}^\infty dt\,e^{iE(t-t_0)}\,\rho_S(t)\\
\label{laplace_rho}
&=&\,\mbox{Tr}_{res}\,{i\over E-L}\,\rho_S(t_0)\rho_{res}\\
\nonumber
\tilde{\langle I^\gamma \rangle}(E)\,&=&\,\int_{t_0}^\infty dt\,e^{iE(t-t_0)}\,
\langle I^\gamma \rangle (t)\\
\label{laplace_current}
&=&\mbox{Tr}_S\,\mbox{Tr}_{res}\,L_{I^\gamma}\,{1\over E-L}\,\rho_S(t_0)\rho_{res}\quad.
\end{eqnarray}

\begin{figure}
  \includegraphics[scale=0.3]{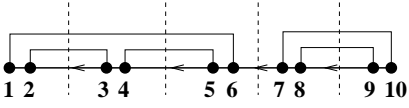}
  \caption{Example of a diagram for the reduced density matrix of the dot. The time
direction is to the left. The dots represent the interaction vertex $G$ between
the local quantum system and the reservoirs. Each vertex corresponds to two adjacent dots
indicating that two reservoir field operators are associated with each vertex.
The horizontal lines connecting the
vertices denote the free time propagation of the quantum system, leading to the resolvents
${1\over E+X_i-L_S^{(0)}}$ in Laplace space. The lines
connecting the dots are the reservoir contractions 
arising from the application of Wick's theorem. The dashed vertical lines between the vertices 
are auxiliary lines to determine the energy argument $X_i$ of the resolvents.}
\label{fig:diagram_example}
\end{figure}

{\it Perturbative expansion.} The next step is to expand expressions (\ref{laplace_rho}) and
(\ref{laplace_current}) in the interacting part $L_V=[V,\cdot]_-$
of the Liouvillian and to integrate out the reservoir part. As outlined in detail in
Ref.~\onlinecite{RTRG_FS}, this leads to a diagrammatic representation in Liouville space. We
shortly summarize this procedure here. First, $L_V$ can be written in the
form
\begin{equation}
\label{coupling_product}
L_V\,=\,{1\over 2}\,p'\,G^{pp'}_{11'}\,
:J^{p}_1 J^{p'}_{1'}:\quad,
\end{equation}
where $J^p_1$ is a quantum field superoperator in Liouville space for the reservoirs, 
defined by ($A$ is an arbitrary reservoir operator)
\begin{equation}
\label{liouville_field_operators}
J_1^p\,A \,=\,
\left\{
\begin{array}{cl}
a_1\,A\, &\mbox{for }p=+ \\
A\,a_1\, &\mbox{for }p=-
\end{array}
\right.\quad.
\end{equation}
$p=\pm$ is the Keldysh index indicating whether the field operator is acting on the
upper or the lower part of the Keldysh contour. $G^{pp'}_{11'}$ is
a superoperator acting in Liouville space of the quantum dot, and is defined
by ($A$ is an arbitrary operator of the quantum dot)
\begin{equation}
\label{G_vertex_liouville}
G^{pp'}_{11'}\,A\,=\,
\delta_{pp'}\,
\left\{
\begin{array}{cl}
g_{11'}\,A\, &\mbox{for }p=+ \\
-A \,g_{11'}\, &\mbox{for }p=-
\end{array}
\right.\quad.
\end{equation}
Inserting the form (\ref{coupling_product}) into Eqs.~(\ref{laplace_rho}) and (\ref{laplace_current}),
expanding in $L_V$, and shifting all reservoir field superoperators $J_1^p$ to the right, one
can show that each term of perturbation theory can be written as a product of a dot part and
an average over a sequence of field superoperators of the reservoirs with respect to $\rho_{res}$.
Evaluating the latter with the help of Wick's theorem, one can represent each
term of the Wick decomposition by a diagram, see, e.g., Fig.~\ref{fig:diagram_example}
describing a certain process for the time evolution of the reduced density matrix
of the dot. 
Each process consists of a sequence of interaction vertices $G_{11'}^{pp'}$ between the
dot and the reservoirs, and a free time propagation of the dot in between (leading to
resolvents in Laplace space). Since the reservoirs have been integrated out, the vertices 
are connected by reservoir contractions (the solid lines without
arrows in Fig.~\ref{fig:diagram_example}).
This means that the various diagrams represent terms for the effective time evolution of 
the dot in the presence of dissipative reservoirs. Each diagram for the reduced density matrix 
has the form
\begin{eqnarray}
\nonumber
\tilde{\rho}_S(E) &\,\rightarrow\,&\\
\nonumber && \hspace{-1.5cm}
{i\over S} \, (-1)^{N_p} \, \left(\prod\gamma\right)
\,{1\over E-L_S^{(0)}}\,G\,{1\over E+X_1-L_S^{(0)}}\,\\
\label{value_diagram}
&& \hspace{-1.5cm}
\cdot\,G\,\dots \,G\,
{1\over E+X_r-L_S^{(0)}}\,G\,{1\over E-L_S^{(0)}}\,\rho_S(t_0)
\,\,,
\end{eqnarray}
where 
\begin{equation}
\label{L_S_(0)}
L_S^{(0)}=[H_S,\cdot]_-\quad,
\end{equation}
$G\equiv G_{ij}^{p_i p_j}$ indicates an interaction vertex, and
$\gamma\equiv\gamma^{p_i p_j}_{ij}$ is a contraction between the reservoir field 
superoperators, defined by 
\begin{eqnarray}
\nonumber
\gamma_{11'}^{pp'}\,&=&\,{J_1^p\,J_{1'}^{p'}
  \begin{picture}(-20,11) 
    \put(-22,8){\line(0,1){3}} 
    \put(-22,11){\line(1,0){12}} 
    \put(-10,8){\line(0,1){3}}
  \end{picture}
  \begin{picture}(20,11) 
  \end{picture}
}
\,=\,p'\,\mbox{Tr}_{res}\,J_1^p J_{1'}^{p'}\,\rho_{res}\\
\label{liouville_contraction}
\,&=&\,\delta_{1\bar{1}'}\,\rho(\omega)\,p'
\,f_\alpha(\eta p' \omega) \,\,.
\end{eqnarray}
To factorize the
Wick decomposition, a fermionic sign has to be assigned to each permutation of reservoir
field superoperators, indicated by the sign factor $(-1)^{N_p}$ in Eq.~(\ref{value_diagram}).
For each pair of vertices connected by two reservoir lines, a combinatorical factor ${1\over 2}$
occurs, leading to the prefactor ${1\over S}$ in Eq.~(\ref{value_diagram}). The value of the  
frequencies $X_i$ in the resolvents between the interaction vertices is determined by
the sum over all variables $x=\eta(\omega+\mu_\alpha)$ of those indices belonging to the
reservoir lines which are crossed by a vertical line at the position of the resolvent (see
the dashed lines in Fig.~\ref{fig:diagram_example}). Thereby, the index of the left vertex has
to be taken of the corresponding reservoir line, e.g., the diagram of Fig.~\ref{fig:diagram_example}
is given by (the obvious dependence on the Keldysh indices has been omitted for simplicity,
i.e., $\gamma_{ij}\equiv\gamma_{ij}^{p_i p_j}$ and $G_{ij}\equiv G_{ij}^{p_i p_j}$) 
\begin{eqnarray}
\nonumber
&&i\,{1\over E-L_S^{(0)}}
\left({\over}
\gamma_{16}\gamma_{23}\gamma_{45}\,
G_{12}\,\Pi_{12}\,G_{34}\,\Pi_{14}\,G_{56}\right)
{1\over E-L_S^{(0)}}\\
\label{diagram_example}
&& \hspace{0.5cm}
\left({1\over 2}
\gamma_{7,10}\gamma_{89}\,
G_{78}\,\Pi_{78}\,G_{9,10}\right)
{1\over E-L_S^{(0)}}\rho_S(t_0)\quad,
\end{eqnarray}
where the resolvents are defined by
\begin{equation}
\label{pi_bare}
\Pi_{1\dots n}\,=\,{1\over E_{1\dots n}+\bar{\omega}_{1\dots n}-L_S^{(0)}}\quad,
\end{equation}
with
\begin{eqnarray}
\label{E_omega_definition}
E_{1\dots n}\,&=&\,E\,+\,\sum_{i=1}^n\,\bar{\mu}_i\quad,\quad
\bar{\omega}_{1\dots n}\,=\,\sum_{i=1}^n\,\bar{\omega}_i\quad,\\
\label{mu_omega_bar}
\bar{\mu}_i\,&=&\,\eta_i\,\mu_{\alpha_i}\quad,\quad
\bar{\omega}_i\,=\,\eta_i\,\omega_i \quad.
\end{eqnarray}
As can be seen from example (\ref{diagram_example}), each diagram consists of a sequence
of irreducible blocks (where a vertical line always cuts at least one reservoir line) and
free resolvents $1/(E-L_S^{(0)})$ in between. Similiar to Dyson equations one can formally resum
this series with the result
\begin{equation}
\label{reduced_dm}
\tilde{\rho}_S(E)\,=\,{i\over E-L_S^{eff}(E)}\,\rho_S(t_0)\quad,
\end{equation}
with 
\begin{equation}
\label{L_eff}
L_S^{eff}(E)\,=\,L_S^{(0)}\,+\,\Sigma(E)\quad,
\end{equation}
where the kernel $\Sigma(E)$ contains the sum over all irreducible diagrams. A similiar
procedure can be used to calculate the average (\ref{laplace_current}) of the current
operator with the result
\begin{eqnarray}
\nonumber
\tilde{\langle I^\gamma \rangle}(E)\,&=&\,\text{Tr}_S\,\Sigma_\gamma(E)\,
{1\over E-L_S^{eff}(E)}\,\rho_S(t_0)\\
\label{current}
&=&\,-i\,\text{Tr}_S\,\Sigma_\gamma(E)\,\tilde{\rho}_S(E)\quad,
\end{eqnarray}
where the current kernel $\Sigma_\gamma(E)$ is defined similiarly to $\Sigma(E)$, but the first vertex $G$
is replaced by the current vertex $I^\gamma$, defined by
\begin{equation}
\label{current_liouvillian_vertex}
(I^\gamma)^{pp'}_{11'}\,=\,c^\gamma_{11'}\,\delta_{pp'}\,p\,G^{pp}_{11'}\quad,
\end{equation}
such that, in analogy to Eq.~(\ref{coupling_product}),
\begin{equation}
\label{current_decomposition}
L_{I^\gamma}\,=\,{1\over 2}\,p'\,(I^\gamma)^{pp'}_{11'}\,
:J^{p}_1 J^{p'}_{1'}:\quad.
\end{equation}

Using Eq.~(\ref{value_diagram}), a certain diagram for the kernels has to be translated
according to
\begin{eqnarray}
\nonumber
\left\{
\begin{array}{cl}
\Sigma(E)
\\
\Sigma_\gamma(E)
\end{array}
\right\}
\,&\rightarrow&\, 
{1\over S} \, (-1)^{N_p} \, \left(\prod\gamma\right)_{irr}\\
\label{value_kernel}
&& \hspace{-3cm}
\left\{
\begin{array}{cl}
G \\ I^\gamma
\end{array}
\right\}
\,{1\over E+X_1-L_S^{(0)}}\,G\,\dots \,G\,
{1\over E+X_r-L_S^{(0)}}\,G\,\,,
\end{eqnarray} 
where the subindex $irr$ indicates that only irreducible diagrams are allowed where
any vertical line between the vertices cuts through at least one reservoir contraction.

The stationary solutions for the reduced density matrix and the current follow from the
Laplace transform by
$\rho^{st}_S=\lim_{E\rightarrow 0^+}(-iE)\tilde{\rho}_S(E)$ and 
$\langle I^\gamma\rangle^{st}=\lim_{E\rightarrow 0^+}(-iE)\langle\tilde{I}^\gamma\rangle(E)$, 
and can be calculated from
\begin{eqnarray}
\label{rd_stationary}
L_S^{eff}(i0^+)\,\rho_S^{st}\,&=&\,0\quad,\\
\label{current_stationary}
\langle I^\gamma\rangle^{st}\,&=&\,
-i\,\text{Tr}_S\,\Sigma_\gamma(0^+)\,\rho^{st}_S\quad.
\end{eqnarray}

In addition to previous formulations \cite{RTRG_FS} of the perturbation series, we note
that the diagrammatic series can be partially resummed by taking all closed subdiagrams between 
two fixed vertices together which contain only contractions connecting vertices between the two fixed
ones. This has the effect that the resolvents in Eq.~(\ref{value_kernel}) are replaced by 
\begin{equation}
\label{resolvent_replacement}
{1\over E+X_i-L_S^{(0)}}\rightarrow
{1\over E+X_i-L_S^{eff}(E+X_i)}\quad,
\end{equation}
i.e., the full effective Liouville operator occurs in the denominator. This means that 
Eqs.~(\ref{L_eff}) and (\ref{value_kernel}) turn into self-consistent equations for $L_S^{eff}(E)$
for any approximation. Of course, the number of diagrams is reduced in this
formulation. No diagrams are allowed anymore which contain closed subdiagrams between two 
vertices.

When calculating diagrams with the replacement (\ref{resolvent_replacement}), one faces the
problem that the frequency integrations cannot be performed analytically because the energy dependence
of the effective Liouvillian is not known. This would require the solution of a complicated
self-consistent integral equation. To avoid this, it is useful to formulate an appropriate
approximation for the resolvents which can be improved systematically. 
To define this approximation, we write the resolvents
in terms of the eigenvectors and eigenvalues of the Liouvillian $L_S^{eff}(z)$,
\begin{equation}
\label{pi_spectral_representation}
\Pi(z)={1\over z-L_S^{eff}(z)}=\sum_i{1\over z-\lambda_i(z)}\,P_i(z)
\end{equation}
where the projectors are defined by
\begin{equation}
\label{def_projectors}
P_i(z)\,=\,|x_i(z)\rangle\langle \bar{x}_i(z)|\quad,
\end{equation}
and $|x_i(z)\rangle$ and $\langle\bar{x}_i(z)|$ are the right and left eigenvectors of $L_S^{eff}(z)$,
\begin{eqnarray}
\label{def_right_eigenvectors}
L_S^{eff}(z)|x_i(z)\rangle\,&=&\,\lambda_i(z)|x_i(z)\rangle\quad,\\
\label{def_left_eigenvectors}
\langle\bar{x}_i(z)|L_S^{eff}(z)\,&=&\,\lambda_i(z)\langle\bar{x}_i(z)|\quad,
\end{eqnarray}
with eigenvalues $\lambda_i(z)$.
Assuming that $|x_i(z)\rangle$ and $\langle\bar{x}_i(z)|$ have no poles (or poles with very large 
negative imaginary part so that they influence only the short-time behaviour), the poles $z_i$ of the 
resolvent follow from the self-consistent equation
\begin{equation}
\label{pole_equation}
z_i\,=\,\lambda_i(z_i)
\end{equation}
for all values of $i$. Expanding $\lambda_i(z)$, $|x_i(z)\rangle$, and $\langle\bar{x}_i(z)|$
around $z=z_i$, we see that the nonanalytic part of the resolvent is given by
\begin{equation}
\label{Pi_nonanalytic_part}
\Pi(z)\approx\sum_i{a_i\over z-z_i}\,P_i(z_i)\quad,
\end{equation}
with residua (also called $Z$ factors) given by
\begin{equation}
\label{residuum}
a_i\,=\,{1\over 1-{d\lambda_i\over dz}(z_i)}\quad.
\end{equation}
Equation~(\ref{Pi_nonanalytic_part}) defines our approximation which is the appropriate one to 
describe especially line shapes at resonance (analytic parts are expected to have no special
features at resonance and will only lead to an overall perturbative shift of the background). 
To avoid the summation index $i$, we will write the approximation in the more compact form
\begin{equation}
\label{Pi_general_approximation}
\Pi(z)\,\approx \,{\tilde{Z}\over z-\tilde{L}_S}\quad,
\end{equation}
where we use the convention that any function of $\tilde{Z}$ and $\tilde{L}_S$ is interpreted as
\begin{equation}
\label{tilde_LZ_convention}
f(\tilde{Z},\tilde{L}_S)\,\equiv\,
\sum_i f({1\over 1-{d\lambda_i\over dz}(z_i)},z_i)\,P_i(z_i) \quad.
\end{equation}
The eigenvalues $z_i$ can be decomposed into real and imaginary parts
\begin{equation}
\label{effective_h_Gamma}
z_i\,=\,\tilde{h}_i\,-\,i\tilde{\Gamma}_i\quad,\quad 
\tilde{\Gamma}_i\,>\,0\quad,
\end{equation}
and are the poles of the original full resolvent $\Pi(z)$. According to Eq.~(\ref{reduced_dm}), 
this resolvent describes the reduced density matrix in Laplace space. Therefore, the resolvent must
be analytic in the upper half of the complex plane since otherwise solutions would
exist in time space which are exponentially increasing and no stationary state can
be reached. Thus, the negative imaginary parts $\tilde{\Gamma}_i$ must be strictly
positive and describe the various relaxation and dephasing rates of the different modes
described by the eigenvectors $|x_i(z_i)\rangle$. Correspondingly, the real parts
$\tilde{h}_i$ describe the oscillation frequencies of the modes, e.g., the effective 
magnetic field for the Kondo problem.

The renormalization group treatment described in the next section can be set up within
the original perturbation series (\ref{value_kernel}) or the partially resummed series
using the replacement (\ref{resolvent_replacement}).
Since we aim at a weak coupling expansion, the partially resummed series makes only
sense if the full Liouvillian is expanded in the same parameter as the renormalized
vertices. Therefore, we will make use of the resummed series only at the end of the
RG flow where perturbation theory in the renormalized couplings at a fixed physical
cutoff scale can be used. For other problems like quantum dots in the charge fluctuation
regime or systems in the strong coupling regime, it might be helpful to use the resummed
series from the very beginning.

Finally, we note some useful symmetry properties for the vertices and the Liouvillian 
(see Ref.~\onlinecite{RTRG_FS} for the proof),
\begin{eqnarray}
\label{G_symmetry}
\bar{G}_{12}\,&=&\,-\,\bar{G}_{21}\,,\\
\label{I_symmetry}
\bar{I}^\gamma_{12}\,&=&\,-\,\bar{I}^\gamma_{21}\,,\\
\label{L_property}
\mbox{Tr}_S\,L_S^{eff}(z)\,&=&\,0\,,\\
\label{G_property}
\mbox{Tr}_S\,\bar{G}_{12}\,&=&\, 0 \,,\\
\label{L_c_transform}
L_S^{eff}(z)^c\,&=&\,-L_S^{eff}(-z^*) \,,\\
\label{Sigma_I_c_transform}
\Sigma_\gamma(z)^c\,&=&\,-\Sigma_\gamma(-z^*) \,,\\
\label{G_c_transform}
(\bar{G}_{12})^c\,&=&\,-\,\bar{G}_{\bar{2}\bar{1}}(-z^*)\,,\\
\label{I_c_transform}
(\bar{I}^\gamma_{12})^c\,&=&\,-\,\bar{I}^\gamma_{\bar{2}\bar{1}}(-z^*)\,,
\end{eqnarray}
where 
\begin{equation}
\label{vertex_bar}
\bar{G}_{11'}\,=\,\sum_p\,G^{pp}_{11'}\quad,\quad
\bar{I}^\gamma_{11'}\,=\,\sum_p\,(I^{\gamma})^{pp}_{11'}\quad,
\end{equation}
and the $c$ transform $A^c$ of any dot operator $A$ in Liouville space is defined by
\begin{equation}
\label{c_transformation}
(A^c)_{ss',\bar{s}\bar{s}'}\,=\,A_{s's,\bar{s}'\bar{s}}^*\quad.
\end{equation}
Properties (\ref{L_c_transform}) and (\ref{Sigma_I_c_transform})
are important to show in time space that the reduced density matrix of the dot 
stays hermitian and the current stays real. Property (\ref{L_property})
leads to conservation of probability, i.e., the normalization of the reduced
density matrix stays constant. From this property it also follows that 
$L_S^{eff}(z)$ has an eigenvector with zero eigenvalue:
\begin{equation}
\label{right_zero_eigenvector}
L_S^{eff}(z)\,|x_0(z)\rangle \,=\,0\quad.
\end{equation}
This eigenvector corresponds to the stationary state for
$z\rightarrow i0^+$ and depends on the physical system under consideration. In
contrast, the corresponding left eigenvector is unique and, according to
Eq.~(\ref{L_property}), is given by
\begin{equation}
\label{left_eigenvector_zero}
\langle \bar{x}_0(z)|ss'\rangle\,=\,\delta_{ss'}\quad.
\end{equation}
As a consequence, in combination with property (\ref{G_property}),
we obtain zero if the left eigenvector for zero eigenvalue acts from the left on the vertex
averaged over the Keldysh indices:
\begin{equation}
\label{zero_eigenvector_G_property}
\langle \bar{x}_0(z)|\,\bar{G}_{12}\,=\,0\quad.
\end{equation}
Therefore, by decomposing the vertex according to 
\begin{equation}
\label{vertex_decomposition}
G_{11'}^{pp}\,=\,{1 \over 2}(\bar{G}_{11'}+p\tilde{G}_{11'})\quad,\quad
\tilde{G}_{11'}\,=\,\sum_p\,p\,G^{pp}_{11'}\quad,
\end{equation}
we see that the zero eigenvalue of $L_S^{eff}(z)$ can only occur in the
resolvents when the part $\tilde{G}_{11'}$ of the vertex is standing right
to the resolvent. Therefore, to avoid this zero eigenvalue in the RG treatment, 
we will first use a certain perturbative treatment to eliminate
the part $\tilde{G}_{11'}$ of the vertex from the very beginning. 
This is described in the next section.

\subsection{RG equations}
\label{sec:rg_equations}

\begin{figure}
  \includegraphics[scale=0.3]{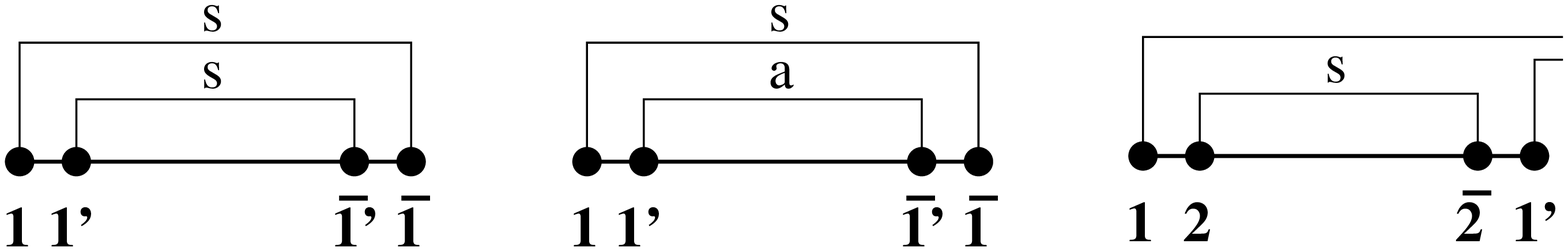}
  \caption{The lowest order diagrams for the effective Liouvillian and the
effective vertex when the symmetric part of the contraction is integrated out. 
$s$ ($a$) denotes the symmetric (antisymmetric) contraction $\gamma^s$ ($\gamma^a$).}
\label{fig:discrete_LG}
\end{figure}

{\it First RG step.} The first discrete RG step consists in integrating
out the symmetric part ${1\over 2}[f_\alpha(\omega)+f_\alpha(-\omega)]={1\over 2}$ 
of the Fermi function in the contraction (\ref{liouville_contraction}). This
part depends only weakly on the frequency and creates no logarithmic
divergencies in perturbation theory. Furthermore, as explained in detail in 
Ref.~\onlinecite{RTRG_FS}, it is the symmetric part of the Fermi function which allows
the zero eigenvalue of the effective Liouvillian $L_S^{eff}(E)$ to occur in the
resolvents between the vertices. This part should be integrated out before starting
the continuous RG in order to show that the renormalization of the vertices is cut off by
relaxation and dephasing rates. To get rid of the symmetric part, one decomposes the
contraction (\ref{liouville_contraction}) according to
\begin{eqnarray}
\label{contraction_decomposition}
\gamma_{11'}^{pp'}\,&=&\delta_{1\bar{1}'}\,p'\,\gamma^s_1
\,+\,\delta_{1\bar{1}'}\,\gamma^a_1\quad,\\
\label{sym_antisym_contraction}
\gamma^s_1\,&=&\,{1\over 2}\,\rho(\bar{\omega})\quad,\quad 
\gamma^a_1\,=\,\rho(\bar{\omega})\,\left[f_\alpha(\bar{\omega})-{1\over 2}\right] \,\,,
\end{eqnarray}
with $\bar{\omega}\,\equiv\,\eta\,\omega$. Using this decomposition in
Eq.~(\ref{value_kernel}), one finds that each diagram, which is irreducible with
respect to the full contraction $\gamma$, decomposes into a series of 
blocks which are irreducible with respect to the symmetric part $\gamma^s$ (i.e., any
vertical line hits at least one symmetric contraction) and connected
to each other by antisymmetric contractions $\gamma^a$. The blocks which are
irreducible with respect to $\gamma^s$ can be formally resummed into
an effective Liouvillian $L^a(E)$ or into effective vertices $G^a(E)$, which obtain
an additional energy variable to account for the reservoir contractions which
cross over the effective quantities. The lowest-order 
diagrams for $L^a$ and $G^a$ are shown in Fig.~\ref{fig:discrete_LG}.
The first two diagrams correspond to the effective Liouvillian (no free lines) and
the third one to the effective vertex (since two lines are free). Using the diagrammatic
rules (\ref{value_kernel}) together with Eq.~(\ref{contraction_decomposition}) and the
convention (\ref{E_omega_definition}), we obtain for the first two diagrams,
\begin{equation}
\nonumber
\gamma_1^s\,\left({1\over 2}\gamma_{1'}^s+p'\gamma_{1'}^a\right)
\,G^{pp}_{11'}\,{1\over E_{11'}\,+\,\bar{\omega}_{11'}\,-\,L_S^{(0)}}
\,G^{p'p'}_{\bar{1}'\bar{1}}\quad,
\end{equation}
and for the third one (including the interchange $1\leftrightarrow 1'$)
\begin{equation}
\nonumber
p'\,\gamma^s_2\,G^{pp}_{12}\,
{1\over E_{12}\,+\,\bar{\omega}_{12}\,-\,L_S^{(0)}}\, 
G^{p'p'}_{\bar{2}1'}\,-\,(1\leftrightarrow 1') \quad.
\end{equation}
We use here the original perturbation series (\ref{value_kernel}) so that the 
unperturbed Liouvillian $L_S^{(0)}$ occurs in the resolvents. Performing the frequency integrations
and omitting terms of order $O(1/D)$, we
obtain the following perturbative result for the effective Liouvillian and the
effective vertex containing the symmetric part of the contraction:
\begin{eqnarray}
\label{L_initial}
L_S^a(E)\,&=&\,L_S\,+\,\Sigma^a(E),\\
\nonumber
\Sigma^a(E)\,&=&-\,i\,{\pi^2 \over 16}\,D\,\bar{G}_{11'}\,\bar{G}_{\bar{1}'\bar{1}}
\,-\,{\pi \over 4}\,D\,\bar{G}_{11'}\,\tilde{G}_{\bar{1}'\bar{1}}\\
\nonumber
&& 
\,+\,{\pi^2 \over 32}\,\bar{G}_{11'}\,(E_{11'}\,-\,L_S^{(0)})\,
\bar{G}_{\bar{1}'\bar{1}}\\
\label{Sigma_initial}
&&
\,-\,i\,{\pi\over 4}\,\bar{G}_{11'}\,(E_{11'}\,-\,L_S^{(0)})\,\tilde{G}_{\bar{1}'\bar{1}}
\quad,\\
\label{G_initial}
\bar{G}^a_{11'}\,&=&\,\bar{G}_{11'}
\,-\,i\,{\pi\over 2}\,\left(\bar{G}_{12}\,\tilde{G}_{\bar{2}1'}\,-\,
\bar{G}_{1'2}\,\tilde{G}_{\bar{2}1}\right).
\end{eqnarray}
Analog equations hold for the effective current kernel $\Sigma_\gamma^a(E)$
and for the effective current vertex $I^{\gamma,a}$. These are obtained by replacing the first vertex $G$
by the current vertex $I^\gamma$ in Eqs.~(\ref{Sigma_initial}) and (\ref{G_initial}).

After integrating out the symmetric part of the Fermi function in this way, we 
obtain a new diagrammatic series for the kernels analog to Eq.~(\ref{value_kernel}), but
the Liouvillian and the vertices have to be replaced by the effective ones and the
contractions between the effective vertices contain only the antisymmetric part
$\gamma^a$. Furthermore, since the effective quantities have become energy dependent
(also the effective vertex $\bar{G}^a$ becomes energy dependent in higher order perturbation theory),
one has to replace 
\begin{equation}
\nonumber
{1\over E+X_i-L_S^{(0)}}\,G\,\rightarrow\,
{1\over E+X_i-L_S^{a}(E+X_i)}\,\bar{G}^a(E+X_i)
\end{equation}
in Eq.~(\ref{value_kernel}). Since the antisymmetric part of the contraction 
(\ref{contraction_decomposition}) does not depend on the Keldysh indices, only
the effective vertex $\bar{G}^a$ averaged over the Keldysh indices occurs in the
new perturbative series. As a consequence [see Eq.~(\ref{zero_eigenvector_G_property})], 
the zero eigenvalue of the effective Liouvillian can no longer
occur in the denominator of the resolvents.

\begin{figure}
  \includegraphics[scale=0.28]{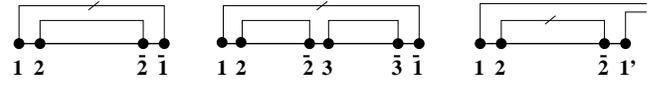}
  \caption{RG diagrams for the renormalization of the Liouvillian in $O(G^2)$ and
$O(G^3)$ (first two diagrams) and the one-loop renormalization of the vertex in $O(G^2)$ (last
diagram). The slash indicates the contraction where the Fermi function has to be
replaced by $-d\Lambda{df^\Lambda_\alpha\over d\Lambda}$.}
\label{fig:RG_Liouvillian_G}
\end{figure}
\begin{figure}
  \includegraphics[scale=0.25]{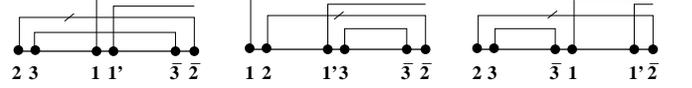}
  \caption{RG diagrams for the renormalization of the vertex in $O(G^3)$ (two-loop).}
\label{fig:RG_G_2_loop}
\end{figure}
{\it Second RG step.} The task of the second continuous RG procedure is to integrate out
the antisymmetric part of the Fermi distribution function step by step. In each 
infinitesimal step, a small energy shell is integrated out and is incorporated into
renormalizations of the vertices and the Liouvillian. However, instead of integrating
out the energies on the real axis, it has turned out to be more efficient to integrate
out the Matsubara poles of the Fermi distribution function on the
imaginary axis,\cite{jakobs_meden_hs_PRL07,RTRG_FS}  i.e., in each RG
step one integrates out one
Matsubara pole starting from high energies. To obtain a continuum version at finite 
temperatures, one introduces a formal cutoff dependence into the antisymmetric part of
the Fermi distribution by
\begin{equation}
\label{Fermi_cutoff}
f^\Lambda_\alpha(\omega)\,=\,
-\,T_\alpha\,\sum_n\,{1\over \omega-i\omega^\alpha_n}
\,\,\theta_{T_\alpha}(\Lambda-|\omega_n^\alpha|)\quad,
\end{equation}
where $\omega_n^\alpha=(2n+1)\pi T_\alpha$ are the Matsubara frequencies corresponding
to the temperature of reservoir $\alpha$, and
\begin{equation}
\label{theta_T}
\theta_T(\omega)\,=\,
\left\{
\begin{array}{cl}
\theta(\omega) &\quad\mbox{for }|\omega|\,>\,\pi T \\
{1\over 2}+{\omega \over 2\pi T} &\quad\mbox{for }|\omega|\,<\,\pi T
\end{array}
\right.
\end{equation}
is a theta function smeared by temperature. For $\Lambda=\infty$, 
Eq.~(\ref{Fermi_cutoff}) yields the full antisymmetric part $f_\alpha(\omega)-{1\over 2}$
of the Fermi distribution. In each RG step, one reduces the cutoff $\Lambda$
by $d\Lambda$, and integrates out the infinitesimal part 
$f_\alpha^\Lambda-f_\alpha^{\Lambda-d\Lambda}=d\Lambda{df^\Lambda_\alpha\over d\Lambda}$
of the Fermi distribution. The new effective Liouvillian and
the new effective vertices at scale $\Lambda-d\Lambda$,
\begin{eqnarray}
\label{L_new}
L_S^{\Lambda-d\Lambda}(E)\,&=&\,L_S^{\Lambda}(E)-dL_S^{\Lambda}(E)\,,\\
\bar{G}_{11'}^{\Lambda-d\Lambda}(E)\,&=&\,\bar{G}_{11'}^{\Lambda}(E)-d\bar{G}_{11'}^{\Lambda}(E)\,,
\end{eqnarray}
can be calculated technically in the same way as for the first
discrete RG step. The only difference is that an infinitesimal small part is integrated out
so that the RG diagrams contain only one contraction
involving the part $d\Lambda{df^\Lambda_\alpha\over d\Lambda}$. Furthermore, since
the diagrams should be irreducible with respect to this part, this contraction must 
connect the first with the last vertex of the diagram.

Up to $O(G^3)$ (which we call two-loop here \footnote{Our convention is that all terms on
the r.h.s. of the RG equation which are of $O(G^2)$ [$O(G^3)$] are called one-loop (two-loop)
terms. For the RG of the vertices, this is in agreement with the conventional classification 
but the two RG diagrams in Fig.~\ref{fig:RG_Liouvillian_G} for the Liouvillian are also called two-loop and
three-loop terms in the literature.}), the RG diagrams for the Liouvillian and the vertices
are shown in Figs.~\ref{fig:RG_Liouvillian_G} and~\ref{fig:RG_G_2_loop}. Using the 
definition
\begin{equation}
\label{gamma_cutoff}
\gamma^\Lambda_1\,=\,\rho(\bar{\omega})\,f_\alpha^\Lambda(\bar{\omega})\quad,
\end{equation}
together with the convention 
$L_S\equiv L_S^{\Lambda}$, $\bar{G}_{11'}\equiv \bar{G}^{\Lambda}_{11'}$ and
\begin{equation}
\label{pi_energy}
\Pi_{1\dots n}\,=\,{1\over E_{1\dots n}+\bar{\omega}_{1\dots n}
-L_S(E_{1\dots n}+\bar{\omega}_{1\dots n})}\quad,
\end{equation}
we obtain the following RG equations:
\begin{widetext}
\begin{eqnarray}
\nonumber
{dL_S(E)\over d\Lambda}\,&=&\,
-\,{d\gamma^\Lambda_{1}\over d\Lambda}\,\gamma^\Lambda_{2}\,\bar{G}_{12}(E)\,\Pi_{12}\,
\bar{G}_{\bar{2}\bar{1}}(E_{12}+\bar{\omega}_{12})\\
\label{rg_L}
&& - \,{d\gamma^\Lambda_{1}\over d\Lambda}\,\gamma^\Lambda_{2}\,\gamma^\Lambda_{3}
\,\bar{G}_{12}(E)\,\Pi_{12}\,\bar{G}_{\bar{2}3}(E_{12}+\bar{\omega}_{12})\,\Pi_{13}\,
\bar{G}_{\bar{3}\bar{1}}(E_{13}+\bar{\omega}_{13}) 
\end{eqnarray}
for the Liouvillian, and
\begin{eqnarray}
\nonumber
{d\bar{G}_{11'}(E)\over d\Lambda}\,&=&\,
-\,\left\{{d\gamma^\Lambda_{2}\over d\Lambda}\,\bar{G}_{12}(E)\,\Pi_{12}\,
\bar{G}_{\bar{2}1'}(E_{12}+\bar{\omega}_{12})
\,\, - \,\, (1\leftrightarrow 1')\right\}\\
\nonumber
&& \hspace{-1.5cm} - \, {d\gamma^\Lambda_{2}\over d\Lambda}\,\gamma^\Lambda_{3}
\,\bar{G}_{23}(E)\,\Pi_{23}\,\bar{G}_{11'}(E_{23}+\bar{\omega}_{23})\,\Pi_{11'23}\,
\bar{G}_{\bar{3}\bar{2}}(E_{11'23}+\bar{\omega}_{11'23})\\ 
\nonumber
&& \hspace{-1.5cm} + \,\left\{ {d\gamma^\Lambda_{2}\over d\Lambda}\,\gamma^\Lambda_{3}
\,\bar{G}_{12}(E)\,\Pi_{12}\,\bar{G}_{1'3}(E_{12}+\bar{\omega}_{12})\,\Pi_{11'23}\,
\bar{G}_{\bar{3}\bar{2}}(E_{11'23}+\bar{\omega}_{11'23})
\,\, - \,\,  (1\leftrightarrow 1')\right\} \\ 
\label{rg_G}
&& \hspace{-1.5cm} - \, \left\{{d\gamma^\Lambda_{2}\over d\Lambda}\,\gamma^\Lambda_{3}
\,\bar{G}_{23}(E)\,\Pi_{23}\,\bar{G}_{\bar{3}1}(E_{23}+\bar{\omega}_{23})\,\Pi_{12}\,
\bar{G}_{1'\bar{2}}(E_{12}+\bar{\omega}_{12})
\,\, - \,\, (1\leftrightarrow 1')\right\} 
\end{eqnarray}
\end{widetext}
for the vertex. Similiar RG equations hold for the current kernel $\Sigma_\gamma(E)$
and the current vertex $I^\gamma_{11'}(E)$ by replacing the first vertex in all terms
on the r.h.s. of the RG equation by the current vertex. The initial conditions of the
RG equations are given by Eqs.~(\ref{L_initial})--(\ref{G_initial}) from the first discrete RG step.
Since $\gamma_1^{\Lambda=0}=0$, the final solution at $\Lambda=0$ provides the result 
for the effective Liouvillian and the current kernel
\begin{eqnarray}
\label{L_eff_final}
L_S^{eff}(E)\,&=&\,L_S(E)|_{\Lambda=0}\quad,\\
\label{Sigma_I_final}
\Sigma_\gamma(E)\,&=&\,\Sigma_\gamma(E)|_{\Lambda=0}\quad,
\end{eqnarray}
from which the reduced density matrix and the current can be calculated in Laplace
space via Eqs.~(\ref{reduced_dm}) and (\ref{current}). 

We note that one can stop at each step of the RG and use the perturbative series 
(\ref{value_kernel}) with the contractions $\gamma_1^\Lambda$ at scale $\Lambda$ 
together with the replacement 
\begin{equation}
\nonumber
{1\over E+X_i-L_S^{(0)}}\,G\,\rightarrow\,
{1\over E+X_i-L_S^\Lambda(E+X_i)}\,\bar{G}^\Lambda(E+X_i)
\end{equation}
for the resolvents and the vertices, where $L_S^\Lambda(E)$ and $\bar{G}^\Lambda(E)$ 
are the renormalized quantities at scale $\Lambda$ [in higher
order in the coupling, also vertices $\bar{G}^\Lambda_{1\dots n}(E)$ with more than two indices can be
generated]. If the exact RG equations in all orders are used, this perturbative series gives
the full kernels at each scale $\Lambda$. Therefore, it is possible at each step
of the RG to resum all closed subdiagrams between two vertices, leading to the replacement
\begin{equation}
\label{rg_resolvent_replacement}
{1\over E+X_i-L_S^\Lambda(E+X_i)}\rightarrow
{1\over E+X_i-L_S^{eff}(E+X_i)}\quad,
\end{equation}
where $L_S^{eff}(E)$ is the full effective Liouvillian at the end of the RG flow at scale $\Lambda=0$.
Using the perturbative series in a certain approximation, one can set up a self-consistent
equation for $L_S^{eff}(E)$ at each step of the RG. However, this is only possible if the
perturbation theory in the renormalized coupling is well defined. We will see that this is
only possible at a certain scale $\Lambda_c$ where some physical cutoff scale is reached,
see Sec.~\ref{sec:generic_2_loop}. Up to this scale, we will always use the renormalized 
Liouvillian $L_S^\Lambda(E)$ in the denominator of the resolvents.

{\it RG in Matsubara space.}
Using the fact that the resolvents and the vertices on the r.h.s. of the RG equations
are analytic functions in all frequencies $\bar{\omega}_i$ in the upper half of the 
complex plane, all frequency integrations can be calculated analytically by closing 
the contour in the upper half of the complex plane. The only
poles occurring there are the poles of the contractions and their derivatives, given by
\begin{eqnarray}
\label{gamma_lambda}
\gamma_1^\Lambda\,&=&\,-\rho(\bar{\omega})\,T_\alpha\,
\sum_n\,{1\over \bar{\omega}-i\omega_n^\alpha}\,\theta_{T_\alpha}(\Lambda-|\omega_n^\alpha|),\\
\label{gamma_lambda_derivative}
{d\over d\Lambda}\gamma_1^\Lambda\,&=&\,-\rho(\bar{\omega})\,{1\over 2\pi}\,
\left({1\over \bar{\omega}-i\Lambda_{T_\alpha}}\,+\,{1\over \omega+i\Lambda_{T_\alpha}}\right).
\end{eqnarray}
$\Lambda_{T_\alpha}$ denotes the Matsubara frequency $\omega_n^\alpha$
which lies closest to the cutoff $\Lambda$.
After performing the integration we find that, due to the presence of the cutoff function 
$\rho(\bar{\omega})={D^2\over D^2+\bar{\omega}^2}$,
the r.h.s. of the RG equations gives a negligible
contribution for $\Lambda\gg D$. Therefore, we can start the RG at $\Lambda_0\sim D$ and
omit the cutoff function $\rho(\bar{\omega})$ (finally, the precise ratio between $\Lambda_0$ and
$D$ is determined such that no linear terms in $D$ are generated, see below). As a consequence,
only the Matsubara poles of the Fermi function in the upper half of the complex plane will
contribute to the frequency integrations and all real frequencies are replaced by Matsubara
frequencies. From now on, we write the frequency dependence explicitly and define the analytic
continuation of the Liouvillian and the vertices in imaginary frequency space by
\begin{eqnarray}
\label{L_matsubara}
L_S(E,\omega)\,&=&\,L_S(E+i\omega)\quad,\\
\label{G_matsubara}
\bar{G}_{\eta_1\nu_1,\eta_2\nu_2}(E,\omega,\omega_1,\omega_2)\,&=&\,
\bar{G}_{12}(E+i\omega)|_{\bar{\omega}_i\rightarrow i\omega_i}\,,
\end{eqnarray}
where $\omega\equiv \omega_n^\alpha$, $\omega_i\equiv\omega_{n_i}^{\alpha_i}$ correspond to
Matsubara frequencies. With the definition 
\begin{equation}
\label{pi_matsubara}
\Pi(E,\omega)\,=\,{1\over E+i\omega-L_S(E,\omega)}\quad,
\end{equation}
the RG equations in Matsubara space can be written as
\begin{widetext}
\begin{eqnarray}
\nonumber
{dL_S(E,\omega)\over d\Lambda}\,&=&\,
\bar{G}_{12}(E,\omega,\Lambda_{T_{\alpha_1}},\omega_2)\,
\Pi(E_{12},\Lambda_{T_{\alpha_1}}+\omega+\omega_2)\,
\bar{G}_{\bar{2}\bar{1}}(E_{12},\Lambda_{T_{\alpha_1}}+\omega+\omega_2,-\omega_2,-\Lambda_{T_{\alpha_1}})\\
\label{rg_L_matsubara}
&& - \,i\,\bar{G}_{12}\,\Pi(E_{12},\Lambda_{T_{\alpha_1}}+\omega+\omega_2)\,
\bar{G}_{\bar{2}3}\,\Pi(E_{13},\Lambda_{T_{\alpha_1}}+\omega+\omega_3)\,
\bar{G}_{\bar{3}\bar{1}}\quad, \\
\nonumber
{d\bar{G}_{11'}(E,\omega,\omega_1,\omega_1^\prime)\over d\Lambda}\,&=&\,
i\,\left\{\bar{G}_{12}(E,\omega,\omega_1,\Lambda_{T_{\alpha_2}})\,
\Pi(E_{12},\Lambda_{T_{\alpha_2}}+\omega+\omega_1)\,
\bar{G}_{\bar{2}1'}(E_{12},\Lambda_{T_{\alpha_2}}+\omega+\omega_1,-\Lambda_{T_{\alpha_2}},\omega_1^\prime)
\,\, - \,\, (1\leftrightarrow 1')\right\}\\
\nonumber
&& \hspace{-1.5cm} +  
\,\bar{G}_{23}\,\Pi(E_{23},\Lambda_{T_{\alpha_2}}+\omega+\omega_3)\,
\bar{G}_{11'}\,\Pi(E_{11'23},\Lambda_{T_{\alpha_2}}+\omega+\omega_1+\omega_1^\prime+\omega_3)\,
\bar{G}_{\bar{3}\bar{2}}  \\
\nonumber
&& \hspace{-1.5cm} - \,\left\{ 
\,\bar{G}_{12}\,\Pi(E_{12},\Lambda_{T_{\alpha_2}}+\omega+\omega_1)\,
\bar{G}_{1'3}\,\Pi(E_{11'23},\Lambda_{T_{\alpha_2}}+\omega+\omega_1+\omega_1^\prime+\omega_3)\,
\bar{G}_{\bar{3}\bar{2}}\,\, - \,\,  (1\leftrightarrow 1')\right\} \\
\label{rg_G_matsubara}
&& \hspace{-1.5cm} + \, \left\{
\,\bar{G}_{23}\,\Pi(E_{23},\Lambda_{T_{\alpha_2}}+\omega+\omega_3)\,
\bar{G}_{\bar{3}1}\,\Pi(E_{12},\Lambda_{T_{\alpha_2}}+\omega+\omega_1)\,\bar{G}_{1'\bar{2}}
\,\, - \,\, (1\leftrightarrow 1')\right\}\quad.
\end{eqnarray}
\end{widetext}
In these equations, the index $1\equiv \eta\nu$ includes no longer the frequency variable, and
we implicitly sum over all indices and Matsubara frequencies on the r.h.s. of the RG equations
which do not occur on the left hand side (l.h.s.). Only positive Matsubara frequencies smaller than the cutoff 
$\Lambda$ are allowed and each sum has to be written as
\begin{equation}
\label{matsubara_sum}
2\pi T_\alpha\,\sum_{n}\,\theta_{T_\alpha}(\Lambda-\omega_n^\alpha)\,\theta(\omega_n^\alpha)
\end{equation}
which reduces to an integral $\int_0^\Lambda d\omega$ for zero temperature.
The frequency arguments of the vertices in the terms of $O(G^3)$ in Eqs.~(\ref{rg_L_matsubara}) and
(\ref{rg_G_matsubara}) have been omitted since they are not needed for the weak coupling 
analysis up to two-loop order, see below.

The RG equations in Matsubara space are the final result of this section and are the starting
point for the analytical solution in the weak coupling regime presented in the next section.
Similiar RG equations hold for the current kernel and the current vertex in Matsubara space by
replacing the first vertex in all terms on the r.h.s. by the current vertex.
Using Eqs.~(\ref{L_eff_final}), (\ref{Sigma_I_final}), and (\ref{L_matsubara}), the effective 
Liouvillian and the current kernel follow from
\begin{eqnarray}
\label{final_result_L_matsubara}
L_S^{eff}(E)\,&=&\,L_S(E,\omega=0)|_{\Lambda=0}\quad,\\
\label{final_result_Sigma_I_matsubara}
\Sigma_\gamma(E)\,&=&\,\Sigma_\gamma(E,\omega=0)|_{\Lambda=0}\quad.
\end{eqnarray}

Finally, we note that all symmetry properties stated in 
Eqs.~(\ref{G_symmetry})--(\ref{I_c_transform}) are preserved under the RG flow 
(see Ref.~\onlinecite{RTRG_FS} for the proof),
\begin{eqnarray}
\label{rg_G_symmetry}
\hspace{-0.5cm}
\bar{G}_{12}(E,\omega,\omega_1,\omega_2)\,&=&\,
-\,\bar{G}_{21}(E,\omega,\omega_2,\omega_1)\,,\\
\label{rg_I_symmetry}
\hspace{-0.5cm}
\bar{I}^\gamma_{12}(E,\omega,\omega_1,\omega_2)\,&=&\,
-\,\bar{I}^\gamma_{21}(E,\omega,\omega_2,\omega_1)\,,\\
\label{rg_L_property}
\hspace{-0.5cm}
\mbox{Tr}_S\,L_S(E,\omega)\,&=&\,0\,,\\
\label{rg_G_property}
\hspace{-0.5cm}
\mbox{Tr}_S\,\bar{G}_{12}(E,\omega,\omega_1,\omega_2)\,&=&\,0 \,,\\
\label{rg_L_c_transform}
\hspace{-0.5cm}
L_S(E,\omega)^c\,&=&\,-L_S(-E,\omega) \,,\\
\label{rg_Sigma_I_c_transform}
\hspace{-0.5cm}
\Sigma_\gamma(E,\omega)^c\,&=&\,-\Sigma_\gamma(-E,\omega) \,,\\
\label{rg_G_c_transform}
\hspace{-0.5cm}
\bar{G}_{12}(E,\omega,\omega_1,\omega_2)^c\,&=&\,
-\,\bar{G}_{\bar{2}\bar{1}}(-E,\omega,\omega_2,\omega_1)\,,\\
\label{rg_I_c_transform}
\hspace{-0.5cm}
\bar{I}^\gamma_{12}(E,\omega,\omega_1,\omega_2)^c\,&=&\,
-\,\bar{I}^\gamma_{\bar{2}\bar{1}}(-E,\omega,\omega_2,\omega_1)\,,
\end{eqnarray}
where all energy variables are real.

Similiar to the discussion at the end of Sec.~\ref{sec:generic_model}, properties 
(\ref{rg_L_property}) and (\ref{rg_G_property}) have the consequence that we obtain zero 
if the left eigenvector $\langle \bar{x}_0^\Lambda(E,\omega)|$ of the effective 
Liouvillian $L_S^\Lambda(E,\omega)$ for zero eigenvalue acts from the left on the vertex
\begin{equation}
\label{rg_zero_eigenvector_G_property}
\langle \bar{x}_0(E,\omega)|\,\bar{G}_{12}\,=\,0\quad,
\end{equation}
compare with Eq.~(\ref{zero_eigenvector_G_property}).
Therefore, it is not allowed that the zero eigenvalue of $L_S^\Lambda(E,\omega)$ occurs in the
resolvents on the r.h.s. of the RG equations, proving that the RG is always cut off
by relaxation and dephasing rates (this property holds in all orders of perturbation
theory and within all truncation schemes). As we will see in the next section, this property 
is essential to prove that a systematic weak coupling analysis can be carried out 
in the generic case.

\subsection{Two-loop analysis}
\label{sec:generic_2_loop}

In this section, we solve the two-loop RG equations (\ref{rg_L_matsubara}) and (\ref{rg_G_matsubara})
analytically in the weak coupling regime up to two-loop order. Weak coupling is defined by the
condition that the renormalized vertices $\bar{G}_{12}(E,\omega,\omega_1,\omega_2)$ stay small 
compared to one throughout the RG flow so that a systematic expansion is possible on the
r.h.s. of the RG equations. This condition is fulfilled if the various cutoff scales occurring
in the resolvents $\Pi(E,\omega)$ are larger than the energy scale $T_K$ at which the vertices
would diverge in the absence of any cutoff scales (the so-called Kondo temperature for the Kondo model). 
From the form of the RG equations, we see that the resolvents at scale $\Lambda$ have the form
\begin{equation}
\label{form_resolvent}
{1\over z-L_S^\Lambda(z)}\quad,
\end{equation}
with 
\begin{equation}
\label{form_z}
z=E_{1\dots n}+i\omega+i(\Lambda_{T_{\alpha_1}}+\omega_2+\dots +\omega_n)\quad,
\end{equation}
and positive Matsubara frequencies $0<\omega_k<\Lambda$ for $k=2,\dots,n$. Here, $E+i\omega$ is 
the original Laplace variable at which we want to calculate the final effective
Liouvillian $L_S^{eff}(E+i\omega)$, and the index $1$ corresponds to the 
contraction connecting the first with the last vertex of the RG diagram (where the Matsubara
frequency is replaced by the cutoff $\Lambda_{T_\alpha}$).  

Expanding the resolvent (\ref{form_resolvent}) around its poles analog to the discussion at the
end of Sec.~\ref{sec:generic_model}, we arrive at the approximation 
(\ref{Pi_general_approximation}) which contains the most important terms leading to logarithmic 
enhancements 
\begin{equation}
\label{Pi_leading}
\Pi^\Lambda(z)\,\approx \,{\tilde{Z}^\Lambda\over z-\tilde{L}_S^\Lambda}\,=\,
\sum_i{a_i^\Lambda\over z-z_i^\Lambda}|x_i^\Lambda(z_i^\Lambda)\rangle\langle \bar{x}_i^\Lambda(z_i^\Lambda)|\,,
\end{equation}
with
\begin{eqnarray}
\label{form_single_poles}
\hspace{0.5cm}
{1\over z-z_i^\Lambda}&=&\\
\nonumber
&&\hspace{-2.5cm}
={1\over i\Lambda_{T_{\alpha_1}}+E_{1\dots n}-\tilde{h}_i^\Lambda
+i(\omega+\omega_2+\dots +\omega_n)+i\tilde{\Gamma}_i^\Lambda}\quad,
\end{eqnarray}
where $z_i^\Lambda=\tilde{h}_i^\Lambda-i\tilde{\Gamma}_i^\Lambda$, $\tilde{\Gamma}_i^\Lambda>0$, 
denote the positions of the 
nonzero poles of the resolvent (we assume single poles here, but the following discussion holds 
also for other cases; note that the zero pole of the stationary solution can not occur in the
resolvent as discussed at the end of the last section).  
Since all Matsubara frequencies and the relaxation/dephasing rates are positive, we see that the
resolvents can not become large. Using $0<\omega_k<\Lambda$ for $k=2,\dots,n$, and setting $\omega=0$,
we find that the resolvent is cut off at the scale
\begin{equation}
\label{cutoff_resolvent}
\Lambda\,\sim\,\text{max}\{T_\alpha,|E_{1\dots n}-\tilde{h}_i^\Lambda|,\tilde{\Gamma}_i^\Lambda\}\quad,
\end{equation}
where the maximum is taken over all values of the occurring indices. Here, temperature is
a trivial cutoff parameter, because, for $\Lambda<2\pi T_\alpha$, 
the sum (\ref{matsubara_sum}) over the Matsubara frequencies
for reservoir $\alpha$ reduces to one term $n=0$ and the cutoff $\Lambda_{T_\alpha}=\pi T_\alpha$
becomes independent of $\Lambda$. Therefore, temperature is a unique cutoff for all terms on the
r.h.s. of the RG equations, like in equilibrium problems. This trivial cutoff is set to zero in
the following, i.e., $T_\alpha=0$, and we discuss only the nontrivial dependence on the other 
cutoff scales. The minimal cutoff scale occurs for $E_{1\dots n}-\tilde{h}_i^\Lambda=0$ and is given 
by the relaxation 
or dephasing rates $\tilde{\Gamma}_i^\Lambda$. These points define the positions of resonances where 
renormalization-group-induced logarithmic enhancements or suppressions have to be expected. 
However, as we will show in the following, these logarithmic terms can be calculated systematically
by perturbation theory in the renormalized couplings, provided that the weak coupling condition
\begin{equation}
\label{Lambda_c}
\Lambda_c\,\equiv\,\text{max}\{|E|,|\mu_\alpha|,|\tilde{h}_i|\}\,\gg\,T_K
\end{equation} 
is fulfilled, where by convention $\tilde{h}_i\equiv\tilde{h}_i^{\Lambda=0}$ denotes the final
(physical) renormalized oscillation frequency at scale $\Lambda=0$ (we will see that the difference between
$\tilde{h}_i^\Lambda$ and $\tilde{h}_i^{\Lambda=0}$ is proportional to the final renormalized 
coupling, i.e., only a small perturbative correction). $\Lambda_c$ is an important energy scale 
separating two energy regions where the RG equations are solved in a different way. It is given by the 
maximum of the 
Laplace variable $E$, the chemical potentials of the reseroirs (giving some voltage $V$),
and the oscillation frequencies $\tilde{h}_i$ of the different physical modes (e.g., the renormalized 
magnetic field in the Kondo problem). This is roughly the maximum value the various cutoff scales 
$|E_{1\dots n}-\tilde{h}_i^\Lambda|=|E+\sum_k\eta_k\mu_{\alpha_k}-\tilde{h}_i^\Lambda|$ 
of the resolvents can take, see Eq.~(\ref{cutoff_resolvent}).
Thus, for $\Lambda>\Lambda_c\gg T_K$, the cutoff scales do not play an important role and we get
$J_{\Lambda}\ll 1$, where $J_{\Lambda}$ is the order of magnitude
of the vertex at scale $\Lambda$. Since $\Lambda$ is the relevant energy scale in this regime, the 
order of magnitude of the relaxation/dephasing rates is given by 
$\tilde{\Gamma}_i^\Lambda\sim\Lambda J_\Lambda^2\ll\Lambda$ (note that the RG for the Liouvillian
starts in second order in $J$). Therefore, the relaxation 
and dephasing rates are small perturbative corrections in the denominators of the resolvents and
do not lead to any cutoff of the RG in the regime $\Lambda>\Lambda_c$. Since all vertices are
small, we can systematically truncate the hierarchy of RG equations and expand the solution 
systematically around the poor man scaling solution (i.e., the lowest-order solution for the vertex
in the absence of any cutoff scales $\Lambda_c=T_\alpha=0$). This gives
a certain initial condition for the RG at $\Lambda=\Lambda_c$ presented as a power series in 
$J_c=J_{\Lambda=\Lambda_c}$.

In the second regime $0<\Lambda<\Lambda_c$, the RG for $\tilde{\Gamma}_i^\Lambda$ is very weak and will be
roughly cut off by $\Lambda_c$. The reason is that there are many terms on the r.h.s. of the 
RG equation involving different cutoff scales $|E_{1\dots n}-\tilde{h}_i^\Lambda|$, but usually one 
of them will be given by $\Lambda_c$ already in second order in $J$. In this case (see a comment below for the
other cases), we get $\tilde{\Gamma}_i^\Lambda\sim\Lambda_c J_c^2$ for all $0<\Lambda<\Lambda_c$ because
$\tilde{\Gamma}_i^\Lambda$ becomes smaller for decreasing $\Lambda$. 
This means that even at resonance, the minimal cutoff scale is given by $\Gamma\sim\Lambda_c J_c^2$.
Therefore, by expanding the solution for the vertices systematically in $J_c$, we get in the worst
case at resonance a series of the schematic form
\begin{equation}
\label{worst_case}
J_\Lambda\,\sim\,J_c\,\left(1\,+\,J_c\,\ln{\Lambda_c \over |\Lambda+i\Gamma|}\,+\,\dots\,\right)\quad.
\end{equation}
For $\Lambda\rightarrow 0$, the logarithmic term becomes maximally of the order $\sim J_c\ln{J_c}\ll 1$,
which is a perturbative correction in the weak coupling case $J_c\ll 1$. Therefore, under condition
(\ref{Lambda_c}), we stay in the weak-coupling regime and the RG equations can be
solved perturbatively in $J_c$ in the whole regime 
$0<\Lambda<\Lambda_c$. We note that this fact relies essentially on the condition that all resolvents 
on the r.h.s. of the RG equations contain some relaxation/dephasing rate $\tilde{\Gamma}_i^\Lambda$.
As explained at the end of the last section, our RG approach gives this property in the 
generic case in all orders of $J$ by construction.

In case that the RG equation for $\tilde{\Gamma}_i^\Lambda$ contains a smaller cutoff scale
$\Lambda_c^\prime\ll\Lambda_c$ in second order in $J$, we expect
$\tilde{\Gamma}^\Lambda_i\sim\Lambda_c^\prime (J_c^\prime)^2$ for $\Lambda<\Lambda_c^\prime$,
with $J_c^\prime=J_{\Lambda=\Lambda_c^\prime}$. In this case, the logarithmic term in Eq.~(\ref{worst_case})
leads to contributions $\sim J_c\ln{\Lambda_c\over\Lambda_c^\prime}$ for $\Lambda\rightarrow 0$, 
giving rise to additional enhancements and sharper features at resonance.
However, even for $\Lambda_c^\prime\rightarrow 0$, there is no divergence since the cutoff scale 
$\Lambda_c$ will certainly occur in some higher order term on the r.h.s. of the RG equation for
$\tilde{\Gamma}_i^\Lambda$. Thus, the minimal cutoff scale will be of order $\sim\Lambda_c J_c^k$ 
with $k>2$. This gives a maximal value $\sim k\,J_c\ln J_c$ for the logarithm which is again a 
perturbative correction just enhanced by a factor of $k$. This shows that the height of logarithmic
enhancements at resonance are expected to be increasable only by factors of $O(1)$, but the sharpness
of features at resonance (which are controlled by $\tilde{\Gamma}_i$) can become orders of magnitude smaller.

As a consequence, we have seen that for $0<\Lambda<\Lambda_c$, we can perform a perturbation theory
in $J_c$, which is the order of the vertex at scale $\Lambda_c$. This means that we can equivalently
stop the RG at $\Lambda=\Lambda_c$ and use the perturbative series (\ref{value_kernel}) with the
contraction $\gamma_1^{\Lambda_c}$ at scale $\Lambda_c$ together with the replacement
\begin{eqnarray}
\nonumber
{1\over E+X_i-L_S^{(0)}}\,G\,&\rightarrow&\,\\
\nonumber
&&\hspace{-2cm}
\rightarrow\,{1\over E+X_i-L_S^{\Lambda_c}(E+X_i)}\,\bar{G}^{\Lambda_c}(E+X_i)
\end{eqnarray}
for the resolvents and the vertices. In contrast to the perturbative series at scales $\Lambda\gg\Lambda_c$,
this perturbative series at scale $\Lambda_c$ is well defined and can be used alternatively to the
RG approach. Furthermore, as explained in Sec.~\ref{sec:generic_model}, the perturbation series can
be partially resummed, leading to the replacement
\begin{equation}
\label{rg_Lambda_c_resolvent_replacement}
{1\over E+X_i-L_S^{\Lambda_c}(E+X_i)}\rightarrow
{1\over E+X_i-L_S^{eff}(E+X_i)}\quad,
\end{equation}
i.e., the final full effective Liouvillian can be written in the denominator. This series has the
advantage that the oscillation frequencies $\tilde{h}_i$, defining the resonance positions
\begin{equation}
\label{resonance_positions}
E_{1\dots n}\,=\,E+\sum_k\eta_k\mu_{\alpha_k}\,=\,\tilde{h}_i\quad,
\end{equation}
and the relaxation/dephasing rates $\tilde{\Gamma}_i$, cutting off the logarithmic enhancements
at resonance, are the final physical ones at scale $\Lambda=0$. This is expected on physical grounds
and leaves no question open what the precise prefactor of these energy scales is. Using the
replacement (\ref{rg_Lambda_c_resolvent_replacement}), one can either write down directly the perturbative
series or one can use the RG equations [again using the replacement 
(\ref{rg_Lambda_c_resolvent_replacement}) to define the resolvents] and solve them perturbatively in $J_c$.
Both approaches give the same because the RG equations are formally exact.

Having shown that a weak coupling analysis is well defined for all cutoff scales under the condition
(\ref{Lambda_c}), we proceed to show analytically the perturbative solution of the RG equations in 
all details for the two regimes $\Lambda>\Lambda_c$ and $0<\Lambda<\Lambda_c$ in the generic case.

\subsubsection{RG above $\Lambda_c$}
\label{sec:generic_above_Lambda_c}

{\it Lowest order.} For $\Lambda>\Lambda_c$, we define the reference solution for the vertex by considering 
only the first term on the r.h.s. of RG equation (\ref{rg_G_matsubara}) with 
$\Pi(E_{12},\Lambda+\omega+\omega_1)\rightarrow {1\over i\Lambda}$, i.e., by setting all frequencies
to zero and omitting the cutoff scales from $E$, $\bar{\mu}_{12}$, and $L_S(E)$. This defines the 
leading order RG equations
\begin{eqnarray}
\label{G_reference_solution}
{d\bar{G}_{11'}^{(1)}\over d\Lambda}\,&=&\,
{1\over\Lambda}\,\left\{\bar{G}^{(1)}_{12}\,\bar{G}^{(1)}_{\bar{2}1'}\,-\,
(1\leftrightarrow 1')\right\}\quad,\\
\label{I_reference_solution}
{d\bar{I}_{11'}^{\gamma (1)}\over d\Lambda}\,&=&\,
{1\over\Lambda}\,\left\{\bar{I}^{\gamma (1)}_{12}\,\bar{G}^{(1)}_{\bar{2}1'}\,-\,
(1\leftrightarrow 1')\right\}\, \quad.
\end{eqnarray}
The initial condition for these RG equations is the 
bare vertex. The order of magnitude of the leading order solution is denoted by the
dimensionless parameter $J\sim \bar{G}^{(1)}_{12}$. The connection to conventional poor man scaling
is established by recognizing that the leading order vertices have the same form in Liouville
space as the original vertices, given by Eqs.~(\ref{vertex_decomposition}), (\ref{G_vertex_liouville}), 
and (\ref{current_liouvillian_vertex}), i.e., one can prove that \cite{RTRG_FS}
\begin{eqnarray}
\label{G_bar}
\bar{G}^{(1)}_{12}\,&=&\,[g_{12},\cdot]_-\quad,\\
\label{G_tilde}
\tilde{G}^{(1)}_{12}\,&=&\,[g_{12},\cdot]_+\quad,\\
\label{I_bar}
\bar{I}^{\gamma (1)}_{12}\,&=&\,c^\gamma_{12}\,\tilde{G}^{(1)}_{12}\quad,
\end{eqnarray}
with
\begin{equation}
\label{rg_g}
{dg_{11'}\over d\Lambda}\,=\,
{1\over\Lambda}\,\left\{g_{12}\,g_{\bar{2}1'}\,-\,(1\leftrightarrow 1')\right\} \quad.
\end{equation}
Thereby, the form (\ref{I_bar}) of the current vertex can only be proven if one takes
the $\text{Tr}_S$ over the local quantum system, i.e., it holds only for the combination
$\text{Tr}_S \bar{I}^{\gamma (1)}_{12}$. Implicitly, for all following equations, we will 
always consider this combination for the current vertex because this is finally needed for
the calculation of the average of the current. Equation (\ref{rg_g}) is the usual poor man
scaling equation which can also be derived on a pure Hamiltonian level by, e.g., leaving the 
$t$ matrix invariant.\cite{hewson}

Next we set up the lowest order RG equation for the Liouvillian $L_S(E,\omega)$ by 
considering the first term on the r.h.s. of Eq.~(\ref{rg_L_matsubara}) and replacing 
the vertices by the leading order ones. Furthermore, we replace $L_S(E,\omega)$ by
$L_S^{(0)}$ in the resolvent. This gives
\begin{equation}
\label{rg_L_approx}
{dL_S(E,\omega)\over d\Lambda}\,\approx\,
-i\,\bar{G}^{(1)}_{12}\,{\cal{K}}_\Lambda(E_{12}+i\omega-L_S^{(0)})
\,\bar{G}^{(1)}_{\bar{2}\bar{1}}\quad,
\end{equation}
with 
\begin{equation}
\label{K_function}
{\cal{K}}_\Lambda(z)\,=\,\ln\left({2\Lambda-iz \over \Lambda-iz}\right)\quad.
\end{equation}
To extract the lowest-order term $L_S^{(1)}\sim J$ from this equation we treat the terms
$\sim z/\Lambda$ of ${\cal{K}}_\Lambda(z)$ separately by the decomposition
\begin{equation}
\label{K_tilde_function}
{\cal{K}}_\Lambda(z)\,=\,{\tilde{\cal{K}}}_\Lambda(z)\,+\,{iz \over 2\Lambda}\quad
\end{equation}
so that ${\tilde{\cal{K}}}_\Lambda(z)$ is integrated by the function
\begin{eqnarray}
\label{L_F}
{\tilde{\cal{K}}}_\Lambda(z)\,&=&\,{d\over d\Lambda}\,\tilde{F}_\Lambda(z)\,,\\
\nonumber
\tilde{F}_\Lambda(z)\,&=&\,\Lambda\,\ln\left({2\Lambda-iz \over \Lambda-iz}\right)\\
\label{F_tilde_function}
&&-\,{iz\over 2}\,\left(\ln{(2\Lambda-iz)\Lambda\over 2(\Lambda-iz)^2}\,+\,1\right)\,,
\end{eqnarray}
with the following asymptotic behavior:
\begin{equation}
\label{F_tilde_asymptotic}
\tilde{F}_\Lambda(z)\,\rightarrow\,
\Lambda\,\left[\ln{2}\,+\,O({z\over\Lambda})^2\right]
\quad \mbox{for }\Lambda \gg |z| \,.
\end{equation}
Using Eq.~(\ref{rg_L_approx}), the second term on the r.h.s. of Eq.~(\ref{K_tilde_function}) leads 
to the following RG equation for the Liouvillian in leading order:
\begin{equation}
\label{rg_L_first_order}
{d L^{(1)}_S(E,\omega)\over d\Lambda}\,=\,{1\over 2\Lambda}
\,\bar{G}^{(1)}_{12}\,(E_{12}+i\omega-L_S^{(0)})\,\bar{G}^{(1)}_{\bar{2}\bar{1}}\quad,
\end{equation} 
with the initial condition
\begin{equation}
L_S^{(1)}(E,\omega)|_{\Lambda=\Lambda_0}\,=\,0\quad.
\end{equation}
When integrated, we obtain a contribution $L_S^{(1)}(E,\omega)\sim J$, i.e., one power less
than expected due to the factor ${1\over\Lambda}$ [compare with RG equation
(\ref{G_reference_solution}), where the same happens]. In contrast, the contributions
from ${\tilde{\cal{K}}}_\Lambda(z)$ are of second and third order in $J$, as will be
discussed below. We write
\begin{equation}
\label{L_first_decomposition}
L_S^{(1)}(E,\omega)\,=\,L_S^{(1)}\,-\,(E+i\omega)\,Z^{(1)}\,
\end{equation}
with $L_S^{(1)}\equiv L_S^{(1)}(E=0,\omega=0)$, and
\begin{eqnarray}
\label{L_first_const}
{d L^{(1)}_S \over d\Lambda}\,&=&\,{1\over 2\Lambda}
\,\bar{G}^{(1)}_{12}\,(\bar{\mu}_{12}-L_S^{(0)})\,\bar{G}^{(1)}_{\bar{2}\bar{1}}\quad,\\
\label{Z_first}
{d Z^{(1)}\over d\Lambda}\,&=&\,-\,{1\over 2\Lambda}
\,\bar{G}^{(1)}_{12}\,\bar{G}^{(1)}_{\bar{2}\bar{1}}\quad.
\end{eqnarray}
We note that ${1\over 1+Z^{(1)}}$ can be interpreted as the $Z$ factor in Liouville space
at scale $\Lambda$, and it can be shown that $L^{(1)}_S$ and $Z^{(1)}$ are hermitian operators. 
Similiar equations can be set up for the current kernel $\Sigma_\gamma^{(1)}(E,\omega)$ 
in leading order by replacing the first vertex on the r.h.s. of Eqs.~(\ref{L_first_const}) and
(\ref{Z_first}) by the current vertex.

When integrating the RG equations (\ref{L_first_const}) and (\ref{Z_first}) up
to $\Lambda_c$, we obtain a linear contribution in the renormalized coupling $J_c$ at scale
$\Lambda_c$. Since the perturbative treatment in $J_c$ for the regime $0<\Lambda<\Lambda_c$ can
only give corrections $\sim J_c^2$ to the Liouvillian, we know that the final effective 
Liouvillian up to $O(J_c)$ is given by
\begin{eqnarray}
\label{L_eff_1_order}
L_S^{eff}(E,\omega)\,&=&\,\\
\nonumber
&&\hspace{-2cm}
=\,L_S^{(0)}\,+\,L_S^{(1)c}\,-\,(E+i\omega)Z^{(1)c}\,
+\,O(J_c^2)\quad,
\end{eqnarray}
with $L_S^{(1)c}=L_S^{(1)}|_{\Lambda=\Lambda_c}$ and $Z^{(1)c}=Z^{(1)}|_{\Lambda=\Lambda_c}$.

{\it Second order.}
With the vertex and the Liouvillian in leading order, we can now expand the full RG equations
systematically around these reference solutions and calculate the higher orders. The
Liouvillian and the vertex are written as an expansion in $J$,
\begin{eqnarray}
\nonumber
&&\hspace{-0.8cm}
L_S(E,\omega)\,=\\
\label{L_expansion}
&&\hspace{0cm}
=\,L_S^{(0)}\,+\,L_S^{(1)}(E,\omega)
\,+\,L_S^{(2)}(E,\omega)\,+\,\dots,\\
\nonumber
&&\hspace{-0.8cm}
\bar{G}_{12}(E,\omega,\omega_1,\omega_2)\,=\\
\label{G_expansion}
&&\hspace{0cm}
=\,\bar{G}_{12}^{(1)}\,+\,
\bar{G}_{12}^{(2)}(E,\omega,\omega_1,\omega_2)\,+\,\dots\quad,
\end{eqnarray}
where $L_S^{(n)},\bar{G}_{12}^{(n)}\sim J^n$ (possibly with additional factors $\sim \ln^k{J}$
with $k<n$, see below).

To get an RG equation for $L_S^{(2)}(E,\omega)$ and $\bar{G}^{(2)}_{12}(E,\omega,\omega_1,\omega_2)$, 
we insert the expansions (\ref{L_expansion}) and (\ref{G_expansion}) into the r.h.s. of the full RG 
equations (\ref{rg_L_matsubara}) and (\ref{rg_G_matsubara}). For the RG of the Liouvillian (vertex),
the resolvents are expanded such that we collect all terms in $O[({\Delta\over\Lambda})^k J^2]$
\{$O[{1\over\Lambda}({\Delta\over\Lambda})^k J^2]$\}, with $k>1$, and $O({\Delta \over \Lambda}J^3)$  
[$O({1\over \Lambda}J^3)$] on the r.h.s. of the RG equation, where $\Delta$ is some cutoff scale 
arising from $E_{12}$, $L_S(E)$, or the frequencies. To
achieve this, we leave out $L_S^{(2)}(E,\omega)$ in the resolvents [leading to terms of $O(J^4)$ on the
r.h.s. of the RG equation], and use Eq.~(\ref{L_first_decomposition}). This gives for the resolvent
\begin{eqnarray}
\nonumber
i\,\Pi(E,\omega)\,&\approx &\,{1\over \omega-iE+iL_S^{(0)}+iL_S^{(1)}+(\omega-iE)Z^{(1)}}\\
\nonumber
&&\hspace{-1.5cm}
\approx \,{1\over 1+{Z^{(1)}\over 2}}\,\cdot\,{1\over \omega-iE+iL_S^{(0)}+
i\tilde{L}_S^{(1)}}\,\cdot\,{1\over 1+{Z^{(1)}\over 2}},\\
\label{resolvent_form}
\end{eqnarray}
with the hermitian operator
\begin{eqnarray}
\nonumber
\tilde{L}_S^{(1)}\,&=&\,{1\over 1+{Z^{(1)}\over 2}}\,L_S^{(1)}
\,{1\over 1+{Z^{(1)}\over 2}}\\
\label{tilde_L}
&&\hspace{-1cm}
\approx \,L_S^{(1)}-{1\over 2}(Z^{(1)}\,L_S^{(0)}+L_S^{(0)}\,Z^{(1)})\,.
\end{eqnarray}

{\it Vertex in second order.}
We start with the RG equation (\ref{rg_G_matsubara}) for the vertex (or equivalently
the current vertex by replacing the first vertex by the current vertex in all equations).
For the resolvent in the first term on the r.h.s. of this RG equation,
we expand (\ref{resolvent_form}) in the following way:
\begin{eqnarray}
\nonumber
i\,\Pi(E_{12},\Lambda+\omega+\omega_1)\,&\approx &\\
\nonumber
&&\hspace{-4cm}
\approx \,{1-Z^{(1)}\over\Lambda}\,
+\left({1\over \Lambda+\omega+\omega_1-iE_{12}+iL_S^{(0)}}-{1\over\Lambda}\right)\\
\label{resolvent_expansion}
&&\hspace{-4cm}= \, {1-Z^{(1)}\over\Lambda}+
{d\over d\Lambda}\ln{\Lambda+\omega+\omega_1-iE_{12}+iL_S^{(0)}\over\Lambda}.
\end{eqnarray}
Together with the two vertices, this approximation contains systematically all terms of 
$O({1\over\Lambda}J^2)$,
$O({1\over\Lambda}J^3)$, and $O[{1\over\Lambda}({\Delta\over\Lambda})^k J^2]$, with $k>1$,
which are important to calculate the vertex up to $O(J^2)$. The last three terms on the r.h.s.
of Eq.~(\ref{rg_G_matsubara}) are already $\sim J^3$, therefore we replace the resolvents by their lowest 
order term $\sim{1\over\Lambda}$:
\begin{eqnarray}
\nonumber
i\,\Pi(E_{23},\Lambda+\omega+\omega_3)\,&\approx &\,{1\over \Lambda+\omega_3}\,,\\
\nonumber
i\,\Pi(E_{11'23},\Lambda+\omega+\omega_1+\omega_1^\prime+\omega_3)\,
&\approx &\,{1\over \Lambda+\omega_3}\,,\\
\nonumber
i\,\Pi(E_{12},\Lambda+\omega+\omega_1)\,
&\approx &\,{1\over \Lambda}\,.
\end{eqnarray}
Note that $\omega_3$ is an integration variable and has to be kept in the resolvent.
As a result, we get the following RG equation for the vertex in second
order in $J$:
\begin{widetext}
\begin{eqnarray}
\nonumber
{d\bar{G}_{11'}^{(2)}(E,\omega,\omega_1,\omega_1^\prime)\over d\Lambda}\,&=&\,\\
\label{G_second_term_1}
&&\hspace{-3cm}
=\,\left\{\bar{G}_{12}^{(1)}\left(-{Z^{(1)}\over\Lambda}+{d\over d\Lambda}
\ln{\Lambda+\omega+\omega_1-iE_{12}+iL_S^{(0)}\over \Lambda}\right)
\bar{G}_{\bar{2}1'}^{(1)}-(1\leftrightarrow 1')\right\}\\
\label{G_second_term_2}
&&\hspace{-3cm}
+\,{1\over\Lambda}\left\{\bar{G}_{12}^{(1)}
\bar{G}_{\bar{2}1'}^{(2)}(E_{12},\Lambda+\omega+\omega_1,-\Lambda,\omega_1^\prime)
+\bar{G}_{12}^{(2)}(E,\omega,\omega_1,\Lambda)\bar{G}_{\bar{2}1'}^{(1)}
-(1\leftrightarrow 1')\right\}\\
\label{G_second_term_3}
&& \hspace{-5cm} 
-\,\int_0^\Lambda d\omega_3\,\left({1\over \Lambda+\omega_3}\right)^2\,
\bar{G}_{23}^{(1)}\bar{G}_{11'}^{(1)}\bar{G}_{\bar{3}\bar{2}}^{(1)}
+{1\over\Lambda}\int_0^\Lambda d\omega_3\,{1\over\Lambda+\omega_3}\,
\left\{\bar{G}_{12}^{(1)}\bar{G}_{1'3}^{(1)}\bar{G}_{\bar{3}\bar{2}}^{(1)}
-\bar{G}_{23}^{(1)}\bar{G}_{\bar{3}1}^{(1)}\bar{G}_{1'\bar{2}}^{(1)}
- (1\leftrightarrow 1')\right\}\quad.
\end{eqnarray}
\end{widetext}
The first term [Eq.~(\ref{G_second_term_1})] on the r.h.s. containing the logarithm
induces the frequency dependence of the vertex. If we neglect in this term
the derivative of $\bar{G}^{(1)}$ with respect to $\Lambda$ (giving rise to terms
of $O[{1\over\Lambda}({\Delta\over\Lambda})^k J^3]$, with $k>1$, contributing to $\bar{G}^{(3)}$),
we can decompose the vertex in the following way to solve the above RG equation in $O(J^2)$:
\begin{equation}
\label{G_second_decomposition}
\bar{G}_{11'}^{(2)}(E,\omega,\omega_1,\omega_1^\prime)\,=\,
\bar{G}_{11'}^{(2a)}\,+\,
\bar{G}_{11'}^{(2b)}(E,\omega,\omega_1,\omega_1^\prime)\,,
\end{equation}
with
\begin{eqnarray}
\nonumber
\bar{G}_{11'}^{(2b)}(E,\omega,\omega_1,\omega_1^\prime)\,&=&\,\\
\nonumber
&&\hspace{-3cm}
=\,\bar{G}_{12}^{(1)}
\ln{\Lambda+\omega+\omega_1-iE_{12}+iL_S^{(0)}\over \Lambda}
\bar{G}_{\bar{2}1'}^{(1)}\\
\label{G_second_frequency}
&&\hspace{-3cm}
-\,\bar{G}_{1'2}^{(1)}
\ln{\Lambda+\omega+\omega_1^\prime-iE_{1'2}+iL_S^{(0)}\over \Lambda}
\bar{G}_{\bar{2}1}^{(1)}\,.
\end{eqnarray}
Using this solution in the second term [Eq.~(\ref{G_second_term_2})] on the r.h.s., and neglecting
again terms of $O[{1\over\Lambda}({\Delta\over\Lambda})^k J^3]$, we can use the approximations
\begin{eqnarray}
\nonumber
\bar{G}_{\bar{2}1'}^{(2b)}(E_{12},\Lambda+\omega+\omega_1,-\Lambda,\omega_1^\prime)
&\approx & -\ln{2}\,\bar{G}_{1'3}^{(1)}\bar{G}_{\bar{3}\bar{2}}^{(1)}\\
\nonumber
\bar{G}_{12}^{(2b)}(E,\omega,\omega_1,\Lambda)
&\approx & -\ln{2}\,\bar{G}_{23}^{(1)}\bar{G}_{\bar{3}1}^{(1)}
\end{eqnarray}
in Eq.~(\ref{G_second_term_2}), which lead to two terms cancelling precisely the 
second term of Eq.~(\ref{G_second_term_3}). Thus, we obtain the following RG equation
for the frequency-independent part $\bar{G}^{(2a)}_{11'}$ of the vertex in $O(J^2)$:
\begin{widetext}
\begin{eqnarray}
\nonumber
{d\bar{G}_{11'}^{(2a)}\over d\Lambda}\,&=&\,
{1\over \Lambda}\left\{\bar{G}_{12}^{(1)}\bar{G}_{\bar{2}1'}^{(2a)}
+\bar{G}_{12}^{(2a)}\bar{G}_{\bar{2}1'}^{(1)}-(1\leftrightarrow 1')\right\}\\
\label{RG_G_2a}
&&-\,{1\over \Lambda}\left\{\bar{G}_{12}^{(1)}Z^{(1)}\bar{G}_{\bar{2}1'}^{(1)}
-(1\leftrightarrow 1')\right\}\,
-\,{1\over 2\Lambda}\bar{G}_{23}^{(1)}\bar{G}_{11'}^{(1)}\bar{G}_{\bar{3}\bar{2}}^{(1)}\quad,
\end{eqnarray}
\end{widetext}
where the initial condition is given by the second term of the inital condition 
(\ref{G_initial}) for the vertex. It can be shown \cite{RTRG_FS} that the form of the initial 
condition is preserved [with the vertices given by the leading-order solutions (\ref{G_bar}) and
(\ref{G_tilde})] if one considers only the first term on the r.h.s. of Eq.~(\ref{RG_G_2a}).
Therefore, we decompose the frequency-independent part of the vertex
in second order as
\begin{equation}
\label{G_2a_decomposition}
\bar{G}^{(2a)}_{11'}\,=\,i\,\bar{G}^{(2a_1)}_{11'}\,+\,\bar{G}^{(2a_2)}_{11'}\quad,
\end{equation}
with
\begin{equation}
\label{G_2a_1}
\bar{G}^{(2a_1)}_{11'}\,=\,
-\,{\pi\over 2}\,\left(\bar{G}^{(1)}_{12}\,\tilde{G}^{(1)}_{\bar{2}1'}\,-\,
\bar{G}^{(1)}_{1'2}\,\tilde{G}^{(1)}_{\bar{2}1}\right)\quad,
\end{equation}
and $\bar{G}_{11'}^{(2a_2)}$ fulfils the same RG equation (\ref{RG_G_2a}) as 
$\bar{G}_{11'}^{(2a)}$, but with zero initial condition.
Since no explicit imaginary factors occur in Eq.~(\ref{RG_G_2a}), the
decomposition (\ref{G_2a_decomposition}) can also be viewed as a decomposition 
of the vertex into real and complex parts (provided there are no complex terms in the initial
condition for the original quantities). Therefore, the two parts have a completely different
physical meaning. Whereas the part $\bar{G}_{11'}^{(2a_1)}$ is even important to calculate
the rates in second order in $J$ (see below), the part $\bar{G}_{11'}^{(2a_2)}$ denotes
a renormalization of the coupling constants in two-loop order, which can lead to logarithmic
corrections of the form $\sim J^2\ln{J\over J_0}$, where $J_0$ denotes the original coupling
constant (see Sec.~\ref{sec:kondo_above_Lambda_c}, where such terms are explicitly calculated
for the Kondo model). Such terms are not well defined in the scaling limit
$J_0\rightarrow 0$ and, therefore, should be taken together with the leading order vertex
by redefining certain characteristic low-energy scales (like the Kondo temperature for the
Kondo model). Thus, in the following we will redefine the leading-order vertex by
the replacement
\begin{equation}
\label{rg_G_replacement}
\bar{G}^{(1)}_{11'}\,\rightarrow\,\bar{G}^{(1)}_{11'}\,+\,\bar{G}^{(2a_2)}_{11'}\quad,
\end{equation}
and will consider only the part $\bar{G}^{(2a_1)}_{11'}$ explicitly in all equations. We note that
it can not generically be shown that the replacement (\ref{rg_G_replacement}) can be accounted for
by just renormalizing the Kondo temperature in the lowest-order vertices. For the Kondo model, this
can be shown in all orders \cite{andrei_furuya_lowenstein_RMP83} because only one coupling constant 
remains in the scaling limit $J_0\rightarrow 0$,
$D\rightarrow\infty$, such that $T_K$ stays constant. However, for more complicated models including
orbital degrees of freedoms, interference effects, etc., it can happen that the matrix structure
of $\bar{G}^{(2a_2)}_{11'}$ in Liouville space is different from that of $\bar{G}^{(1)}_{11'}$. In 
this case, new terms can finally arise from the second-order part of the vertices, and their influence
might be quite nontrivial.

We note that all equations also hold for the
current vertex $\bar{I}^{\gamma(2)}$ by replacing the first vertex in all terms on the r.h.s. 
of the RG equations by the current vertex. Thereby Eq.~(\ref{G_2a_1}) is only valid 
if the trace $\text{Tr}_S$ over the local quantum system is taken from the left
(which we always implicitly assume for the current vertex and the current kernel). 

{\it Liouvillian in second order.}
We proceed with RG equation (\ref{rg_L_matsubara}) for the Liouvillian 
(or equivalently the current kernel by replacing the first vertex by the 
current vertex in all equations).
Inserting the expansion (\ref{G_expansion}) into this RG equation, using Eqs.~(\ref{G_second_decomposition}),
(\ref{G_2a_decomposition}), and (\ref{rg_G_replacement}), and neglecting all terms of $O(J^4)$ on the r.h.s., 
we obtain
\begin{widetext}
\begin{eqnarray}
\label{L_expansion_term_1}
{dL_S(E,\omega)\over d\Lambda}\,&\approx&
\bar{G}_{12}^{(1)}\,\Pi(E_{12},\Lambda+\omega+\omega_2)\,
\bar{G}_{\bar{2}\bar{1}}^{(1)}\,\\
\label{L_expansion_term_2}
&&\hspace{-1cm}
+\,i\,\bar{G}_{12}^{(1)}\,\Pi(E_{12},\Lambda+\omega+\omega_2)\,
\bar{G}_{\bar{2}\bar{1}}^{(2a_1)}
\,+\,i\,\bar{G}_{12}^{(2a_1)}\,\Pi(E_{12},\Lambda+\omega+\omega_2)\,
\bar{G}_{\bar{2}\bar{1}}^{(1)}\\
\label{L_expansion_term_3}
&&\hspace{-3cm}
+\,\bar{G}_{12}^{(1)}\,\Pi(E_{12},\Lambda+\omega+\omega_2)\,
\bar{G}_{\bar{2}\bar{1}}^{(2b)}(E_{12},\Lambda+\omega+\omega_2,-\omega_2,-\Lambda)
+\,\bar{G}_{12}^{(2b)}(E,\omega,\Lambda,\omega_2)\,
\Pi(E_{12},\Lambda+\omega+\omega_2)\,
\bar{G}_{\bar{2}\bar{1}}^{(1)}\\
\label{L_expansion_term_4}
&&\hspace{-2cm}
 - \,i\,\bar{G}_{12}^{(1)}\,\Pi(E_{12},\Lambda+\omega+\omega_2)\,
\bar{G}_{\bar{2}3}^{(1)}\,\Pi(E_{13},\Lambda+\omega+\omega_3)\,
\bar{G}_{\bar{3}\bar{1}}^{(1)}\quad.
\end{eqnarray}
\end{widetext}
For the terms (\ref{L_expansion_term_1}), (\ref{L_expansion_term_2}), and
(\ref{L_expansion_term_4}) we need the resolvent integrated over frequency, 
which, by using Eqs.~(\ref{resolvent_form}), (\ref{tilde_L}), (\ref{K_function}) 
and (\ref{K_tilde_function}), can be expanded as
\begin{widetext}
\begin{eqnarray}
\nonumber
i\,\int_0^\Lambda\,d\omega_2\,\Pi(E_{12},\Lambda+\omega+\omega_2)\,&\approx&\,
\left(1-{Z^{(1)}\over 2}\right)\,{\cal{K}}_\Lambda(E_{12}+i\omega-L_S^{(0)}-\tilde{L}_S^{(1)})\,
\left(1-{Z^{(1)}\over 2}\right)\\
\nonumber
&&\hspace{-5cm}
\approx\,{i\over 2\Lambda}\,(1-{Z^{(1)}\over 2})\,(E_{12}+i\omega-L_S^{(0)}-\tilde{L}_S^{(1)})\,
(1-{Z^{(1)}\over 2})\,+\,{\tilde{\cal{K}}}_\Lambda(E_{12}+i\omega-L_S^{(0)}-\tilde{L}_S^{(1)})\\
\label{integral_resolvent_expansion}
&&\hspace{-5cm}
\approx\,{i\over 2\Lambda}\,(E_{12}+i\omega-L_S^{(0)})\,
-\,{i\over 2\Lambda}\,
\left((E_{12}+i\omega)Z^{(1)}+L_S^{(1)}-Z^{(1)}L_S^{(0)}-L_S^{(0)}Z^{(1)}\right)\,
\,+\,{\tilde{\cal{K}}}_\Lambda(E_{12}+i\omega-L_S^{(0)}).
\end{eqnarray}
\end{widetext}
The first term is of order $O({\Delta\over\Lambda})$, the second one of order 
$O({\Delta\over\Lambda}J)$, and the last one contains terms of $O(1)$ and $O[({\Delta\over\Lambda})^k]$,
with $k>1$. Therefore, when multiplied with $J^n$ and integrated, the first term gives
a contribution of order $O(\Delta J^{n-1})$, the second one leads to $O(\Delta J^n)$, and
the last one gives $O(\Lambda_0 J^n)$ or $O(\Delta J^n)$. Thereby, the terms 
$\sim \Lambda_0 J^n$ are cancelled by the initial condition from the first RG step (see below).
Therefore, for the calculation of ${d\over d\Lambda}L_S^{(2)}(E,\omega)$, the terms
(\ref{L_expansion_term_1}) and (\ref{L_expansion_term_2}) give rise to
\begin{widetext}
\begin{eqnarray}
\label{L_2_term_a}
{d L_S^{(2)}(E,\omega)\over d\Lambda}\,&\rightarrow &\,
-i\,{d\over d\Lambda}\left\{\bar{G}_{12}^{(1)}\tilde{F}_\Lambda(E_{12}+i\omega-L_S^{(0)})
\bar{G}_{\bar{2}\bar{1}}^{(1)}\right\}\\
\label{L_2_term_b}
&&+\,{i\over 2\Lambda}\,\left\{\bar{G}_{12}^{(1)}(E_{12}+i\omega-L_S^{(0)})
\bar{G}_{\bar{2}\bar{1}}^{(2a_1)}\,
+\,\bar{G}_{12}^{(2a_1)}(E_{12}+i\omega-L_S^{(0)})\bar{G}_{\bar{2}\bar{1}}^{(1)}\right\}\\
\label{L_2_term_c}
&&-\,{1\over 2\Lambda}\,\bar{G}_{12}^{(1)}\left\{(E_{12}+i\omega)Z^{(1)}+L_S^{(1)}-
Z^{(1)}L_S^{(0)}-L_S^{(0)}Z^{(1)}\right\}\bar{G}_{\bar{2}\bar{1}}^{(1)}.
\end{eqnarray}
\end{widetext}
As is shown in Appendix \ref{sec:appendix_A}, the other two terms, Eqs.~(\ref{L_expansion_term_3})
and (\ref{L_expansion_term_4}), have nearly no effect when expanded systematically, one just has to
replace the function $\tilde{F}_\Lambda(z)$ in Eq.~(\ref{L_2_term_a}) by the function
\begin{equation}
\label{tilde_F_prime}
\tilde{F}_\Lambda^\prime(z)\,=\,
\tilde{F}_\Lambda(z)\,-\,{iz\over 2}\,\ln{2}\quad.
\end{equation}
According to the three terms Eqs.~(\ref{L_2_term_a})--(\ref{L_2_term_c}), we
decompose
\begin{eqnarray}
\nonumber
L_S^{(2)}(E,\omega)\,&=&\,L_S^{(2a)}(E,\omega)\,+\,L_S^{(2b)}\,+\,L_S^{(2c)}\,\\
\label{L_2_decomposition}
&&-\,(E+i\omega)\,(Z^{(2b)}\,+\,Z^{(2c)})\quad,
\end{eqnarray}
with
\begin{eqnarray}
\label{L_2a}
\hspace{-1cm}
L_S^{(2a)}(E,\omega)\,&=&\,
-i\bar{G}_{12}^{(1)}\tilde{F}^\prime_\Lambda(E_{12}+i\omega-L_S^{(0)})
\bar{G}_{\bar{2}\bar{1}}^{(1)}\,,\\
\nonumber
\hspace{-1cm}
{d L_S^{(2b)}\over d\Lambda}\,&=&\,
{i\over 2\Lambda}\left\{\bar{G}_{12}^{(1)}(\bar{\mu}_{12}-L_S^{(0)})
\bar{G}_{\bar{2}\bar{1}}^{(2a_1)}+\right.\\
\label{L_2b}
\hspace{-1cm}
&&\left.+\bar{G}_{12}^{(2a_1)}(\bar{\mu}_{12}-L_S^{(0)})\bar{G}_{\bar{2}\bar{1}}^{(1)}\right\}\,,\\
\label{Z_2b}
\hspace{-1cm}
{d Z^{(2b)}\over d\Lambda}\,&=&\,
-{i\over 2\Lambda}\left\{\bar{G}_{12}^{(1)}\bar{G}_{\bar{2}\bar{1}}^{(2a_1)}
+\bar{G}_{12}^{(2a_1)}\bar{G}_{\bar{2}\bar{1}}^{(1)}\right\}\,,\\
\nonumber
\hspace{-1cm}
{d L_S^{(2c)}\over d\Lambda}\,&=&\,
-{1\over 2\Lambda}\bar{G}_{12}^{(1)}\left\{(\bar{\mu}_{12}Z^{(1)}+L_S^{(1)}-\right.\\
\label{L_2c}
\hspace{-1cm}
&&\left. -Z^{(1)}L_S^{(0)}-L_S^{(0)}Z^{(1)}\right\}\bar{G}_{\bar{2}\bar{1}}^{(1)}\,,\\
\label{Z_2c}
\hspace{-1cm}
{d Z_S^{(2c)}\over d\Lambda}\,&=&\,
{1\over 2\Lambda}\bar{G}_{12}^{(1)}Z^{(1)}\bar{G}_{\bar{2}\bar{1}}^{(1)}\,.
\end{eqnarray}
Note that, according to Eq.~(\ref{F_tilde_asymptotic}), the initial value of
$L_S^{(2a)}(E,\omega)$ is given by
\begin{equation}
\label{L_2a_initial}
L_S^{(2a)}(E,\omega)|_{\Lambda=\Lambda_0}\,=\,-i\,\ln(2)\,\Lambda_0\,
\bar{G}_{12}^{(1)}\,\bar{G}_{\bar{2}\bar{1}}^{(1)}|_{\Lambda=\Lambda_0}\,,
\end{equation}
which coincides with the first term of the inital condition (\ref{Sigma_initial})  
from the first RG step if we choose
\begin{equation}
\label{Lambda_initial}
\Lambda_0\,=\,{\pi^2\over 16\ln(2)}\,D\,.
\end{equation}

One can group the various terms occurring in Eqs.~(\ref{L_2_decomposition})--(\ref{Z_2c}) 
regarding the occurence of the complex factor $i$. The RG equations (\ref{L_2c}) and (\ref{Z_2c})
do not contain an explicit factor $i$ and integrating them
up to $\Lambda_c$, one generates a second-order contribution which is negligible 
compared to the first-order terms (\ref{L_first_const}) and (\ref{Z_first}) (up to this value
of the cutoff parameter, no logarithmic contributions are generated, i.e., the terms just have
one factor more in $J$). In contrast, the second-order terms containing the factor $i$ are
not negligible because they do not occur in first order. They are generated by
Eq.~(\ref{L_2a}) [with $F^\prime_\Lambda(z)$ replaced by its real value] and by Eqs.~(\ref{L_2b})
and (\ref{Z_2b}). This is a general rule that the terms containing the factor $i$ are always 
generated one order higher in $J$ compared to those without this factor.

\subsubsection{RG below $\Lambda_c$}
\label{sec:generic_below_Lambda_c}

The two-loop RG until $\Lambda_c$ has resummed all leading
and subleading logarithmic contributions in $\ln{D\over\Lambda_c}$ into the renormalized vertices.
This means that we have considered all terms of the form
\begin{equation}
\label{log_terms}
J_0^k\,\ln^l{D\over\Lambda_c}\quad,\quad l=k-1,k-2\quad,
\end{equation} 
for the Liouvillian and the vertices, where $J_0$ denotes the original
coupling constant. The terms with $l=k-1$ and $l=k-2$ are defined
in our terminology as one-loop and two-loop terms, respectively, irrespective of the topology of the 
diagrams, which is not a unique definition and depends on the formalism used. Roughly speaking, at 
$\Lambda=\Lambda_c$, the band width $D$ has been replaced by an effective band width 
$\Lambda_c$ and the bare coupling constant is replaced by a renormalized one $J_c$ (including
one-loop and two-loop renormalizations). This eliminates all the logarithmic contributions of
Eq.~(\ref{log_terms}), and a simple power series in $J_c$ remains, which is well defined for
$J_c\ll 1$, as described in detail at the beginning of Sec.~\ref{sec:generic_2_loop}.

As a consequence, we solve the RG equations perturbatively in $J_c$ in the regime $0<\Lambda<\Lambda_c$. 
Furthermore, we replace the Liouvillian in the resolvents by the full effective Liouvillian 
$L_S^{eff}(E,\omega)$, and we use the approximation
(\ref{Pi_general_approximation})
\begin{equation}
\label{Pi_Lambda_c_approximation}
\Pi(E,\omega)\,\approx \,{\tilde{Z}\over E+i\omega-\tilde{L}_S}\quad,
\end{equation}
together with the convention (\ref{tilde_LZ_convention}).
The corrections to this approximation are at least of order $O(J_c)$ and contain no poles in the
variable $E+i\omega$ (at least if we assume that the projectors on the eigenvectors do not contain
poles or do not contribute). Therefore, when inserted in the RG equations, these corrections lead to terms
of order $O(J_c^3)$ without any logarithmic enhancement, which we neglect in the following.
The eigenvalues of $\tilde{L}_S$ are given by $\tilde{h}_i-i\tilde{\Gamma}_i$ with $\tilde{\Gamma}_i>0$.
These eigenvalues are finally calculated self-consistently from Eq.~(\ref{pole_equation}). In contrast,
the $Z$~factor is expanded as
\begin{equation}
\label{Z_expansion}
\tilde{Z}\,=\,1\,+\,\tilde{Z}^{(1)}\,+\,O(J_c^2)\quad,
\end{equation}
where $\tilde{Z}^{(1)}\sim J_c\ll 1$. This first-order correction to the $Z$~factor can in principle
be calculated from the effective Liouvillian up to first order in $J_c$, given by the result
(\ref{L_eff_1_order}). If the eigenvalues of $L_S^{(0)}$ are separated well compared to the 
first-order corrections to the effective Liouvillian, we get 
\begin{equation}
\label{Z_tilde_1}
\tilde{Z}^{(1)}\,=\,-Z^{(1)c}_d\quad,
\end{equation} 
where $Z^{(1)c}_d$ is the diagonal part of $Z^{(1)c}$ with respect to the eigenbasis of $L_S^{(0)}$.
However, we will see later that $\tilde{Z}^{(1)}$ does not enter our final result so its precise value is 
of no relevance.

In the RG equations, the variable $E$ is replaced by $E_{1\dots n}$ and $\omega$ is
the imaginary part of the Laplace variable plus the integration variables (the cutoff 
$\Lambda$ and some Masubara frequencies). Thus, in the end the low-energy cutoff 
will be given by an expression of the form
\begin{equation}
\label{low_energy_cutoff}
\Delta_{1\dots n}\,=\,E_{1\dots n}\,+\,i\omega\,-\,\tilde{L}_S\quad,
\end{equation}
where from now on $\omega$ is the imaginary part of the original Laplace variable. 
Resonance positions are defined by $E_{1\dots n}=\tilde{h}_i$, see Eq.~(\ref{resonance_positions}).
At these points, logarithmic terms of the form
\begin{equation}
\label{log_terms_with_gamma}
J_c^k\,\ln^l{\Lambda_c\over|\Delta|}\quad,\quad 
l=k-1,k-2,\dots,0\quad,
\end{equation} 
are generated with $\Delta\equiv E_{1\dots n}-\tilde{h}_i+i\tilde{\Gamma}_i$.
At resonance, the logarithmic terms are of order $J_c^k\ln^l J_c$ since 
$\tilde{\Gamma}_i\sim \Lambda_c J_c^2$, leading to enhanced, but still perturbative corrections.

Our first aim is to collect all terms of the form (\ref{log_terms_with_gamma})
with $k\le 3$ and $l=k-1,k-2$, i.e., all terms of the form
\begin{eqnarray}
\nonumber
&&O(1)\,,\,O(J_c)\,,\,O(J_c^2)\,,\,O(J_c^2\ln{\Lambda_c\over|\Delta+i\Gamma|}),\\
\label{log_terms_considered}
&&O(J_c^3\ln{\Lambda_c\over|\Delta+i\Gamma|})
\,,\,\,O(J_c^3\ln^2{\Lambda_c\over|\Delta+i\Gamma|})\quad.
\end{eqnarray}
Only terms of $O(J_c^3)$ without a logarithmic factor are neglected [note that 
approximation (\ref{Pi_Lambda_c_approximation}) for the resolvent is already neglecting
such terms]. Therefore, it is not necessary
to calculate the Liouvillian in third order in $J$ for $\Lambda>\Lambda_c$ since no
logarithmic contributions are generated up to $\Lambda_c$. Furthermore, for $\Lambda<\Lambda_c$,
we see that the lowest order term on the r.h.s of the RG equation (\ref{rg_L_matsubara}) for the
Liouvillian is already of $O(J_c^2)$. Therefore, we need the vertices only up to $O(J_c^2)$, and
it is not necessary to calculate the vertices up to third order for $\Lambda>\Lambda_c$.

{\it Vertices.} We start with the perturbative expansion for the vertices. Up to order
$O(J_c^2)$, we get from the RG equation (\ref{rg_G_matsubara}) together with 
Eqs.~(\ref{Pi_Lambda_c_approximation}), (\ref{Z_expansion}), and (\ref{low_energy_cutoff})
\begin{eqnarray}
\nonumber
\hspace{-1cm}
{d\bar{G}_{11'}(E,\omega,\omega_1,\omega_1^\prime)\over d\Lambda}\,&=&\,\\
\label{rg_G_second_order}
&&\hspace{-3cm}
=\,\left\{\bar{G}_{12}^{(1)c}
{1\over\Lambda+\omega_1-i\Delta_{12}} 
\bar{G}_{\bar{2}1'}^{(1)c}-(1\leftrightarrow 1')\right\}\,.
\end{eqnarray}
Integrating this equation from $\Lambda=\Lambda_c$ to $\Lambda$, we obtain
in $O(J_c^2)$
\begin{eqnarray}
\nonumber
\bar{G}_{11'}^{(2)}(E,\omega,\omega_1,\omega_1^\prime)\,&=&\,
\bar{G}_{11'}^{(2)c}(E,\omega,\omega_1,\omega_1^\prime)\,-\\
\nonumber
&&\hspace{-3cm}
-\,\left\{\bar{G}_{12}^{(1)c}
\ln{\Lambda_c+\omega_1-i\Delta_{12} \over \Lambda+\omega_1-i\Delta_{12}} 
\bar{G}_{\bar{2}1'}^{(1)c}-(1\leftrightarrow 1')\right\}\,,
\end{eqnarray}
where $\bar{G}_{11'}^{(2)c}(E,\omega,\omega_1,\omega_1^\prime)$ is the value
of the vertex in second order at $\Lambda=\Lambda_c$. Inserting this value
from Eqs.~(\ref{G_second_decomposition}), (\ref{G_second_frequency}), and 
(\ref{G_2a_decomposition}), we find up to $O(J_c^2)$ 
\begin{eqnarray}
\nonumber
\bar{G}_{11'}^{(2)}(E,\omega,\omega_1,\omega_1^\prime)\,&=&\,
i\bar{G}_{11'}^{(2a_1)c}\,+\\
\label{G_second_order}
&&\hspace{-3.5cm}
+\,\left\{\bar{G}_{12}^{(1)c}
\ln{\Lambda+\omega_1-i\Delta_{12} \over \Lambda_c} 
\bar{G}_{\bar{2}1'}^{(1)c}-(1\leftrightarrow 1')\right\}\,,
\end{eqnarray}
where we have used that $\Delta_{12}=E_{12}+i\omega-L_S^{(0)}+O(J_c)$. Note that the 
part $\bar{G}^{(2a_2)}_{11'}$ is always taken together with $\bar{G}^{(1)}_{11'}$, 
see Eq.~(\ref{rg_G_replacement}), therefore it does not occur in Eq.~(\ref{G_second_order}).

{\it Liouvillian.} We proceed with the perturbative expansion for the Liouvillian
given by the RG equation (\ref{rg_L_matsubara}). In the second term on
the r.h.s. of this RG equation, we can replace the vertex by the lowest-order term $\bar{G}^{(1)c}$
and the resolvent by $\Pi(E,\omega)\rightarrow {1\over E+i\omega-\tilde{L}_S}$,
according to Eq.~(\ref{Pi_Lambda_c_approximation}). In the first term on the r.h.s., we have
to consider the first-order correction $\tilde{Z}^{(1)}$ of the $Z$~factor as well, and we 
have to use the second-order term for the vertex given by Eq.~(\ref{G_second_order}), which gives
\begin{widetext}
\begin{eqnarray}
\label{G_second_1}
\bar{G}_{12}^{(2)}(E,\omega,\Lambda,\omega_2)\,&=&\,
i\bar{G}_{12}^{(2a_1)c}\,+\,
\bar{G}_{13}^{(1)c}\ln{2\Lambda-i\Delta_{13} \over \Lambda_c} 
\bar{G}_{\bar{3}2}^{(1)c}
\,-\,\bar{G}_{23}^{(1)c}\ln{\Lambda+\omega_2-i\Delta_{23} \over \Lambda_c} 
\bar{G}_{\bar{3}1}^{(1)c}\,,\\
\label{G_second_2}
\bar{G}_{\bar{2}\bar{1}}^{(2)}(E_{12},\Lambda+\omega+\omega_2,-\omega_2,-\Lambda)\,&=&\,
i\bar{G}_{\bar{2}\bar{1}}^{(2a_1)c}\,+\,
\bar{G}_{\bar{2}3}^{(1)c}\ln{2\Lambda-i\Delta_{13} \over \Lambda_c} 
\bar{G}_{\bar{3}\bar{1}}^{(1)c}
\,-\,\bar{G}_{\bar{1}3}^{(1)c}\ln{\Lambda+\omega_2-i\Delta_{23} \over \Lambda_c} 
\bar{G}_{\bar{3}\bar{2}}^{(1)c}\,.
\end{eqnarray}
For the frequency integral over the resolvent, we use
\begin{equation}
\label{resolvent_integral}
\int_0^\Lambda d\omega_2{1\over i\Lambda+i\omega_2+\Delta}=
-i{\cal{K}}_\Lambda(\Delta)=-i\ln{2\Lambda-i\Delta \over \Lambda-i\Delta}\quad.
\end{equation}

Using these replacements in Eq.~(\ref{rg_L_matsubara}) and collecting the various terms
according to their order in $J_c$, we obtain up to $O(J_c^3)$
\begin{equation}
\label{L_expansion_below_c}
L_S(E,\omega)\,=\,L_S^{(0)}\,+\,L_S^{(1)c}(E,\omega)\,+\,
L_S^{(2)}(E,\omega)\,+\,L_S^{(3a)}(E,\omega)\,+\,L_S^{(3b)}(E,\omega)\quad,
\end{equation}
with
\begin{eqnarray}
\label{rg_L_2}
{dL_S^{(2)}(E,\omega)\over d\Lambda}&=&
-i\bar{G}_{12}^{(1)c}{\cal{K}}_\Lambda(\Delta_{12})\bar{G}_{\bar{2}\bar{1}}^{(1)c}\,,\\
\nonumber
{dL_S^{(3a)}(E,\omega)\over d\Lambda}&=&
-{i\over 2}\bar{G}_{12}^{(1)c}
\left\{\tilde{Z}^{(1)}{\cal{K}}_\Lambda(\Delta_{12})+
{\cal{K}}_\Lambda(\Delta_{12})\tilde{Z}^{(1)}\right\}
\bar{G}_{\bar{2}\bar{1}}^{(1)c}\\
\label{rg_L_3a}
&&
+\bar{G}_{12}^{(1)c}{\cal{K}}_\Lambda(\Delta_{12})
\bar{G}_{\bar{2}\bar{1}}^{(2a_1)c}
+\bar{G}_{12}^{(2a_1)c}{\cal{K}}_\Lambda(\Delta_{12})
\bar{G}_{\bar{2}\bar{1}}^{(1)c}\,,\\
\nonumber
{dL_S^{(3b)}(E,\omega)\over d\Lambda}&=&
-i\bar{G}_{12}^{(1)c}{\cal{K}}_\Lambda(\Delta_{12})
\bar{G}_{\bar{2}3}^{(1)c}\ln{2\Lambda-i\Delta_{13} \over \Lambda_c}
\bar{G}_{\bar{3}\bar{1}}^{(1)c}
-i\bar{G}_{12}^{(1)c}\ln{2\Lambda-i\Delta_{12} \over \Lambda_c}
\bar{G}_{\bar{2}3}^{(1)c}{\cal{K}}_\Lambda(\Delta_{13})
\bar{G}_{\bar{3}\bar{1}}^{(1)c}\\
\nonumber
&&\hspace{-2cm}
-i \bar{G}_{12}^{(1)c}\int_\Lambda^{2\Lambda} d\omega_2\left\{
{1\over \omega_2-i\Delta_{12}}
\bar{G}_{\bar{2}3}^{(1)c}\ln{\omega_2-i\Delta_{13} \over \Lambda_c}
+\ln{\omega_2-i\Delta_{12} \over \Lambda_c}
\bar{G}_{\bar{2}3}^{(1)c}{1\over \omega_2-i\Delta_{13}}
\right\}\bar{G}_{\bar{3}\bar{1}}^{(1)c}\\
\label{rg_L_3b}
&&
+i\bar{G}_{12}^{(1)c}{\cal{K}}_\Lambda(\Delta_{12})
\bar{G}_{\bar{2}3}^{(1)c}{\cal{K}}_\Lambda(\Delta_{13})
\bar{G}_{\bar{3}\bar{1}}^{(1)c}\,.
\end{eqnarray}
Using
\begin{equation}
\label{integral}
\int_\Lambda^{2\Lambda}dx\left\{{1\over x+a}\left(\ln{x+b\over \Lambda_c}\right)
+\left(\ln{x+a\over \Lambda_c}\right){1\over x+b}\right\}=
\ln{2\Lambda+a\over\Lambda_c}\ln{2\Lambda+b\over\Lambda_c}
-\ln{\Lambda+a\over\Lambda_c}\ln{\Lambda+b\over\Lambda_c}
\end{equation}
for the third term on the r.h.s. of Eq.~(\ref{rg_L_3b}), we obtain after some
manipulations
\begin{equation}
\label{rg_L_3b_simplified}
{dL_S^{(3b)}(E,\omega)\over d\Lambda}=
-2i\bar{G}_{12}^{(1)c}\left\{\ln{2\Lambda-i\Delta_{12}\over\Lambda_c}
\bar{G}_{\bar{2}3}^{(1)c}\ln{2\Lambda-i\Delta_{13}\over\Lambda_c}
-\ln{\Lambda-i\Delta_{12}\over\Lambda_c}
\bar{G}_{\bar{2}3}^{(1)c}\ln{\Lambda-i\Delta_{13}\over\Lambda_c}\right\}
\bar{G}_{\bar{3}\bar{1}}^{(1)c}\,.
\end{equation}
Equations~(\ref{rg_L_2}) and (\ref{rg_L_3a}) can be integrated easily from
$\Lambda=\Lambda_c$ to $\Lambda=0$ by using 
\begin{equation}
\label{F_function}
{\cal{K}}_\Lambda(\Delta)\,=\,{d\over d\Lambda}\,F_\Lambda(\Delta)\quad,\quad
F_\Lambda(\Delta)\,=\,\tilde{F}^\prime_\Lambda(\Delta)\,+\,{i\Delta\over 2}\,
\left(\ln{\Lambda\over -2i\Delta}\,+\,1\right)\,,
\end{equation}
where $\tilde{F}^\prime_\Lambda(\Delta)$ is defined by Eqs.~(\ref{tilde_F_prime}) and
(\ref{F_tilde_function}). Using $F_{\Lambda=0}(\Delta)=-{i\Delta\over 2}\ln{2}$, we
get 
\begin{equation}
\label{cal_L_integral}
\int_{\Lambda_c}^0\,d\Lambda\,{\cal{K}}_\Lambda(\Delta)\,=\,
-\tilde{F}^\prime_{\Lambda_c}(\Delta)\,-\,{i\Delta\over 2}
\left(\ln{\Lambda_c\over -i\Delta}\,+\,1\right)\,.
\end{equation}
The value of the Liouvillian at $\Lambda=\Lambda_c$ in second order has to be
taken from Eqs.~(\ref{L_2_decomposition})--(\ref{Z_2b}). When 
integrating Eq.~(\ref{rg_L_2}), the part from the first term on the r.h.s. of 
Eq.~(\ref{cal_L_integral}) cancels term (\ref{L_2a}) for 
$L_S^{(2a)c}(E,\omega)$ if we neglect the difference between $L_S^{(0)}$ and 
$\tilde{L}_S$ [leading to terms of order $O(J_c^3)$]. Thus, for $\Lambda=0$, we obtain the contributions
\begin{eqnarray}
\label{L_1_final}
L_S^{(1)}(E,\omega)_{\Lambda=0}&=&L_S^{(1)c}-(E+i\omega)Z^{(1)c}\quad,\\
\label{L_2_final}
L_S^{(2)}(E,\omega)_{\Lambda=0}&=&
L_S^{(2b)c}+L_S^{(2c)c}-
(E+i\omega)(Z^{(2b)c}+Z^{(2c)c})
-{1\over 2}\bar{G}_{12}^{(1)c}\Delta_{12}\left(\ln{\Lambda_c\over -i\Delta_{12}}+1\right)
\bar{G}_{\bar{2}\bar{1}}^{(1)c}\,,\\
\nonumber
L_S^{(3a)}(E,\omega)_{\Lambda=0}&=&
-{1\over 4}\bar{G}_{12}^{(1)c}
\left\{\tilde{Z}^{(1)}\Delta_{12}\left(\ln|{\Lambda_c\over \Delta_{12}}|\right)+
\Delta_{12}\left(\ln|{\Lambda_c\over \Delta_{12}}|\right)\tilde{Z}^{(1)}\right\}
\bar{G}_{\bar{2}\bar{1}}^{(1)c}\\
\label{L_3a_final}
&&-i{1\over 2}\bar{G}_{12}^{(1)c}\Delta_{12}\left(\ln|{\Lambda_c\over \Delta_{12}}|\right)
\bar{G}_{\bar{2}\bar{1}}^{(2a_1)c}
-i{1\over 2}\bar{G}_{12}^{(2a_1)c}\Delta_{12}\left(\ln|{\Lambda_c\over \Delta_{12}}|\right)
\bar{G}_{\bar{2}\bar{1}}^{(1)c}\,.
\end{eqnarray}
Thereby, we have only included the logarithmic terms in $L_S^{(3a)}$, i.e., 
terms $\sim J_c^3$ without a logarithmic factor are neglected. 

Finally, integrating Eq.~(\ref{rg_L_3b_simplified}), we obtain
\begin{equation}
\label{L_3b_final}
L_S^{(3b)}(E,\omega)_{\Lambda=0}\,=\,
-i\Lambda_c\int_0^1\,dx\,\bar{G}_{12}^{(1)c}\ln(x-i{\Delta_{12} \over \Lambda_c})\,
\bar{G}_{\bar{2}3}^{(1)c}\,\ln(x-i{\Delta_{13} \over \Lambda_c})\,
\bar{G}_{\bar{3}\bar{1}}^{(1)c}\,,
\end{equation}
Neglecting again terms $\sim J_c^3$, the double logarithmic integral can be replaced by
(with $a\equiv -i{\Delta_{12}\over\Lambda_c}$ and $b\equiv -i{\Delta_{13}\over\Lambda_c}$)
\begin{equation}
\label{double_logarithmic_integral}
\int_0^1\,dx\,\ln(x+a)\,\ln(x+b)
\,\approx\,-{1\over 2}a\ln^2{a}-{1\over 2}b\ln^2{b}
+a\ln{a}+b\ln{b}
+{1\over 2}a\ln^2{a\over b}-a\ln{a\over b}\quad,\quad \text{for}\quad 
|a|<|b|
\quad,
\end{equation}
and analog for $|a|>|b|$ by interchanging $a\leftrightarrow b$. For
some special cases,~(\ref{double_logarithmic_integral}) can be written
as 
\begin{equation}
\label{double_logarithmic_integral_special}
\int_0^1\,dx\,\ln(x+a)\,\ln(x+b)\,\approx\,
\left\{
\begin{array}{cl}
-{1\over 2}a\ln^2{a}-{1\over 2}b\ln^2{b}+a\ln{a}+b\ln{b} \\ -a\ln{a}\ln{b}  
\end{array}
\right.
\begin{array}{cl}
\text{for}\quad |a|\sim |b|\ll 1 \\
\text{for}\quad |a|\ll |b|\sim 1 \\
\end{array}
\quad.
\end{equation}

Equations~(\ref{L_1_final})--(\ref{L_3b_final}) together with 
Eqs.~(\ref{double_logarithmic_integral})--(\ref{double_logarithmic_integral_special})
are the final results for the effective Liouvillian
(or the current kernel if the first vertex in all terms is replaced by the current vertex) 
for a generic model of a quantum dot in the Coulomb blockade regime. The vertices
$\bar{G}^{(1)c}_{11'}$ and $\bar{G}^{(2a_1)c}_{11'}$ at $\Lambda=\Lambda_c$ follow from 
Eqs.~(\ref{G_reference_solution}) and (\ref{G_2a_1}). The first-order
quantities $L_S^{(1)c}$ and $Z^{(1)c}$ follow from the solution of the RG equations
(\ref{L_first_const}) and (\ref{Z_first}) at $\Lambda=\Lambda_c$, and the second-order terms
$L^{(2b)c}_S$, $L^{(2c)c}_S$, $Z^{(2b)c}$, and $Z^{(2c)c}$ are determined by the RG equations
(\ref{L_2b})--(\ref{Z_2c}). 

Looking back at Eqs.~(\ref{rg_L_2})--(\ref{rg_L_3b}), we see that the third-order terms involving
three vertices on the r.h.s. of the RG equations enter only explicitly via the last term on the r.h.s.
of Eq.~(\ref{rg_L_3b}). This part is of $O(J_c^3)$ and does not contain any logarithmic contribution
since $K_\Lambda(\Delta_{12})\rightarrow i{\Lambda\over\Delta_{12}}$ for $\Lambda\rightarrow 0$. 
This means that the logarithmic contributions are only generated by
the terms in second order in
the renormalized vertices, but including their corrections in second order in $J_c$ 
via Eq.~(\ref{G_second_order}). Both the imaginary parts $i\bar{G}^{(2a_1)c}$ and the 
frequency dependence generate logarithmic contributions. Implicitly, third-order terms in the
vertices are also present in the last term on the r.h.s. of the two-loop RG equation
(\ref{RG_G_2a}), which determines the contribution $\bar{G}^{(2a_2)}$, see Eq.~(\ref{rg_G_replacement}). 
As already mentioned after Eq.~(\ref{rg_G_replacement}), it may happen for more complicated models
than the Kondo model that this part can change the leading-order vertex in a nontrivial way (i.e.,
not only by a change of the Kondo temperature) leading to new logarithmic contributions not
expected from the matrix structure of the leading order vertices.

For $\omega=0$ and away from resonance $|\Delta_{12}|,|\Delta_{13}|\gg\tilde{\Gamma}$, we have
$a=-i\bar{a}$ and $b=-i\bar{b}$, with $\bar{a}$ and $\bar{b}$ being real. In this case, 
we get for Eqs.~(\ref{double_logarithmic_integral}) and (\ref{double_logarithmic_integral_special})
\begin{eqnarray}
\nonumber
-i\int_0^1\,dx\,\ln(x-i\bar{a})\,\ln(x-i\bar{b})\,&\approx&\,
{1\over 2}\bar{a}\ln^2|\bar{a}|+{1\over 2}\bar{b}\ln^2|\bar{b}|
-\bar{a}\ln|\bar{a}|-\bar{b}\ln|\bar{b}|-{1\over 2}\bar{a}\ln^2|{\bar{a}\over\bar{b}}|\\
\label{double_logarithmic_omega=0}
&&
-i{\pi\over 2}\left(|\bar{a}|\ln|\bar{a}|+|\bar{b}|\ln|\bar{b}|
-|\bar{a}|(1-\text{sign}\,{\bar{a}}\,\text{sign}\,{\bar{b}})\ln|{\bar{a}\over \bar{b}}|\right)
\,\,,\,\, \text{for}\,\,|\bar{a}|< |\bar{b}|\\
\label{double_logarithmic_special_omega=0}
&&\hspace{-4cm}\approx\,
\left\{
\begin{array}{cl}
{1\over 2}\bar{a}\ln^2|\bar{a}|+{1\over 2}\bar{b}\ln^2|\bar{b}|
-\bar{a}\ln|\bar{a}|-\bar{b}\ln|\bar{b}|
-i{\pi\over 2}\left(|\bar{a}|\ln|\bar{a}|+|\bar{b}|\ln|\bar{b}|\right) 
& \text{for}\quad |\bar{a}|\sim |\bar{b}|\ll 1 \\ 
-i{\pi\over 2}\,\bar{a}\,\text{sign}\,\bar{b}\,\ln|\bar{a}|
& \text{for}\quad |\bar{a}|\ll |\bar{b}|\sim 1 
\end{array}
\right.
\quad.
\end{eqnarray}

For $\omega=0$, we see that the terms containing explicitly the complex factor $i$ start 
always one power less in the logarithm compared to those without the factor $i$. 
The terms without the factor $i$ start with the logarithmic contribution
$\sim J_c^2\ln{\Lambda_c\over|\Delta+i\tilde{\Gamma}|}$, whereas the ones 
with the factor $i$ start with terms $\sim i J_c^3\ln{\Lambda_c\over|\Delta+i\tilde{\Gamma}|}$. 
Therefore, in the following section, where
we apply the general formalism to the Kondo model, we will restrict ourselves only to those terms.
This means that for $\omega=0$ we take into account all terms of order
$O(1)$, $O(J_c)$ and $O(J_c^2\ln{\Lambda_c\over|\Delta+i\tilde{\Gamma}|})$ without the factor $i$, and
all terms of order $O(J_c^2)$ and $O(J_c^3\ln{\Lambda_c\over|\Delta+i\tilde{\Gamma}|})$ with the factor $i$.
Thus, we can neglect the contributions from $L_S^{(2c)c}$ and $Z^{(2c)c}$ in Eq.~(\ref{L_2_final})
[according to Eqs.~(\ref{L_2c}) and (\ref{Z_2c})], and the first two terms on the r.h.s. of Eq.~(\ref{L_3a_final}).
Therefore, the first-order correction $\tilde{Z}^{(1)}$ is not important for the final result.
Furthermore, we can neglect all real terms in Eqs.~(\ref{double_logarithmic_omega=0}) and 
(\ref{double_logarithmic_special_omega=0}). This leads to the final equations
\begin{eqnarray}
\label{L_1_final_omega=0}
L_S^{(1)}(E)_{\Lambda=0}&=&L_S^{(1)c}-E Z^{(1)c}\quad,\\
\label{L_2_final_omega=0}
L_S^{(2)}(E)_{\Lambda=0}&\approx&
L_S^{(2b)c}-E Z^{(2b)c}
-i{\pi\over 4}\bar{G}_{12}^{(1)c}|\Delta_{12}|\bar{G}_{\bar{2}\bar{1}}^{(1)c}
-{1\over 2}\bar{G}_{12}^{(1)c}\Delta_{12}\left(\ln|{\Lambda_c\over \Delta_{12}}|\right)
\bar{G}_{\bar{2}\bar{1}}^{(1)c}\,,\\
\label{L_3a_final_omega=0}
L_S^{(3a)}(E)_{\Lambda=0}&\approx&
-i{1\over 2}\bar{G}_{12}^{(1)c}\Delta_{12}\left(\ln|{\Lambda_c\over \Delta_{12}}|\right)
\bar{G}_{\bar{2}\bar{1}}^{(2a_1)c}
-i{1\over 2}\bar{G}_{12}^{(2a_1)c}\Delta_{12}\left(\ln|{\Lambda_c\over \Delta_{12}}|\right)
\bar{G}_{\bar{2}\bar{1}}^{(1)c}\,,\\
\label{L_3b_final_omega=0}
L_S^{(3b)}(E)_{\Lambda=0}\,&\approx&\,
-i\Lambda_c\int_0^1\,dx\,\bar{G}_{12}^{(1)c}\ln(x-i{\Delta_{12} \over \Lambda_c})\,
\bar{G}_{\bar{2}3}^{(1)c}\,\ln(x-i{\Delta_{13} \over \Lambda_c})\,
\bar{G}_{\bar{3}\bar{1}}^{(1)c}\,,
\end{eqnarray}
with
\begin{equation}
\label{delta_definition}
\Delta_{ij}=E_{ij}-\tilde{L}_S\quad,
\end{equation}
and
\begin{equation}
\label{double_logarithmic_omega=0_approx}
-i\int_0^1\,dx\,\ln(x-i\bar{a})\,\ln(x-i\bar{b})\,\approx\,
-i{\pi\over 2}\left\{
\begin{array}{cl}
|\bar{a}|\ln|\bar{a}|+|\bar{b}|\ln|\bar{b}| 
-|\bar{a}|(1-\text{sign}\,{\bar{a}}\,\text{sign}\,{\bar{b}})\ln|{\bar{a}\over \bar{b}}|
& \text{for}\quad |\bar{a}|< |\bar{b}|\\
|\bar{a}|\ln|\bar{a}|+|\bar{b}|\ln|\bar{b}| 
& \text{for}\quad |\bar{a}|\sim |\bar{b}|\ll 1 \\
\bar{a}\,\text{sign}\,\bar{b}\,\ln|\bar{a}|
& \text{for}\quad |\bar{a}|\ll |\bar{b}|\sim 1 
\end{array}
\right.\quad.
\end{equation}
Close to resonance, where the effect of $\tilde{\Gamma}$ can not be neglected, we have to consider
the imaginary parts of the eigenvalues of $\tilde{L}_S$ as well. If an eigenvalue $\tilde{h}-i\tilde{\Gamma}$
occurs, we have to replace $\Delta_{ij}\rightarrow E_{ij}-\tilde{h}+i\tilde{\Gamma}$, and
\begin{equation}
\label{replacements_resonance}
|\Delta_{ij}| \,\rightarrow\, (E_{ij}-\tilde{h})\,
{2\over\pi}\arctan{E_{ij}-\tilde{h}\over\tilde{\Gamma}}\quad,\quad
\text{sign}\,\Delta_{ij} \,\rightarrow\, {2\over\pi}\arctan{E_{ij}-\tilde{h}\over\tilde{\Gamma}}\quad,\quad
\ln|\Delta_{ij}| \,\rightarrow\, \ln|E_{ij}-\tilde{h}+i\tilde{\Gamma}|\quad.
\end{equation}
\end{widetext}

\section{The Kondo model}
\label{sec:kondo}

\subsection{Model and algebra in Liouville space}
\label{sec:kondo_model_algebra}
{\it Model.} Now the RG formalism developed in the previous section is
applied to the anisotropic spin-$\tfrac12$ Kondo model in an external
magnetic field $h_0>0$. In this case, we have
\begin{eqnarray}
\label{H_S_kondo}
\hspace{-0.5cm}
H_S&=&h_0\,S^z \quad,\\
\label{g_kondo}
\hspace{-0.5cm}
g_{11'}&=&{1\over 2}
\left\{
\begin{array}{ll}
(J^i_{\alpha\alpha'})_0
S^i\sigma^i_{\sigma\sigma'}
& \text{for $\eta=-\eta'=+$} \\
-(J^i_{\alpha'\alpha})_0
S^i\sigma^i_{\sigma'\sigma}
& \text{for $\eta=-\eta'=-$}
\end{array}
\right.\,,
\end{eqnarray}
where $\alpha$ denotes the reservoir index, $i\in\{x,y,z\}$, and
\begin{equation}
\label{J_xy}
(J^x_{\alpha\alpha'})_0\,=\,(J^y_{\alpha\alpha'})_0\,=\,(J^\perp_{\alpha\alpha'})_0\quad.
\end{equation}
$S^i$ is the $i$ component of the spin-$\tfrac12$ operator of
the quantum dot, $\sigma^i$ is a Pauli matrix, and
$(J^z_{\alpha\alpha'})_0$ $\left[(J^\perp_{\alpha\alpha'})_0\right]$ are the initial
exchange couplings which correspond to processes without (with) spin flip. If we choose the
reservoir states such that the exchange couplings are real, we find from hermiticity 
(\ref{hermiticity}) that the exchange coupling matrices are symmetric,
\begin{equation}
\label{hermiticity_exchange}
(J^{z/\perp}_{\alpha\alpha'})_0\,=\,(J^{z/\perp}_{\alpha'\alpha})_0\quad.
\end{equation}
If one derives the Kondo model via a Schrieffer-Wolff transformation from an Anderson impurity model
(see, e.g., Ref.~\onlinecite{korb_reininghaus_hs_koenig_PRB07}), one finds
\begin{equation}
\label{J_form}
(J^{z/\perp}_{\alpha\alpha'})_0\,=\,2\sqrt{x_\alpha x_{\alpha'}}\,J^{z/\perp}_0\quad,\quad
\sum_\alpha\,x_\alpha\,=\,1\quad,
\end{equation}
which is used sometimes to simplify the calculations. Furthermore, although the general 
formalism and many of the following formulas are also valid for 
an arbitrary number of reservoirs, we will finally consider the case of
two reservoirs only, with $\alpha\equiv L,R\equiv \pm$, and chemical potentials
given by
\begin{equation}
\label{voltage}
\mu_\alpha=\alpha {V\over 2}\quad,
\end{equation}
where $V$ is the applied voltage.

{\it Representation in Liouville space.} In Liouville space, the initial Liouvillian and the 
initial vertex are given by
\begin{eqnarray}
\label{L_initial_kondo}
L_S^{(0)}&=&[H_S,\cdot]_-=h_0\,(L^{+z}+L^{-z})\quad,\\
\nonumber
G^{pp}_{11'}&=&{1\over 2}\,\left\{
\begin{array}{cl}
(J^i_{\alpha\alpha'})_0\,L^{pi}\,\sigma^i_{\sigma\sigma'}\, 
&\mbox{for }\eta=-\eta'=+ \\
-(J^i_{\alpha'\alpha})_0\,L^{pi}\,\sigma^i_{\sigma'\sigma}\, 
&\mbox{for }\eta=-\eta'=- 
\end{array}
\right.\,,\\
\label{G_initial_kondo}
\end{eqnarray}
where the spin superoperators $\underline{L}^p=(L^{px},L^{py},L^{pz})$ 
are defined by their action on an arbitrary operator $A$ in Hilbert space via
\begin{equation}
\label{L_basis_operators}
\underline{L}^+A\,=\,\underline{S}A \quad,\quad
\underline{L}^-A\,=\,-A\underline{S}\quad.
\end{equation}
We will now derive a closed set of basis superoperators to represent the
renormalized Liouvillian and the renormalized vertices (note that
this set will be more complex than the one 
derived in Ref.~\onlinecite{RTRG_FS} for the isotropic Kondo model). Because
the Hilbert space (spanned by the states $|\uparrow\rangle$ and
$|\downarrow\rangle$) is two dimensional, the Liouville space of
operators acting on it is four dimensional, and the superoperators
defined in this section can be represented by
$4\times4$ matrices. This means that we need at most $4\times4=16$
basis superoperators.

We define the four scalar superoperators
\begin{align}
  \label{eq:ScalarSuperoperators}
  \begin{aligned}
    \La&=\tfrac34\cdot\Id+\underline{L}^+\cdot\underline{L}^-, &
    \Lb&=\tfrac14\cdot\Id-\underline{L}^+\cdot\underline{L}^-,
    \\
    \Lc&=\tfrac12\cdot\Id+2L^{+z}L^{-z}, &
    \Lh&=L^{+z}+L^{-z},
  \end{aligned}
\end{align}
where $\Id$ is the identity superoperator, and the vector superoperators
\begin{align}
  \label{eq:VectorSuperoperators}
  \begin{aligned}
    \underline{L}^1&=\tfrac12\left(\underline{L}^+-\underline{L}^-
      -2i\underline{L}^+\times\underline{L}^-\right),\\
    \underline{L}^2&=-\tfrac12\left(\underline{L}^++\underline{L}^-\right),\\
    \underline{L}^3&=\tfrac12\left(\underline{L}^+-\underline{L}^-
      +2i\underline{L}^+\times\underline{L}^-\right).
  \end{aligned}
\end{align}
Because the Hamiltonian is not isotropic, we have to split the vector
superoperators~\eqref{eq:VectorSuperoperators} into their
components and consider these separately. We define
\begin{equation}
  \label{eq:L_i_pm}
  L^i_\pm=L^{ix}\pm iL^{iy} \quad \text{for $i\in\{1,2,3,+,-\}$}
\end{equation}
and find that we get a convenient superoperator basis if we add the
set $\{L^{1z}, L^{3z}, L^1_\pm, L^3_\pm, L^4_\pm, L^5_\pm\}$ to the four
scalar superoperators, where 
\begin{eqnarray}
\label{L_4}
L^4_\pm \,&=&\, L^2_\pm\pm\left(L^+_\pm L^{-z}+L^{+z} L^-_\pm\right)\quad,\\
\label{L_5}
L^5_\pm \,&=&\, L^2_\pm\mp\left(L^+_\pm L^{-z}+L^{+z} L^-_\pm\right)\quad.
\end{eqnarray}
This set of $14$ basis superoperators is sufficient to describe the anisotropic Kondo model with
$J^x=J^y=J^\perp$, where rotational invariance around the $z$ axis holds. In the case of the fully 
anisotropic Kondo model with $J^x\ne J^y$, two additional superoperators
\begin{eqnarray}
\label{L_++}
L_{++}\,=\,L^3_+ L^1_+\,=\,-L^5_+ L^4_+\quad,\\
\label{L_--}
L_{--}\,=\,L^3_- L^1_-\,=\,-L^5_- L^4_-\quad
\end{eqnarray}
have to be used, which we do not need in the following. 

We note the properties
\begin{eqnarray}
\label{traceless_1}
&&\mbox{Tr}_S\,L^\chi\,=\,0\quad,\quad \mbox{for  } \chi=a,c,h,3z\quad,\\
\label{traceless_2}
&&\mbox{Tr}_S\,L^\chi_\pm\,=\,0\quad,\quad \mbox{for  } \chi=3,4,5\quad,\\
\label{tracefull_3}
&&\mbox{Tr}_S\,L^b\,,\,\mbox{Tr}_S\,L^{1z}\,,\,\mbox{Tr}_S\,L^1_\pm\,\ne \,0
\quad,
\end{eqnarray}
together with the transformation under the $c$~transform (\ref{c_transformation})
\begin{eqnarray}
\label{c_transform_1}
(L^\chi)^c\,&=&\,L^\chi\quad,\quad \mbox{for  } \chi=a,b,c,1z,3z\quad,\\
\label{c_transform_2}
(L^h)^c\,&=&\,-L^h \quad,\\
\label{c_transform_3}
(L^\chi_\pm)^c\,&=&\,L^\chi_\mp\quad,\quad \mbox{for  } \chi=1,3\quad,\\
\label{c_transform_4}
(L^\chi_\pm)^c\,&=&\,-L^\chi_\mp\quad,\quad \mbox{for  } \chi=4,5\quad.
\end{eqnarray}

Using spin rotational invariance around the $z$~axis and spin conservation,
together with the properties (\ref{traceless_1})--(\ref{tracefull_3}) and 
$\mbox{Tr}_S L_S(E,\omega)=0$ [see Eq.~(\ref{rg_L_property})],
we find that the Liouvillian and the current kernel can be represented 
as (in each step of the RG)
\begin{eqnarray}
\nonumber
L_S(E,\omega)\,&=&\,h(E,\omega)L^h-i\Gamma^a(E,\omega)L^a\\
\label{L_representation}
&&\hspace{-0.5cm}
-i\Gamma^c(E,\omega)L^c-i\Gamma^{3z}(E,\omega)L^{3z} \,,\\
\label{sigma_current_representation}
\Sigma_\gamma(E,\omega)\,&=&\,i\Gamma^b_\gamma(E,\omega)L^b
+i\Gamma^{1z}_\gamma(E,\omega)L^{1z}\,,
\end{eqnarray}
where as usual we always assume implicitly that the $\mbox{Tr}_S$ is
acting from the left when we consider the current kernel or the current
vertex
\begin{equation}
\nonumber
\Sigma_\gamma \equiv \mbox{Tr}_S\Sigma_\gamma\quad,\quad
\bar{I}^\gamma_{12} \equiv \mbox{Tr}_S \bar{I}^\gamma_{12}\quad.
\end{equation}

From Eqs.~(\ref{rg_L_c_transform}), (\ref{rg_Sigma_I_c_transform}), (\ref{c_transform_1}),
and (\ref{c_transform_2}), we get the following transformation under complex
conjugation:
\begin{eqnarray}
\label{symmetry_gamma_L}
\Gamma^\chi(E,\omega)^*\,&=&\,\Gamma^\chi(-E,\omega)\,\,,\,\,\mbox{for  }\chi=a,c,3z \,,\\
\label{symmetry_gamma_I}
\Gamma^\chi_\gamma(E,\omega)^*\,&=&\,\Gamma^\chi_\gamma(-E,\omega)\,\,,\,\, 
\mbox{for  }\chi=b,1z \,,\\
\label{symmetry_h}
h(E,\omega)^*\,&=&\,h(-E,\omega)\quad.
\end{eqnarray}

The various terms in Eq.~(\ref{L_representation}) can be interpreted if one
analyzes the spectral properties of the renormalized Liouvillian. 
$\L_S(E,\omega)$ has four eigenvalues,
\begin{equation}
\label{eigenvalues}
  \begin{split}
    \lambda_0(E,\omega)&\,=\,0\quad,\\
    \lambda_1(E,\omega)&\,=\,-i\Gamma^a(E,\omega)\quad,\\
    \lambda_\pm(E,\omega)&\,=\,\pm h(E,\omega)-i(\Gamma^a+\Gamma^c)(E,\omega)\quad.
  \end{split}
\end{equation}
$\lambda_0$ corresponds to the stationary state, $\lambda_1$ describes the 
relaxation mode, and $\lambda_\pm$ correspond to the two dephasing modes. The projectors 
$P_i(E,\omega)=|x_i(E,\omega\rangle\langle \bar{x}_i(E,\omega)|$
onto the four eigenspaces are given by
\begin{equation}
\label{projectors}
  \begin{split}
    P_0(E,\omega)&\,=\,L^b-\frac{\Gamma^{3z}(E,\omega)}{\Gamma^a(E,\omega)}L^{3z}\quad,\\
    P_1(E,\omega)&\,=\,L^a-\Lc+\frac{\Gamma^{3z}(E,\omega)}{\Gamma^a(E,\omega)}L^{3z}\quad,\\
    P_\pm(E,\omega)&\,=\,\frac12\left(L^c\pm L^h\right)\quad,
  \end{split}
\end{equation}
and the right and left eigenvectors follow from
\begin{equation}
\label{eigenvectors}
\begin{split}
  \langle \sigma\sigma'|x_0(E,\omega)\rangle\,&=\,\delta_{\sigma\sigma'}
  \left({1\over 2}-\sigma{\Gamma^{3z}(E,\omega)\over 2\Gamma^a(E,\omega)}\right)\,,\\
  \langle \sigma\sigma'|x_1(E,\omega)\rangle\,&=\,\delta_{\sigma\sigma'}\sigma\quad,\\
  \langle \sigma\sigma'|x_\pm(E,\omega)\rangle\,&=\,\delta_{\sigma,-\sigma'}\delta_{\sigma\pm}\quad,\\
  \langle \bar{x}_0(E,\omega)|\sigma\sigma'\rangle\,&=\,\delta_{\sigma\sigma'}\quad,\\
  \langle \bar{x}_1(E,\omega)|\sigma\sigma'\rangle\,&=\,\delta_{\sigma\sigma'}
  \left({1\over 2}\sigma+{\Gamma^{3z}(E,\omega)\over 2\Gamma^a(E,\omega)}\right)\,,\\
  \langle \bar{x}_\pm(E,\omega)|\sigma\sigma'\rangle\,&=\,
  \delta_{\sigma,-\sigma'}\delta_{\sigma\pm}\quad.
\end{split}
\end{equation}
According to Eq.~(\ref{rd_stationary}), the eigenvector $|x_0(0,0^+)\rangle$ corresponds to the
stationary state. Therefore, we get
\begin{equation}
\label{rd_stationary_kondo}
\rho_S^{st}\,=\,{1\over 2}\Id + 2M S^z\quad,\quad
M\,=\,-{\Gamma^{3z}\over 2\Gamma^a}\quad,
\end{equation}
where $M$ denotes the magnetization, which is related to $\Gamma^{3z}$ (if no argument is
written, we implicitly assume $E=0$ and $\omega=0^+$).

The stationary current follows from Eqs.~(\ref{current_stationary}) 
and (\ref{sigma_current_representation}) as
\begin{eqnarray}
\nonumber
\langle I^\gamma\rangle^{st}\,&=&\,
-i\,\text{Tr}_S\,\Sigma_\gamma\,\rho^{st}_S\\
\nonumber
&=&\Gamma^b_\gamma\,\text{Tr}_S\,L^b\,\rho^{st}_S\,+\,
\Gamma^{1z}_\gamma\,\text{Tr}_S\,L^{1z}\,\rho^{st}_S\quad,
\end{eqnarray}
which, by using Eq.~(\ref{rd_stationary_kondo}) together with 
${1\over 2}\text{Tr}_S L^b\Id=\text{Tr}_S L^{1z}S^z=1$ and
$\text{Tr}_S L^b S^z=\text{Tr}_S L^{1z}\Id=0$, gives
\begin{equation}
\label{current_stationary_kondo}
\langle I^\gamma\rangle^{st}\,=\,\Gamma^b_\gamma\,+\,2M\,\Gamma^{1z}_\gamma \quad.
\end{equation}
 
To represent the vertices, we define the reservoir spin matrices
\begin{equation}
\label{spin_matrices}
\sigma^z_\pm\,=\,{1\over 2}(1\pm \sigma^z) \quad,\quad
\sigma_\pm\,=\,{1\over 2}(\sigma_x\pm i\sigma_y)\quad,
\end{equation}
and the following operators in combined reservoir spin space and
Liouville space of the dot
\begin{eqnarray}
\label{L_spin_1}
\hat{L}^\chi_\pm\,&=&\,L^\chi\,\sigma^z_\pm
\,\,,\,\,\mbox{for  }\chi=a,b,c,h,1z,3z \,,\\
\label{L_spin_2}
\hat{L}^\chi_\pm\,&=&\,L^\chi_\pm\,\sigma_\mp
\,\,,\,\,\mbox{for  }\chi=1,3,4,5 \,.
\end{eqnarray}
Using spin rotational invariance around the $z$~axis, together with
the properties (\ref{rg_G_property}) and (\ref{traceless_1})--(\ref{tracefull_3}),
the renormalized vertices can be represented as (for $\eta_1=-\eta_2=+$)
\begin{eqnarray}
\nonumber
\bar{G}_{12}(E,\omega,\omega_1,\omega_2)|_{\eta_1=-\eta_2=+}\,&=&\,\\
\label{G_representation}
&&\hspace{-4cm}
=\,\sum_{s=\pm}\,\,\,\sum_{{\chi=a,c,h,\atop 3z,3,4,5}}\bar{G}^{\chi s}(E,\omega,\omega_1,\omega_2)\,
\hat{L}^\chi_s\,\,,\\
\nonumber
\bar{I}^\gamma_{12}(E,\omega,\omega_1,\omega_2)|_{\eta_1=-\eta_2=+}\,&=&\,\\
\label{I_representation}
&&\hspace{-4cm}
=\,\sum_{s=\pm}\,\,\,\sum_{\chi=b,1z,1}\bar{I}^{\gamma \chi s}(E,\omega,\omega_1,\omega_2)\,
\hat{L}^\chi_s\,\,.
\end{eqnarray}
In these equations, we have used a compact matrix notation in
the reservoir indices $(\alpha_1,\alpha_2)$ and the reservoir spin indices $(\sigma_1,\sigma_2)$.
Whereas $\bar{G}^{\chi s}\equiv (\bar{G}^{\chi s}_{\alpha_1\alpha_2})$ and
$\bar{I}^{\gamma \chi s}\equiv (\bar{I}^{\gamma \chi s}_{\alpha_1\alpha_2})$ denote matrices in the
reservoir indices, the quantities $\hat{L}^\chi_s$ are matrices in the reservoir spin indices
according to definitions (\ref{L_spin_1}) and (\ref{L_spin_2}). Again, we note that 
Eq.~(\ref{I_representation}) holds only if the trace $\mbox{Tr}_S$ is taken from the left.
For $\eta_1=-\eta_2=-$, we use the antisymmetry properties (\ref{rg_G_symmetry})
and (\ref{rg_I_symmetry}), and obtain from Eqs.~(\ref{G_representation}) and (\ref{I_representation})
\begin{eqnarray}
\nonumber
\hspace{-0.5cm}
\bar{G}_{12}(E,\omega,\omega_1,\omega_2)|_{\eta_1=-\eta_2=-}\,&=&\,\\
\label{G_representation_T}
&&\hspace{-4.5cm}
=\,-\sum_{s=\pm}\,\,\,\sum_{{\chi=a,c,h,\atop 3z,3,4,5}}\bar{G}^{\chi s}(E,\omega,\omega_2,\omega_1)^T\,
(\hat{L}^\chi_s)^T\,\,,\\
\nonumber
\hspace{-0.5cm}
\bar{I}^\gamma_{12}(E,\omega,\omega_1,\omega_2)|_{\eta_1=-\eta_2=-}\,&=&\,\\
\label{I_representation_T}
&&\hspace{-4.5cm}
=\,-\sum_{s=\pm}\,\,\,\sum_{\chi=b,1z,1}\bar{I}^{\gamma \chi s}(E,\omega,\omega_2,\omega_1)^T\,
(\hat{L}^\chi_s)^T\,\,,
\end{eqnarray}
where $(\dots)^T$ denotes the transpose only with respect to the reservoir indices or the 
reservoir spin indices
\begin{eqnarray}
\nonumber
(\bar{G}^{\chi s})^T_{\alpha\alpha'}&=&\bar{G}^{\chi s}_{\alpha'\alpha}\quad,\quad
(\bar{I}^{\gamma\chi s})^T_{\alpha\alpha'}=\bar{I}^{\gamma\chi s}_{\alpha'\alpha}\quad,\,\\
\label{transpose}
(\hat{L}^\chi_s)^T_{\sigma\sigma'}&=&(\hat{L}^\chi_s)_{\sigma'\sigma}\quad.
\end{eqnarray}

Using the properties (\ref{rg_G_c_transform}) and (\ref{rg_I_c_transform}) together with
Eqs.~(\ref{c_transform_1})--(\ref{c_transform_4}), we obtain the symmetry relations
\begin{eqnarray}
\nonumber
\bar{G}^{\chi s}_{\alpha\alpha'}(E,\omega,\omega_1,\omega_2)^*\,&=&\,
-\bar{G}^{\chi s}_{\alpha'\alpha}(-E,\omega,\omega_2,\omega_1)\\
\nonumber
&&\hspace{-2cm}\text{for  }\chi=a,b,c,1z,3z\,,\\
\nonumber
\bar{G}^{h s}_{\alpha\alpha'}(E,\omega,\omega_1,\omega_2)^*\,&=&\,
\bar{G}^{h s}_{\alpha'\alpha}(-E,\omega,\omega_2,\omega_1)\,,\\
\nonumber
\bar{G}^{1/3,s}_{\alpha\alpha'}(E,\omega,\omega_1,\omega_2)^*\,&=&\,
-\bar{G}^{1/3,-s}_{\alpha'\alpha}(-E,\omega,\omega_2,\omega_1)\,,\\
\nonumber
\bar{G}^{4/5,s}_{\alpha\alpha'}(E,\omega,\omega_1,\omega_2)^*\,&=&\,
\bar{G}^{4/5,-s}_{\alpha'\alpha}(-E,\omega,\omega_2,\omega_1)\,,\\
\label{G_symmetries_kondo}
\end{eqnarray}
and the same for $\bar{G}\rightarrow \bar{I}^\gamma$.

{\it Algebra.} The results of a multiplication of two of the operators $\hat{L}^\chi_\pm$ are
summarized in the following table (all products not shown are zero):
\begin{widetext}
  \begin{equation}
    \label{algebra_L}
    \begin{array}{c||cccc|cc|cccc}
      & \hat{L}^a_\pm & \hat{L}_\pm^b & \hat{L}_\pm^c & \hat{L}_\pm^h & 
      \hat{L}_\pm^{1z} & \hat{L}_\pm^{3z} & \hat{L}^1_\mp & \hat{L}^3_\mp &
      \hat{L}^4_\mp & \hat{L}^5_\mp   \\ \hline\hline
      \hat{L}_\pm^a & \hat{L}_\pm^a & 0 & \hat{L}_\pm^c & \hat{L}_\pm^h &
      0 & \hat{L}_\pm^{3z} &
      0 & \hat{L}^3_\mp & \hat{L}^4_\mp & \hat{L}^5_\mp
      \\
      \hat{L}_\pm^b & 0 & \hat{L}_\pm^b & 0 & 0 &
      \hat{L}_\pm^{1z} & 0 &
      \hat{L}^1_\mp & 0 & 0 & 0
      \\
      \hat{L}_\pm^c & \hat{L}_\pm^c & 0 & \hat{L}_\pm^c & \hat{L}_\pm^h &
      0 & 0 &
      0 & \hat{L}^3_\mp & 0 &\hat{L}^5_\mp
      \\
      \hat{L}_\pm^h & \hat{L}_\pm^h & 0 & \hat{L}_\pm^h & \hat{L}_\pm^c &
      0 & 0 &
      0 & \mp \hat{L}^3_\mp & 0 & \mp \hat{L}^5_\mp
      \\ \hline
      \hat{L}_\pm^{1z} & \hat{L}_\pm^{1z} & 0 & 0 & 0 &
      0 & \hat{L}_\pm^b &
      0 & 0 & \pm \hat{L}^1_\mp & 0 
      \\
      \hat{L}_\pm^{3z} & 0 & \hat{L}_\pm^{3z} & 0 & 0 &
      \hat{L}_\pm^a - \hat{L}_\pm^c & 0 &
      \pm \hat{L}^4_\mp & 0 & 0 & 0 
      \\ \hline
      \hat{L}^1_\pm & \hat{L}^1_\pm & 0 & \hat{L}^1_\pm & \mp \hat{L}^1_\pm &
      0 & 0 &
      0 & 2\hat{L}_\mp^b & 0 & \mp2\hat{L}_\mp^{1z} \\
      \hat{L}^3_\pm & 0 & \hat{L}^3_\pm & 0 & 0 &
      \pm \hat{L}^5_\pm & 0 &
      \hat{L}_\mp^c\pm \hat{L}_\mp^h & 0 & 0 & 0 \\
      \hat{L}^4_\pm & \hat{L}^4_\pm & 0 & \hat{L}^4_\pm & \mp \hat{L}^4_\pm &
      0 & 0 &
      0 & \mp 2\hat{L}_\mp^{3z} & 0 & 2\hat{L}_\mp^a-2\hat{L}_\mp^c \\
      \hat{L}^5_\pm & \hat{L}^5_\pm & 0 & 0 & 0 &
      0 & \pm \hat{L}^3_\pm &
      0 & 0 & \hat{L}_\mp^c\pm \hat{L}_\mp^h & 0
    \end{array}
  \end{equation}
\end{widetext}
Note that, according to definitions (\ref{L_spin_1}) and (\ref{L_spin_2}), 
the operators $\hat{L}^\chi_\pm$ are matrices in the reservoir spin
indices, where each matrix element is a superoperator in the Liouville space of the
dot. The same algebra holds without the reservoir spin indices if one 
replaces
\begin{eqnarray}
\nonumber
\hat{L}^\chi_\pm &\rightarrow& L^\chi\quad\text{for  }\chi=a,b,c,h,1z,3z\,,\\
\label{algebra_replacement}
\hat{L}^\chi_\pm &\rightarrow& L^\chi_\pm\quad\text{for  }\chi=1,3,4,5\,,
\end{eqnarray}
in Eq.~(\ref{algebra_L}),
with the only difference that also some products not shown are nonzero [these
are the products $L^3_\pm L^1_\pm=-L^5_\pm L^4_\pm$ defining the basis
superoperators $L_{\pm\pm}$ according to Eqs.~(\ref{L_++}) and (\ref{L_--}), which 
are not needed for our case of rotational invariance around the $z$~axis].
Furthermore, we note that the algebra (\ref{algebra_L}) is also
not changed if we replace
\begin{eqnarray}
\nonumber
\hat{L}^\chi_\pm &\rightarrow& (\hat{L}^\chi_\pm)^T \quad\text{for  }\chi=a,b,c,h,1z,3z\,,\\
\nonumber
\hat{L}^\chi_\pm &\rightarrow& (\hat{L}^\chi_\mp)^T\quad\text{for  }\chi=1,3,4,5\,,\\
\label{algebra_replacement_T}
\pm &\rightarrow& \mp \quad \text{for all sign factors}\quad,
\end{eqnarray}
which turns out to be very helpful to consider the two cases $\eta=\pm$ for creation and
annihilation operators, see the representations (\ref{G_representation_T}) and
(\ref{I_representation_T}).

Finally, we note that the spin matrices (\ref{spin_matrices}) fulfil the algebra
(all products not shown are zero)
\begin{equation}
\label{algebra_spin_matrices}
    \begin{array}{c||c|c}
       & \sigma^z_\pm & \sigma_\pm  \\ \hline\hline
      \sigma^z_\pm & \sigma_\pm^z & \sigma_\pm
      \\
      \sigma_\mp & \sigma_\mp & \sigma_\mp^z
    \end{array}
\end{equation}
and the same holds if one replaces
\begin{equation}
\label{algebra_spin_matrices_T}
\sigma^z_\pm\,\rightarrow\,(\sigma^z_\pm)^T \quad,\quad
\sigma_\pm\,\rightarrow\,(\sigma_\mp)^T\quad.
\end{equation}

\subsection{Two-loop analysis}
\label{sec:kondo_2_loop}

\subsubsection{RG above $\Lambda_c$}
\label{sec:kondo_above_Lambda_c}

{\it Initial values.}
We start with the determination of the initial values of the Liouvillian 
and the vertices together with their values after the first RG step, given
by Eqs.~(\ref{L_initial})--(\ref{G_initial}). 

The initial values for the Liouvillian $L_S^{(0)}$ and the vertices 
$\bar{G}_{11'}^{(1)}$ and $\tilde{G}_{11'}^{(1)}$ follow from 
Eqs.~(\ref{L_initial_kondo}) and (\ref{G_initial_kondo}) as
\begin{eqnarray}
\label{L_initial_kondo_h}
L_S^{(0)}\,&=&\,[H_S,\cdot]_-\,=\,h_0\,L^h\quad,\\
\label{G_bar_initial_kondo}
\bar{G}_{11'}^{(1)}\,&=&\,\sum_p G^{pp}_{11'}
=-\hat{J}^i_0\,L^{2i}\,\sigma^i\quad,\\
\label{G_tilde_initial_kondo}
\tilde{G}_{11'}^{(1)}\,&=&\,\sum_p p\,G^{pp}_{11'}
=-{1\over 2}\hat{J}^i_0\,(L^{1i}+L^{3i})\,\sigma^i\quad,
\end{eqnarray}
where we have taken $\eta=-\eta'=+$ and used a matrix notation 
$\hat{J}^i_0\equiv [(J^i_0)_{\alpha\alpha'}]$ for the exchange couplings in
the reservoir indices. Inserting the various definitions of the basis 
superoperators, we find the following representations for $\eta=-\eta'=+$:
\begin{eqnarray}
\label{G_bar_components}
\bar{G}^{(1)}_{11'}&=&{1\over 2}\hat{J}^z_0 s\hat{L}^h_s
-{1\over 2}\hat{J}^\perp_0(\hat{L}^4_s+\hat{L}^5_s)\,,\\
\label{G_tilde_components}
\tilde{G}^{(1)}_{11'}&=&{1\over 2}\hat{J}^z_0 s(\hat{L}^{1z}_s+\hat{L}^{3z}_s)
+{1\over 2}\hat{J}^\perp_0 (\hat{L}^{1}_s+\hat{L}^{3}_s)\,,
\end{eqnarray}
where we sum implicitly over $s=\pm$ on the r.h.s.
According to Eq.~(\ref{current_liouvillian_vertex}), the initial current vertex 
is given by $\bar{I}^{\gamma (1)}_{11'}=c^\gamma_{11'}\tilde{G}^{(1)}_{11'}$.
Using Eq.~(\ref{G_tilde_components}), we obtain for $\eta=-\eta'=+$ 
\begin{equation}
\label{I_components}
\bar{I}^{\gamma (1)}_{11'}={1\over 2}\hat{J}_0^{\gamma z} s\hat{L}^{1z}_s
+{1\over 2}\hat{J}_0^{\gamma\perp} \hat{L}^{1}_s\,,
\end{equation}
where 
\begin{equation}
\label{current_exchange}
(J^{\gamma,z/\perp}_{\alpha\alpha'})_0=
c^\gamma_{\alpha\alpha'}(J^{z/\perp}_{\alpha\alpha'})_0\quad,\quad
c^\gamma_{\alpha\alpha'}=-{1\over 2}(\delta_{\gamma\alpha}-\delta_{\gamma\alpha'})\,.
\end{equation}
We have left out the components $3z$ and $3$ in Eq.~(\ref{I_components}) because we assume implicitly
that the $\text{Tr}_S$ acts from the left on the current vertex. 

To calculate the second-order contributions (\ref{L_initial})--(\ref{G_initial}) from the first RG 
step, we use the following identities for any two superoperators $A_{11'}=A^{\chi s}\hat{L}_s^\chi$
and $B_{11'}=B^{\chi s}\hat{L}_s^\chi$, which can all be proven easily by summing over the two
possibilities $\eta_i=\pm$:
\begin{eqnarray}
\label{product_decompose_1}
&& A_{11'}B_{\bar{1}'\bar{1}}= 2(\text{Tr}_\alpha A^{\chi s}B^{\chi' s'})
\,(\text{Tr}_\sigma \hat{L}^\chi_s \hat{L}^{\chi'}_{s'})\,,\\
\nonumber
&& \hspace{-1cm}A_{11'}(E_{11'}-L_S^{(0)})B_{\bar{1}'\bar{1}}=\\
\label{product_decompose_2}
&&2A^{\chi s}_{\alpha\alpha'}B^{\chi' s'}_{\alpha'\alpha}\,
\,(\text{Tr}_\sigma \hat{L}^\chi_s (E_{\alpha\alpha'}-h_0L^h)\hat{L}^{\chi'}_{s'})\,,
\end{eqnarray}
with $E_{\alpha\alpha'}=E+\mu_\alpha-\mu_{\alpha'}$. Here, $\text{Tr}_\alpha$
and $\text{Tr}_\sigma$ denote the trace with respect to the reservoir indices
and the reservoir spin indices, respectively. Applying these identities to
Eqs.~(\ref{L_initial})--(\ref{G_initial}), inserting the results 
(\ref{G_bar_components})--(\ref{I_components}) for the vertices, and 
applying the algebra of the basis superoperators, we find after some
straightforward calculation the following result for the Liouvillian and the
current kernel after the first RG step:
\begin{eqnarray}
\nonumber
L_S^a(E)&=&-i{\pi^2\over 16}(2D+iE)\left(\text{Tr}\hat{J}^\perp_0\hat{J}^\perp_0\right) L^a\\
\nonumber
&&-i{\pi^2\over 32}(2D+iE)\left(\text{Tr}(\hat{J}^z_0\hat{J}^z_0-\hat{J}^\perp_0\hat{J}^\perp_0)\right)L^c\\
\nonumber
&&+h_0(1-{\pi^2\over 32})\left(\text{Tr}\hat{J}^z_0\hat{J}^z_0\right)L^h\\
\label{Liouvillian_first_RG_step_kondo}
&&-i{\pi\over 2}h_0 \left(\text{Tr}\hat{J}^\perp_0\hat{J}^\perp_0\right)L^{3z}\,,\\
\nonumber
\Sigma_\gamma^a(E)&=&i{\pi\over 4}\delta_{\alpha\gamma}(\mu_\alpha-\mu_{\alpha'}) 
\left((J^z_{\alpha\alpha'})^2+2(J^\perp_{\alpha\alpha'})^2\right)_0 L^b\,.\\
\label{Current_kernel_first_RG_step_kondo}
\end{eqnarray}
Thereby, the terms 
proportional to the band width cancel with terms generated by the second RG step,
see Eq.~(\ref{Lambda_initial}). Therefore, we can omit them together with the real
terms of second order in $J$ [which are not the leading order ones in this order,
see the discussion before Eq.~(\ref{L_1_final_omega=0})]. Thus, we are left with
the following inital condition to solve the RG equations (\ref{L_2b}) and (\ref{Z_2b}):
\begin{eqnarray}
\label{L_2b_initial_kondo}
L_S^{(2b)}|_{\Lambda=\Lambda_0}&=&
-i{\pi\over 2}h_0 \left(\text{Tr}\hat{J}^\perp_0\hat{J}^\perp_0\right)L^{3z}\,,\\
\nonumber
\Sigma_\gamma^{(2b)}|_{\Lambda=\Lambda_0}&=&\\
\nonumber
&&\hspace{-2cm}
=-i{\pi\over 4}c^\gamma_{\alpha\alpha'}(\mu_\alpha-\mu_{\alpha'}) 
\left((J^z_{\alpha\alpha'})^2+2(J^\perp_{\alpha\alpha'})^2\right)_0 L^b\,,\\
\label{Sigma_current_2b_initial_kondo}\\
\label{Z_2b_initial_kondo}
Z^{(2b)}|_{\Lambda=\Lambda_0}&=&Z_\gamma^{(2b)}|_{\Lambda=\Lambda_0}\,=\,0\,.
\end{eqnarray}

We now calculate the form of the vertex (\ref{G_initial}) after the first RG step, which 
leads to the initial condition for the vertices $\bar{G}^{(2a_1)}_{11'}$ and 
$\bar{I}^{\gamma(2a_1)}_{11'}$, given by Eq.~(\ref{G_2a_1}). We use the following identity, 
valid for $\eta=-\eta'=+$:
\begin{eqnarray}
\nonumber
&&\hspace{-1cm}
A_{12}B_{\bar{2}1'}-(1\leftrightarrow 1')=\\
\label{product_decompose_3}
&&\hspace{-1cm}
=A^{\chi s}B^{\chi' s'}\,\hat{L}^\chi_s\hat{L}^{\chi'}_{s'}-
B^{\chi' s'}A^{\chi s}\,\left((\hat{L}^\chi_s)^T\,(\hat{L}^{\chi'}_{s'})^T\right)^T\,.
\end{eqnarray} 
Applying this property to Eq.~(\ref{G_2a_1}), inserting 
Eqs.~(\ref{G_bar_components})--(\ref{I_components}), and 
applying the algebra of the basis superoperators, we find after some straightforward
calculation for $\eta=-\eta'=+$
\begin{eqnarray}
\nonumber
\bar{G}^{(2a_1)}_{11'}|_{\Lambda=\Lambda_0}\,&=&\,
{\pi\over 2}\hat{J}_0^\perp\hat{J}_0^\perp s\hat{L}^{3z}_s \\
\label{G_2a_1_initial_kondo}
&&+{\pi\over 4}(\hat{J}_0^z\hat{J}_0^\perp+\hat{J}_0^\perp\hat{J}_0^z)\hat{L}^{3}_s \,,\\
\nonumber
\bar{I}^{\gamma(2a_1)}_{11'}|_{\Lambda=\Lambda_0}\,&=&\,
-{\pi\over 8}(\hat{J}_0^{\gamma z}\hat{J}_0^z-\hat{J}_0^z\hat{J}_0^{\gamma z}+\\
\label{I_2a_1_initial_kondo}
&&+2\hat{J}_0^{\gamma\perp}\hat{J}_0^\perp-2\hat{J}_0^\perp\hat{J}_0^{\gamma\perp})\hat{L}^{b}_s \,.
\end{eqnarray}

{\it Vertices in first order.} The RG equations for the lowest order vertices $\bar{G}_{11'}^{(1)}$ and
$\bar{I}_{11'}^{(1)}$ are given by Eqs.~(\ref{G_reference_solution}) and (\ref{I_reference_solution})
with initial conditions (\ref{G_bar_components}) and (\ref{I_components}). Using 
Eq.~(\ref{product_decompose_3}), one finds after some algebra that the initial form of the
leading order vertices is preserved, i.e., for $\eta=-\eta'=+$
\begin{eqnarray}
\label{G_bar_leading_kondo}
\bar{G}^{(1)}_{11'}&=&{1\over 2}\hat{J}^z s\hat{L}^h_s
-{1\over 2}\hat{J}^\perp(\hat{L}^4_s+\hat{L}^5_s)\,,\\
\label{I_leading_kondo}
\bar{I}^{\gamma (1)}_{11'}&=&{1\over 2}\hat{J}^{\gamma z} s\hat{L}^{1z}_s
+{1\over 2}\hat{J}^{\gamma\perp} \hat{L}^{1}_s\,,
\end{eqnarray}
with 
\begin{equation}
\label{current_exchange_renormalized}
J^{\gamma,z/\perp}_{\alpha\alpha'}=
c^\gamma_{\alpha\alpha'}J^{z/\perp}_{\alpha\alpha'}\quad,\quad
c^\gamma_{\alpha\alpha'}=-{1\over 2}(\delta_{\gamma\alpha}-\delta_{\gamma\alpha'})\,,
\end{equation}
provided that the renormalized exchange coupling matrices fulfil the
poor man scaling RG equations
\begin{equation}
\label{poor_man_scaling_kondo}
{d\over dl}\hat{J}^z\,=\,(\hat{J}^\perp)^2\quad,\quad
{d\over dl}\hat{J}^\perp\,=\,{1\over 2}(\hat{J}^z\hat{J}^\perp+\hat{J}^\perp\hat{J}^z)\quad,
\end{equation}
where $l=\ln{\Lambda_0/\Lambda}$ is the dimensionless flow parameter and
$\hat{J}^{z/\perp}|_{\Lambda=\Lambda_0}=\hat{J}^{z/\perp}_0$ is the initial condition.

The poor man scaling equations have the invariant
\begin{equation}
\label{invariant}
(2c)^2\,\equiv\,\text{Tr}\,(\hat{J}^z\hat{J}^z-\hat{J}^\perp\hat{J}^\perp)\,=\,\text{const}\quad.
\end{equation}
They can be solved easily if one assumes the form (\ref{J_form}) for the initial couplings.
In this case, we have the same form for the renormalized couplings,
\begin{equation}
\label{J_form_renormalized}
J^{z/\perp}_{\alpha\alpha'}\,=\,2\sqrt{x_\alpha x_{\alpha'}}\,J^{z/\perp}\quad,\quad
\sum_\alpha\,x_\alpha\,=\,1\quad,
\end{equation}
and the two exchange couplings  $J^z$ and $J^{\perp}$ fulfil the RG equations
\begin{equation}
\label{RG_J_kondo}
{d\over dl}{J}^z\,=\,2(J^\perp)^2\quad,\quad
{d\over dl}{J}^\perp\,=\,2{J}^z{J}^\perp\quad.
\end{equation}
These RG equations have two invariants
\begin{equation}
\label{invariant_c_TK}
c^2\,\equiv\,(J^z)^2-(J^\perp)^2\quad,\quad
T_K\,\equiv\,\Lambda\left({J^z-c\over J^z+c}\right)^{1\over 4c}\quad,
\end{equation}
where $c=i\sqrt{|(J^z)^2-(J^\perp)^2|}$ for $|J^z|<|J^\perp|$. $T_K$ denotes
the Kondo temperature, which can also be written as
\begin{equation}
\label{T_K_kondo}
T_K\,=\,\Lambda\,e^{-{1\over 2J^z}{1\over 2\delta}\ln{1+\delta \over 1-\delta}}\quad,
\end{equation}
with
\begin{equation}
\label{delta}
\delta\,=\,\text{sign}(J^z)\,\sqrt{1-({J^\perp\over J^z})^2}\quad,
\end{equation}
and, for $|J^z|<|J^\perp|$ [where $\delta=i\,\text{sign}(J^z)|\delta|$], the imaginary
part of the logarithm is defined such that $-2\pi<\ln(z)<0$ (such that $T_K$ does not
jump during the RG flow and is exponentially small). Thus, for 
$|J^z|<|J^\perp|$, we obtain
\begin{equation}
\label{T_K_small_J_z}
T_K\,=\,\Lambda\,e^{-{1\over 2J^z|\delta|}(\arctan{|\delta|}+{\pi\over 2}(\text{sign}(J^z)-1))}\quad.
\end{equation}
For the isotropic case ($J^z=J^\perp=J$), we get the usual form 
$T_K=\Lambda e^{-{1\over 2J}}$. In terms of the invariants, the solution of
Eq.~(\ref{RG_J_kondo}) can be written as
\begin{equation}
\label{RG_leading_solution}
J^z\,=\,c\,{1+\left({T_K\over\Lambda}\right)^{4c} \over 1-\left({T_K\over\Lambda}\right)^{4c}}
\quad,\quad
J^\perp\,=\,2c\,{\left({T_K\over\Lambda}\right)^{2c} \over 1-\left({T_K\over\Lambda}\right)^{4c}}
\quad.
\end{equation}
In the scaling limit $J^z_0,J^\perp_0\rightarrow 0$ and $\Lambda_0\rightarrow\infty$, such
that ${J^z_0\over J^\perp_0}=\text{const}$ and $T_K=\text{const}$ (which is possible for
$J^z>0$ or $|J^z|<|J^\perp|$), we obtain the well-known isotropic form
\begin{equation}
\label{isotropic_solution}
J^z\,=\,J^\perp\,=\,{1\over 2\ln{\Lambda\over T_K}} \quad,
\end{equation}
with $T_K$ given by Eq.~(\ref{T_K_kondo}) (where the anisotropy of the initial exchange 
couplings still enters).

{\it Liouvillian in first order.} To determine $L_S^{(1)}(E,\omega)$ from 
Eq.~(\ref{L_first_decomposition}), we use 
the solution (\ref{G_bar_leading_kondo}) for $\bar{G}^{(1)}_{11'}$ together with the
identities (\ref{product_decompose_1}) and (\ref{product_decompose_2}) to evaluate 
the r.h.s. of the RG equations (\ref{L_first_const}) and (\ref{Z_first}). We obtain
\begin{eqnarray}
\label{RG_L_first_kondo}
{d\over dl}L_S^{(1)}\,&=&\,{1\over 2}(\text{Tr}\hat{J}^z\hat{J}^z)h_0 L^h\quad,\\
\label{RG_Z_first_kondo}
{d\over dl}Z^{(1)}\,&=&\,(\text{Tr}\hat{J}^\perp\hat{J}^\perp)L^a+2c^2 L^c\quad,
\end{eqnarray}
with the invariant $c$ given by Eq.~(\ref{invariant}).
Using $\text{Tr}\hat{J}^z\hat{J}^z=\text{Tr}\hat{J}^\perp\hat{J}^\perp+4c^2$ and
$\hat{J}^\perp\hat{J}^\perp={d\over dl}\hat{J}^z$ 
according to Eq.~(\ref{poor_man_scaling_kondo}), we find the solution 
(note that the inital conditions are zero in first order in $J$)
\begin{eqnarray}
\label{L_first_E_kondo}
L_S^{(1)}(E,\omega)\,&=&\,L_S^{(1)}\,-\,(E+i\omega)\,Z^{(1)}\quad,\\
\label{L_first_kondo}
L_S^{(1)}\,&=&\,{1\over 2}\text{Tr}(\hat{J}^z-\hat{J}^z_0)h_0 L^h
+2c^2\,l\, h_0 L^h\quad,\\
\label{Z_first_kondo}
Z^{(1)}\,&=&\,\text{Tr}(\hat{J}^z-\hat{J}^z_0)L^a + 2c^2\,l\,L^c\quad,
\end{eqnarray}
where the flow parameter $l=\ln{\Lambda_0\over\Lambda}=\ln{\Lambda_0\over T_K}-\ln{\Lambda\over T_K}$
can be expressed in terms of the exchange couplings by using Eq.~(\ref{invariant_c_TK}):
\begin{equation}
\label{l_J}
l\,=\,{1\over 4c}\left(\ln{J^z-c\over J^z+c}-\ln{J^z_0-c\over J^z_0+c}\right) \quad.
\end{equation} 
Whereas the logarithms in this formula are pure anisotropy terms depending on the ratios of
the exchange couplings, the prefactor ${1\over 4c}$ gives $l\sim O(1/J)$, i.e., the second terms
$\sim c^2\,l$ on the r.h.s. of Eqs.~(\ref{L_first_kondo}) and (\ref{Z_first_kondo}) are of the same
order $O(J)$ as the first terms.

{\it Current kernel in first order.} The current kernel $\Sigma_\gamma^{(1)}(E,\omega)$ can be determined
similarly to $L_S^{(1)}(E,\omega)$, one just has to replace the first vertex on the r.h.s.
of the RG equations (\ref{L_first_const}) and (\ref{Z_first}) by the current vertex 
(\ref{I_components}). Using the identities (\ref{product_decompose_1}) and 
(\ref{product_decompose_2}) together with the algebra of the basis superoperators, we
obtain zero on the r.h.s. of the RG equations, i.e.,
\begin{equation}
\label{sigma_current_first_kondo}
\Sigma_\gamma^{(1)}(E,\omega)\,=\,0\quad.
\end{equation}

{\it Vertices in second order.} The vertices in second order in $J$ follow from the decomposition
(\ref{G_2a_decomposition}), where the first (imaginary) part $i\bar{G}^{(2a_1)}_{11'}$ (or
$i\bar{I}^{\gamma(2a_1)}_{11'}$ for the current kernel) is given by Eq.~(\ref{G_2a_1}). 
Since the renormalized vertices in first order have the same form as the initial ones,
we obtain the same form as Eqs.~(\ref{G_2a_1_initial_kondo}) and (\ref{I_2a_1_initial_kondo})
with $\hat{J}^{z/\perp}_0\rightarrow \hat{J}^{z/\perp}$,
\begin{eqnarray}
\nonumber
\bar{G}^{(2a_1)}_{11'}\,&=&\,
{\pi\over 2}\hat{J}^\perp\hat{J}^\perp s\hat{L}^{3z}_s \\
\label{G_2a_1_kondo}
&&+{\pi\over 4}(\hat{J}^z\hat{J}^\perp+\hat{J}^\perp\hat{J}^z)\hat{L}^{3}_s \,\,,\\
\nonumber
\bar{I}^{\gamma(2a_1)}_{11'}\,&=&\,
-{\pi\over 8}(\hat{J}^{\gamma z}\hat{J}^z-\hat{J}^z\hat{J}^{\gamma z}\\
\label{I_2a_1_kondo}
&&+2\hat{J}^{\gamma\perp}\hat{J}^\perp-2\hat{J}^\perp\hat{J}^{\gamma\perp})\hat{L}^{b}_s \,.
\end{eqnarray}

In contrast, the (real) parts $\bar{G}^{(2a_2)}_{11'}$ and $\bar{I}^{\gamma(2a_2)}_{11'}$ have to
be calculated from the RG equation (\ref{RG_G_2a}) with zero initial condition. We use the same
ansatz (\ref{G_bar_leading_kondo}) and (\ref{I_leading_kondo}) as for the vertices in lowest order,
i.e., for $\eta=-\eta'=+$
\begin{eqnarray}
\label{ansatz_G_2a_2_kondo}
\bar{G}^{(2a_2)}_{11'}&=&{1\over 2}\hat{K}^z s\hat{L}^h_s
-{1\over 2}\hat{K}^\perp(\hat{L}^4_s+\hat{L}^5_s)\,,\\
\label{ansatz_I_2a_2_kondo}
\bar{I}^{\gamma (2a_2)}_{11'}&=&{1\over 2}\hat{K}^{\gamma z} s\hat{L}^{1z}_s
+{1\over 2}\hat{K}^{\gamma\perp} \hat{L}^{1}_s\,,
\end{eqnarray}
with $\hat{K}^{z/\perp},\hat{K}^{\gamma,z/\perp}\sim O(J^2)$. However, instead of 
Eq.~(\ref{current_exchange_renormalized}), we set
\begin{equation}
\label{K_current_K_2a_2}
K^{\gamma,z/\perp}_{\alpha\alpha'}\,=\,c^\gamma_{\alpha\alpha'}K^{z/\perp}_{\alpha\alpha'}
\,+\,R^{\gamma,z/\perp}_{\alpha\alpha'} \quad,
\end{equation}
which can be viewed as a definition of $\hat{R}^{\gamma,z/\perp}$. Inserting 
Eqs.~(\ref{ansatz_G_2a_2_kondo}), (\ref{ansatz_I_2a_2_kondo}), (\ref{G_bar_leading_kondo}), 
(\ref{I_leading_kondo}), and (\ref{Z_first_kondo}) into the RG equation (\ref{RG_G_2a}), 
and using the following properties [analog to (\ref{product_decompose_3})], 
valid for $\eta=-\eta'=+$:
\begin{eqnarray}
\nonumber
&&\hspace{-0.5cm}
A_{12}\,L^\chi\,B_{\bar{2}1'}-(1\leftrightarrow 1')=
A^{\chi' s'}B^{\chi'' s''}\,\hat{L}^{\chi'}_{s'}L^{\chi}\hat{L}^{\chi''}_{s''}-\\
\label{product_decompose_4}
&&\hspace{1cm}
-B^{\chi'' s''}A^{\chi' s'}\,\left((\hat{L}^{\chi'}_{s'})^T\,L^\chi\,
(\hat{L}^{\chi''}_{s''})^T\right)^T\,,\\
\nonumber
&&\hspace{-0.5cm}
A_{23}\,B_{11'}\,C_{\bar{3}\bar{2}}= 
2 B^{\chi s}_{\alpha\alpha'}\,
\left(\text{Tr}_\alpha A^{\chi' s'}C^{\chi'' s''}\right)\cdot \,\\
\label{product_decompose_5}
&&\hspace{2cm}
\cdot\left(\text{Tr}_\sigma \hat{L}^{\chi'}_{s'}(\hat{L}^{\chi}_s)_{\sigma\sigma'}
\hat{L}^{\chi''}_{s''}\right)\,,
\end{eqnarray}
we obtain after some lengthy but straighforward algebra the following RG equations:
\begin{widetext}
\begin{eqnarray}
\label{RG_K_z}
{d \over dl}\hat{K}^z &=& 
-\left(\text{Tr}(\hat{J}^z-\hat{J}^z_0)\right)\hat{J}^\perp\hat{J}^\perp
+{1\over 2}\left(\text{Tr}\hat{J}^z\hat{J}^z\right)\hat{J}^z
+\hat{K}^\perp\hat{J}^\perp+\hat{J}^\perp\hat{K}^\perp\quad,\\
\nonumber
{d \over dl}\hat{K}^\perp &=&  
-\left(\text{Tr}(\hat{J}^z-\hat{J}^z_0)+2c^2\,l\right)
{1\over 2}(\hat{J}^z\hat{J}^\perp+\hat{J}^\perp\hat{J}^z)
+{1\over 2}\left(\text{Tr}\hat{J}^\perp\hat{J}^\perp\right)\hat{J}^\perp\\
\label{RG_K_perp}
&&\hspace{3cm}
+{1\over 2}(\hat{K}^\perp\hat{J}^z+\hat{J}^z\hat{K}^\perp
+\hat{K}^z\hat{J}^\perp+\hat{J}^\perp\hat{K}^z)\quad,\\
\label{RG_R_z}
{d \over dl}\hat{R}^{\gamma z} &=&
-{1\over 2}\left(\text{Tr}\hat{J}^z\hat{J}^z\right)\hat{J}^{\gamma z}
-2c^2\,l\,(\hat{J}^{\gamma\perp}\hat{J}^\perp+\hat{J}^\perp\hat{J}^{\gamma\perp})
+\hat{R}^{\gamma\perp}\hat{J}^\perp+\hat{J}^\perp\hat{R}^{\gamma\perp}\quad,\\
\label{RG_R_perp}
{d \over dl}\hat{R}^{\gamma \perp} &=&  
-{1\over 2}\left(\text{Tr}\hat{J}^\perp\hat{J}^\perp\right)\hat{J}^{\gamma\perp}
+ c^2\,l\,(\hat{J}^{\gamma z}\hat{J}^\perp+\hat{J}^\perp\hat{J}^{\gamma z})
+{1\over 2}(\hat{R}^{\gamma\perp}\hat{J}^z+\hat{J}^z\hat{R}^{\gamma\perp}
+\hat{R}^{\gamma z}\hat{J}^\perp+\hat{J}^\perp\hat{R}^{\gamma z})\quad.
\end{eqnarray}
\end{widetext}
Together with Eq.~\eqref{K_current_K_2a_2}, we have derived here 
the full two-loop equations for the vertex and the current vertex in the most general
case, including anisotropies and all possibilities for the dependencies on the 
reservoir indices. If we define the total vertex by
\begin{equation}
\label{total_vertex}
{\hat{\cal{J}}}^{z/\perp}\,=\,\hat{J}^{z/\perp}\,+\,\hat{K}^{z/\perp}\quad,
\end{equation}
and take the sum of the RG equations (\ref{poor_man_scaling_kondo}),
(\ref{RG_K_z}), and~(\ref{RG_K_perp}), we get, by neglecting terms of $O(J^4)$ and terms
of $O(J_0 J^2,J_0^2 J)$ (which vanish in the scaling limit $J_0\rightarrow 0$), the
following result:
\begin{widetext}
\begin{eqnarray}
\label{RG_total_z}
{d \over dl}{\hat{\cal{J}}}^z &=& \left(1-\text{Tr}{\hat{\cal{J}}}^z\right)({\hat{\cal{J}}}^\perp)^2
+{1\over 2}\left[\text{Tr}({\hat{\cal{J}}}^z)^2\right]{\hat{\cal{J}}}^z\quad,\\
\label{RG_total_perp}
{d \over dl}{\hat{\cal{J}}}^\perp &=&  
\left(1-\text{Tr}{\hat{\cal{J}}}^z\right)
{1\over 2}({\hat{\cal{J}}}^z{\hat{\cal{J}}}^\perp + {\hat{\cal{J}}}^\perp{\hat{\cal{J}}}^z)
+{1\over 2}\left[\text{Tr}({\hat{\cal{J}}}^\perp)^2\right]{\hat{\cal{J}}}^\perp\quad.
\end{eqnarray}
\end{widetext}
For ${\hat{\cal{J}}}^z={\hat{\cal{J}}}^\perp={\hat{\cal{J}}}$ and the case of two reservoirs
with ${\cal{J}}_{LL}={\cal{J}}_{RR}$, these equations reduce to the two-loop equations of 
Ref.~\onlinecite{doyon_andrei_PRB06} [where the nonuniversal parameter $a$ in Eq.~(104) of this reference has
to be chosen as $a=3$]. In equilibrium (i.e., for a single reservoir), we recover the well-known
two-loop RG equations for the anisotropic Kondo model derived in Ref.~\onlinecite{solyom_zawadowski_74}
[up to nonuniversal terms arising from adding terms proportional to the invariant 
$\text{Tr}(\hat{J}^z\hat{J}^z-\hat{J}^\perp\hat{J}^\perp)$].

If the exchange couplings fulfil the relation (\ref{J_form}), the RG equations can be solved
easily. In this case, we have
\begin{equation}
\label{R_gamma_form}
R^{\gamma,z/\perp}_{\alpha\alpha'}\,=\,c^\gamma_{\alpha\alpha'}\,R^{z/\perp}_{\alpha\alpha'}\quad,
\end{equation}
and
\begin{eqnarray}
\label{K_form}
K^{z/\perp}_{\alpha\alpha'}\,&=&\,2\sqrt{x_\alpha x_{\alpha'}}\,K^{z/\perp}\quad,\\
\label{R_form}
R^{z/\perp}_{\alpha\alpha'}\,&=&\,2\sqrt{x_\alpha x_{\alpha'}}\,R^{z/\perp}\quad,
\end{eqnarray}
with
\begin{eqnarray}
\label{nonumber}
{d \over dl}K^z &=& 
-4(J^z-J^z_0)(J^\perp)^2\\
\label{RG_K_z_1}
&&+2(J^z)^3+4J^\perp K^\perp\quad,\\
\nonumber
{d \over dl}K^\perp &=&  
-4(J^z-J^z_0+2c^2\,l)J^z J^\perp \\
\label{RG_K_perp_1}
&&+ 2(J^\perp)^3 + 2(J^z K^\perp+J^\perp K^z)\quad,\\ 
\nonumber
{d \over dl}R^z &=&
-2(J^z)^3 - 4c^2\,l\,(J^\perp)^2 + 2 J^\perp R^\perp\quad,\\
\nonumber
{d \over dl}R^\perp &=&  
-2 (J^\perp)^3 + 2c^2\,l\,J^z J^\perp \\
\label{RG_R_perp_1}
&& + J^z R^\perp + J^\perp R^z\quad.
\end{eqnarray}
Special solutions of these equations are given by
\begin{eqnarray}
\nonumber
K^z &=& {3\over 2}c^2 + 2 l c^2 J^z\\
\label{K_z_special} 
&&+ 4 l J_0^z(J^\perp)^2 -
(J^\perp)^2 \ln{J^\perp\over J_0^\perp} \quad,\\
\label{K_perp_special}
K^\perp &=& l\,c^2 J^\perp + 4 l J^z_0 J^z J^\perp -
J^z J^\perp \ln{J^\perp \over J^\perp_0} \quad,\\
\label{R_special}
R^z &=& -(J^z)^2 - 2c^2 l J^z \,\,,\,\,
R^\perp = -J^z J^\perp \quad,
\end{eqnarray}
and the initial conditions $K^z=K^\perp=R^z=R^\perp=0$ can be fulfilled
by adding linear combinations of the following solutions of the 
homogeneous part of the RG equations:
\begin{eqnarray}
\label{K_homogeneous_1}
K^z=(J^\perp)^2 &,& K^\perp=J^z J^\perp \quad,\\
\label{K_homogeneous_2}
K^z=J^z + 2 l (J^\perp)^2 &,& K^\perp=J^\perp + 2 l J^z J^\perp \quad,\\
\label{R_homogeneous}
R^z=J^z &,& R^\perp=J^\perp \quad.
\end{eqnarray}
As a consequence, the final solutions are given by
\begin{eqnarray}
\nonumber
K^z &=& {3\over 2}J^z(J^z-J^z_0) + 2 l c^2 J^z\\
\label{K_z_solution} 
&& + l J_0^z(J^\perp)^2 -
(J^\perp)^2 \ln{J^\perp\over J_0^\perp} \quad,\\
\nonumber
K^\perp &=& {3\over 2}J^\perp(J^z-J^z_0) + lc^2 J^\perp \\
\label{K_perp_solution}
&&+ l J^z_0 J^z J^\perp -
J^z J^\perp \ln{J^\perp \over J^\perp_0} \quad,\\
\label{R_z_solution}
R^z &=& J^z(J^z_0-J^z) - 2c^2 l J^z \quad,\\
\label{R_perp_solution}
R^\perp &=& J^\perp(J^z_0-J^z)\quad.
\end{eqnarray}

Using Eq.~(\ref{l_J}) and $c^2=(J^z)^2-(J^\perp)^2$, we see that the vertices
in second order contain four different types of terms
\begin{equation}
\label{second_order_vertices_order}
O(J_0 J)\,,\,O(J^2)\,,\,O(J^2\ln{J})\,,\,O(J^2\ln{J_0})\quad,
\end{equation}
in addition to factors which contain ratios of $J^z/J^\perp$ or $J^z_0/J^\perp_0$.
In the scaling limit, the terms of order $O(J_0 J)$ can be neglected. The terms of 
order $O(J^2)$ are just a perturbative correction to the first-order vertices and can
be neglected too. Therefore, we see that the terms $R^{z/\perp}$ are not
important and the current vertex (\ref{K_current_K_2a_2}) in second order 
can be written as
\begin{equation}
\label{K_current_K_2a_2_approx}
K^{\gamma,z/\perp}_{\alpha\alpha'}\,\approx\,c^\gamma_{\alpha\alpha'}K^{z/\perp}_{\alpha\alpha'}\quad.
\end{equation}
The terms of order $O(J^2\ln{J})$ are logarithmic corrections, which are of order $J\ln{J}$ smaller  
compared to the first-order vertices. Since $J\ll 1$ they are also perturbative corrections 
and can be omitted. They lead only to an overall change of the physical quantities 
without any interesting dependence on some physical energy scale (like e.g. the voltage, magnetic field, etc.).
The most important terms are those of order $O(J^2\ln{J_0})$ since they diverge in the scaling limit.
Therefore, they have to be incorporated into the definition of the Kondo temperature.
Using the solution (\ref{RG_leading_solution}) with $T_K\rightarrow T_K^\prime$ and
expanding in $J\ln{T_K^\prime\over T_K}$, we find
\begin{eqnarray}
\nonumber
(J^z)^\prime&=&c{1+\left({T_K^\prime\over\Lambda}\right)^{4c} \over 
1-\left({T_K^\prime\over\Lambda}\right)^{4c}}\approx J^z+2(J^\perp)^2\ln{T_K^\prime\over T_K}\,,\\
\nonumber
(J^\perp)^\prime &=& 2c\,{\left({T_K^\prime\over\Lambda}\right)^{2c} \over 
1-\left({T_K^\prime\over\Lambda}\right)^{4c}}\approx J^\perp+2J^z J^\perp\ln{T_K^\prime\over T_K}\,,
\end{eqnarray}
i.e., the terms of order $O(J^2\ln{J^\perp_0})$ in Eqs.~(\ref{K_z_solution}) and (\ref{K_perp_solution}) can
be acccounted for by the redefinition
\begin{equation}
\label{T_K_2_loop}
T_K^\prime\,=\,\sqrt{J^\perp_0}\,\,T_K \quad.
\end{equation}
As a consequence, the main effect of the two-loop terms for the vertices is the replacement 
$T_K\rightarrow T_K^\prime$, which we will always implicitly assume in the following.

In other (more academic) cases, where the form (\ref{J_form}) of the exchange couplings is not fulfilled,
the part $R^{\gamma z/\perp}_{\alpha\alpha'}$ can generally not be written in the
form (\ref{R_gamma_form}), i.e., $\gamma$ must not necessarily be equal to $\alpha$ or $\alpha'$.
From the RG equations (\ref{RG_R_z}) and (\ref{RG_R_perp}), one can only prove that
\begin{equation}
\label{R_properties}
R^{\gamma z/\perp}_{\alpha\alpha'}\,=\,-R^{\gamma z/\perp}_{\alpha'\alpha} \quad,\quad
\sum_\gamma R^{\gamma z/\perp}_{\alpha\alpha'}\,=\,0 \quad.
\end{equation}
However, in the special case of two reservoirs, these conditions lead again to the form (\ref{R_gamma_form}).
It is an open question of future research to analyze the two-loop equations in all cases and to find
out whether there are interesting situations where the two-loop contributions can not be simply accounted
for by a renormalization of a single parameter. In addition, it is not clear whether the corrections
$R^{\gamma,z/\perp}_{\alpha\alpha'}$ to the current vertex are generically unimportant. In this paper,
we discuss only the physically realizable situation, where the exchange couplings fulfil the
property (\ref{J_form}). 

{\it Liouvillian in second order.} For Eq.~(\ref{L_2_final_omega=0}), we need the following second-order contribution
to the Liouvillian:
\begin{equation}
\label{L_2b_total}
L_S^{(2b)}(E,\omega)\,=\,L_S^{(2b)}\,-(E+i\omega)\,Z^{(2b)} \quad,
\end{equation}
where $L_S^{(2b)}$ and $Z^{(2b)}$ are determined by the RG equations (\ref{L_2b}) and (\ref{Z_2b}) 
with initial values given by Eqs.~(\ref{L_2b_initial_kondo}) and (\ref{Z_2b_initial_kondo}). Inserting
the results (\ref{G_bar_leading_kondo}) and (\ref{G_2a_1_kondo}) for the vertices into
Eqs.~(\ref{L_2b}) and (\ref{Z_2b}), and using the identities (\ref{product_decompose_1}) and 
(\ref{product_decompose_2}), we find after some algebra the RG equations
\begin{eqnarray}
\label{RG_L_2b_kondo}
{d\over dl}L_S^{(2b)}\,&=&\,
-i\,\pi\,h_0\,\left(\text{Tr}\hat{J}^\perp \hat{J}^\perp \hat{J}^z\right)\,L^{3z}\quad,\\
\label{RG_Z_2b_kondo}
{d\over dl}Z^{(2b)}\,&=&\,0\quad.
\end{eqnarray}
Using $(\hat{J}^\perp)^2={d\over dl}\hat{J}^z$ according to Eq.~(\ref{poor_man_scaling_kondo})
together with the inital condition (\ref{L_2b_initial_kondo}), we find the result
\begin{eqnarray}
\label{L_2b_kondo}
L_S^{(2b)}&=&
-i{\pi\over 2}h_0\left(\text{Tr}\hat{J}^\perp \hat{J}^\perp\right)L^{3z}\quad,\\
\label{Z_2b_kondo}
Z_S^{(2b)}&=&0\quad,
\end{eqnarray}
where we have used that 
\begin{equation}
\nonumber
4c^2=\text{Tr}(\hat{J}^z\hat{J}^z-\hat{J}^\perp\hat{J}^\perp)
=\text{Tr}(\hat{J}^z_0\hat{J}^z_0-\hat{J}^\perp_0\hat{J}^\perp_0)
\end{equation}
is an invariant according to Eq.~(\ref{invariant}).

{\it Current kernel in second order.} Analog to $L_S^{(2b)}(E,\omega)$, we evaluate the current kernel
\begin{equation}
\label{sigma_current_2b_total}
\Sigma_\gamma^{(2b)}(E,\omega)\,=\,\Sigma_\gamma^{(2b)}-(E+i\omega)Z^{(2b)}_\gamma
\end{equation}
from the RG equations (\ref{L_2b}) and (\ref{Z_2b}) by using the results 
(\ref{I_leading_kondo}) and (\ref{I_2a_1_kondo}) for the current vertices.
This gives the RG equations
\begin{eqnarray}
\nonumber
{d\over dl}\Sigma_\gamma^{(2b)}\,&=&\,
-i{\pi\over 2}c^\gamma_{\alpha\alpha'}(\mu_\alpha-\mu_{\alpha'})
\left\{J^z_{\alpha\alpha'}(\hat{J}^\perp\hat{J}^\perp)_{\alpha'\alpha}+\right.\\
\label{RG_sigma_current_2b_kondo}
&&\hspace{0.5cm}
\left.+J^\perp_{\alpha\alpha'}(\hat{J}^z\hat{J}^\perp+
\hat{J}^\perp\hat{J}^z)_{\alpha'\alpha}\right\}L^b\quad,
\end{eqnarray}
and ${d\over dl}Z_\gamma^{(2b)}=0$. Using $\hat{J}^\perp\hat{J}^\perp={d\over dl}\hat{J}^z$ and
$\hat{J}^z\hat{J}^\perp+\hat{J}^\perp\hat{J}^z=2{d\over dl}\hat{J}^\perp$ 
according to Eq.~(\ref{poor_man_scaling_kondo}), we find together with the initial 
conditions (\ref{Sigma_current_2b_initial_kondo}) and (\ref{Z_2b_initial_kondo})
the result
\begin{equation}
\label{Sigma_current_2b_kondo}
\begin{split}
  \Sigma_\gamma^{(2b)}&=
  -i{\pi\over 4}c^\gamma_{\alpha\alpha'}(\mu_\alpha-\mu_{\alpha'}) 
  \left[(J^z_{\alpha\alpha'})^2+2(J^\perp_{\alpha\alpha'})^2\right] L^b\,,\\
  Z_\gamma^{(2b)}&=0.
\end{split}
\end{equation}

\subsubsection{RG below $\Lambda_c$}
\label{sec:kondo_below_Lambda_c}

We now start to evaluate the final formulas (\ref{L_1_final_omega=0})--(\ref{L_3b_final_omega=0})
for the effective Liouvillian and the current kernel, which are represented in terms of the basis
superoperators according to the forms (\ref{L_representation}) and (\ref{sigma_current_representation}).
All final quantities are evaluated at $\omega=0$ and are meant at $\Lambda=0$ which is not indicated 
explicitly, i.e., we write from now on
\begin{eqnarray}
\nonumber
L_S(E)\,&\equiv&\,L_S(E,\omega=0)_{\Lambda=0}=\\
\nonumber
&=&h(E)L^h-i\Gamma^a(E)L^a\\
\label{L_representation_final}
&&-i\Gamma^c(E)L^c-i\Gamma^{3z}(E)L^{3z} \,,\\
\nonumber
\Sigma_\gamma(E)\,&\equiv&\,\Sigma_\gamma(E,\omega=0)_{\Lambda=0}=\\
\label{sigma_current_representation_final}
&=&i\Gamma^b_\gamma(E)L^b
+i\Gamma^{1z}_\gamma(E)L^{1z}\,.
\end{eqnarray}

Since only the first-order vertices evaluated at $\Lambda=\Lambda_c$ occur in the final formulas, we will
furthermore use the convention
\begin{equation}
\label{J_c_convention}
J^z_{\alpha\alpha'}\equiv (J^z_{\alpha\alpha'})|_{\Lambda=\Lambda_c}\quad,\quad
J^\perp_{\alpha\alpha'}\equiv (J^\perp_{\alpha\alpha'})|_{\Lambda=\Lambda_c}\quad.
\end{equation}
According to Eq.~(\ref{Lambda_c}), the scale $\Lambda_c$ is defined as the maximum of all physical energy
scales, i.e., in our case of the Kondo problem
\begin{equation}
\label{Lambda_c_kondo}
\Lambda_c\,=\,\text{max}\{E,V,\tilde{h}\}\quad,
\end{equation} 
where $\tilde{h}$ is defined by the renormalized magnetic field or, more precisely, as the real part
of the eigenvalues of the operator $\tilde{L}_S$. Therefore, we have to keep in mind that the 
renormalized exchange couplings (\ref{J_c_convention}) depend implicitly on the variables $E$, 
$V$, and $\tilde{h}$ via the scale $\Lambda_c$.

According to Eq.~(\ref{tilde_LZ_convention}), any function ${\cal{H}}(\Delta_{12})$ 
of the quantities $\Delta_{12}=E_{12}-\tilde{L}_S$ is defined by
\begin{equation}
\label{H_decomposition}
{\cal{H}}(\Delta_{12})\,=\,\sum_i {\cal{H}}(E_{12}-z_i)P_i(z_i)\quad,
\end{equation}
with $i\equiv 0,1,\pm$ and $z_i=\lambda_i(z_i)$. $\lambda_i(z)\equiv\lambda(E,\omega)$ and 
$P_i(z)\equiv P_i(E,\omega)$ are
given by Eqs.~(\ref{eigenvalues}) and (\ref{projectors}), with $z\equiv E+i\omega$. As a consequence,
we obtain the following self-consistent equations for $z_i$:
\begin{eqnarray}
\label{z_0}
z_0\,&=&\,0\quad,\\
\label{z_1}
z_1\,&=&\,-i\Gamma^a(z_1)\quad,\\
\label{z_+-}
z_\pm\,&=&\,\pm h(z_\pm)-i(\Gamma^a+\Gamma^c)(z_\pm)\quad.
\end{eqnarray}
The pole $z_0=0$ corresponds to the stationary state, which does not occur in the
resolvents between the renormalized vertices due to our special construction, where
the eigenvalue zero is perturbatively integrated out during the first discrete RG step.
The poles $z_1$ and $z_{\pm}$ correspond to the spin relaxation and dephasing modes.
Using the symmetry properties (\ref{symmetry_gamma_L}) and (\ref{symmetry_h}), we
obtain under complex conjugation
\begin{eqnarray}
\label{z_1_cc}
z_1^*&=&i\Gamma^a(-z_1^*)\quad,\\
\label{z_+-_cc}
z_\pm^*&=&\pm h(-z_\pm^*)+i(\Gamma^a+\Gamma^c)(-z_\pm^*)\quad,
\end{eqnarray}
which gives 
\begin{eqnarray}
\label{z_1_form}
z_1&=&-z_1^*\equiv -i\tilde{\Gamma}_1\quad,\\
\label{z_+-_form}
z_\pm &=& -z_\mp^*=\pm\tilde{h}-i\tilde{\Gamma}_2\quad.
\end{eqnarray}
As a consequence, Eq.~(\ref{H_decomposition}) becomes
\begin{eqnarray}
\nonumber
{\cal{H}}(\Delta_{12})&=&{\cal{H}}(E_{12})P_0(0) + 
{\cal{H}}(E_{12}+i\tilde{\Gamma}_1)P_1(-i\tilde{\Gamma}_1)\\
\label{delta_decomposition}
&&\hspace{-1.5cm}
+{\cal{H}}(E_{12}-s\tilde{h}+i\tilde{\Gamma}_2)P_s(s\tilde{h}-i\tilde{\Gamma}_2)\,.
\end{eqnarray}
where we sum over $s=\pm$. 

Inserting Eq.~(\ref{delta_decomposition}) for the various
functions occurring in Eq.~(\ref{L_2_final_omega=0})--(\ref{L_3b_final_omega=0}), we see that
the projectors stand always left to $\bar{G}^{(1)c}_{ij}$ or $\bar{G}^{(2a_1)c}_{ij}$. Using
the results (\ref{G_bar_leading_kondo}) and (\ref{G_2a_1_kondo}) for these vertices 
together with the algebra
of the basis superoperators, we conclude that the parts of the projectors (\ref{projectors})
containing either $L^b$
or $L^{3z}$ do not contribute, and we can replace $L^a\rightarrow 1$. Thus, by inserting
Eq.~(\ref{projectors}) in Eq.~(\ref{delta_decomposition}), we obtain the useful identity
\begin{eqnarray}
\nonumber
{\cal{H}}(\Delta_{12})&\rightarrow& {\cal{H}}(E_{12}+i·\tilde{\Gamma}_1)(1-L^c)\\
\label{delta_decomposition_RG}
&&\hspace{-1.5cm}
+{\cal{H}}(E_{12}-s\tilde{h}+i\tilde{\Gamma}_2){1\over 2}(L^c+sL^h)\,,
\end{eqnarray}
which will be used frequently during the following analysis. 
As a consequence, we find that the spin relaxation rate $\tilde{\Gamma}_1$ will
cut off all logarithms where the magnetic field does not occur, and the spin dephasing rates
$\tilde{\Gamma}_2$ cuts off all logarithms where the magnetic field occurs. This is expected since
the spin dephasing rate corresponds to spin flip processes. In addition, we also conclude
from Eq.~(\ref{delta_decomposition_RG}) that the pole of the projector $P_1(E,\omega)$ at $\Gamma^a(E,\omega)=0$ 
does not contribute.

{\it Liouvillian and current kernel in first order.} The Liouvillian
and the current kernel in first order in $J$ have 
already been determined in Eqs.~(\ref{L_first_E_kondo})--(\ref{Z_first_kondo}) and 
(\ref{sigma_current_first_kondo}), which gives
\begin{eqnarray}
\label{h_1}
h^{(1)}(E)\,&=&\,
{1\over 2}\text{Tr}(\hat{J}^z-\hat{J}^z_0)h_0 + 2c^2\,l_c\, h_0 \,,\\
\label{gamma_a_1}
\Gamma^{a(1)}(E)\,&=&\,-i\,E\,\text{Tr}(\hat{J}^z-\hat{J}^z_0)\quad,\\
\label{gamma_c_1}
\Gamma^{c(1)}(E)\,&=&\,-2i\,E\,c^2\,l_c\,\quad,\\
\label{gamma_3z_1}
\Gamma^{3z(1)}(E)\,&=&\,0\quad,
\end{eqnarray}
with $l_c=\ln{\Lambda_0\over\Lambda_c}$, and 
\begin{equation}
\label{gamma_current_b_1z_1}
\Gamma_\gamma^{b(1)}(E)\,=\,\Gamma_\gamma^{1z(1)}(E)\,=\,0\quad.
\end{equation}

{\it Liouvillian and current kernel in second order.} The first two terms on the r.h.s. of Eq.~(\ref{L_2_final_omega=0}) have
already been evaluated in Eqs.~(\ref{L_2b_kondo}), (\ref{Z_2b_kondo}), and (\ref{Sigma_current_2b_kondo}). 
For the evaluation of the last two terms on the r.h.s., we use 
Eq.~(\ref{delta_decomposition_RG}) and the identity [${\cal{H}}(E)$ is any function]
\begin{eqnarray}
\nonumber
&& \hspace{-1cm}
{\cal{H}}(E_{11'}+z)\,A_{11'}\,L^\chi\,B_{\bar{1}'\bar{1}}=\\
\label{product_decompose_6}
&&\hspace{-0.5cm}
=2{\cal{H}}(E_{\alpha\alpha'}+z)\,A^{\chi' s'}_{\alpha\alpha'}B^{\chi'' s''}_{\alpha'\alpha}\,
\,\text{Tr}_\sigma \hat{L}^{\chi'}_{s'}L^\chi\hat{L}^{\chi''}_{s''}\,,
\end{eqnarray}
together with
the results (\ref{G_bar_leading_kondo}) and (\ref{I_leading_kondo}) for the first-order vertices. 
After some algebra, we obtain for the components of the effective
Liouvillian in second order
\begin{eqnarray}
\nonumber
&&\hspace{-1cm}
\text{Re}h^{(2)}(E)=
-{1\over 4}(E_{\alpha\alpha'}-\tilde{h})\mylc{E_{\alpha\alpha'}-\tilde{h}}(J^z_{\alpha\alpha'})^2\\
\label{re_h_2}
&&\hspace{2cm}
+(E\rightarrow -E)\quad,\\
\label{im_h_2}
&&\hspace{-1cm}
\text{Im}h^{(2)}(E)=
-{\pi\over 8}|E_{\alpha\alpha'}-\tilde{h}|(J^z_{\alpha\alpha'})^2
-(E\rightarrow -E)\,,\\
\label{re_gamma_a_2}
&&\hspace{-1cm}
\text{Re}\Gamma^{a(2)}(E)={\pi\over 4}|E_{\alpha\alpha'}-\tilde{h}|(J^\perp_{\alpha\alpha'})^2
+(E\rightarrow -E)\,,\\
\nonumber
&&\hspace{-1cm}
\text{Im}\Gamma^{a(2)}(E)=-{1\over 2}(E_{\alpha\alpha'}-\tilde{h})\mylc{E_{\alpha\alpha'}-\tilde{h}}
(J^\perp_{\alpha\alpha'})^2\\
\label{im_gamma_a_2}
&&\hspace{2cm}
-(E\rightarrow -E)\quad,\\
\nonumber
&&\hspace{-1cm}
\text{Re}[\Gamma^{a(2)}(E)+\Gamma^{c(2)}(E)]=\\
\nonumber
&&\hspace{0cm}
={\pi\over 8}\left\{|E_{\alpha\alpha'}-\tilde{h}|(J^z_{\alpha\alpha'})^2+
|E_{\alpha\alpha'}|(J^\perp_{\alpha\alpha'})^2\right\}\\
\label{re_gamma_ac_2}
&&\hspace{2cm}
+(E\rightarrow -E)\quad,\\
\nonumber
&&\hspace{-1cm}
\text{Im}[\Gamma^{a(2)}(E)+\Gamma^{c(2)}(E)]=\\
\nonumber
&&\hspace{0cm}
=-{1\over 4}\left\{(E_{\alpha\alpha'}-\tilde{h})\mylc{E_{\alpha\alpha'}-\tilde{h}}(J^z_{\alpha\alpha'})^2+\right.\\
\label{im_gamma_ac_2}
&&\hspace{0cm}
\left.+E_{\alpha\alpha'}\myla{E_{\alpha\alpha'}}(J^\perp_{\alpha\alpha'})^2\right\}
-(E\rightarrow -E)\quad,\\
\label{gamma_3z_2}
&&\hspace{-1cm}
\Gamma^{3z(2)}(E)={\pi\over 2}h_0\text{Tr}\hat{J}^\perp\hat{J}^\perp\quad,
\end{eqnarray}
and the following result for the components of the current kernel in
second order:
\begin{eqnarray}
\nonumber
&&\hspace{-0.5cm}
\Gamma^{b(2)}_\gamma(E)=
-{\pi\over 4}(\mu_\alpha-\mu_{\alpha'}) 
\left(J^{\gamma z}_{\alpha\alpha'}J^z_{\alpha\alpha'}+
2J^{\gamma\perp}_{\alpha\alpha'}J^\perp_{\alpha\alpha'}\right)\,,\\
\label{gamma_current_b_2}\\
\nonumber
&&\hspace{-0.5cm}
\text{Re}\Gamma^{1z(2)}_\gamma(E)=
{\pi \over 4}|E_{\alpha\alpha'}-\tilde{h}|J^{\gamma\perp}_{\alpha\alpha'}J^\perp_{\alpha\alpha'}
+(E\rightarrow -E)\,,\\
\label{re_gamma_current_1z_2}\\
\nonumber
&&\hspace{-0.5cm}
\text{Im}\Gamma^{1z(2)}_\gamma(E)=
-{1 \over 2}|E_{\alpha\alpha'}-\tilde{h}|\mylc{E_{\alpha\alpha'}-\tilde{h}}
J^{\gamma\perp}_{\alpha\alpha'}J^\perp_{\alpha\alpha'}\\
\label{im_gamma_current_1z_2}
&&\hspace{3cm}
-(E\rightarrow -E)\quad,
\end{eqnarray}
where we have used the shorthand notation
\begin{equation}
  \label{eq:log}
  \mathcal{L}_i(x):=\ln\frac{\Lambda_c}{\sqrt{x^2+(\tilde{\Gamma}_i)^2}}
\end{equation}
for the logarithmic terms. Note that the results respect the symmetry properties 
(\ref{symmetry_gamma_L})--(\ref{symmetry_h}), i.e., the real parts of all quantities are
symmetric in the Laplace variable $E$, whereas the imaginary parts are antisymmetric in $E$.

As we can see, for finite Laplace variable $E$, logarithmic 
corrections occur already in second order in $J$. This is the case for the real part of $h^{(2)}(E)$
and the imaginary parts of $\Gamma^{a(2)}(E)$, $\Gamma^{c(2)}(E)$, and $\Gamma^{1z(2)}_\gamma(E)$.
Therefore, in the stationary state (i.e., for $E=0$) no logarithmic contributions occur in $O(J^2)$
for the conductance and the magnetization, whereas the time evolution is influenced
by logarithmic contributions already in this order, see Sec.~\ref{sec:relaxation_dephasing_rates} for more
details.

{\it Liouvillian and current kernel in third order.} In third order, there are two contributions to the effective 
Liouvillian and the current kernel, denoted by the superscript $(3a)$ and $(3b)$ and given by the 
two expressions, Eqs.~(\ref{L_3a_final_omega=0}) and
(\ref{L_3b_final_omega=0}). The components $(3a)$ from Eq.~(\ref{L_3a_final_omega=0}) can be evaluated 
similiar to the components in second order from Eq.~(\ref{L_2_final_omega=0}) shown above, just by using 
in addition the results (\ref{G_2a_1_kondo}) and (\ref{I_2a_1_kondo}) for the imaginary part of the 
vertices in second order. We obtain after some algebra for the components of the
effective Liouvillian
\begin{eqnarray}
\label{h_gamma_a_c_3a}
h^{(3a)}(E)&=&\Gamma^{a(3a)}(E)=\Gamma^{c(3a)}(E)=0\,,\\
\nonumber
\Gamma^{3z(3a)}(E)&=&-{\pi\over 4}(E_{\alpha\alpha'}-\tilde{h})\mylc{E_{\alpha\alpha'}-\tilde{h}}\cdot\\
\label{gamma_3z_3a}
&&\hspace{-2cm}
\cdot J^\perp_{\alpha\alpha'}(\hat{J}^z\hat{J}^\perp+\hat{J}^\perp\hat{J}^z)_{\alpha'\alpha}
+(E\rightarrow -E)\,,
\end{eqnarray}
and for the components for the current kernel
\begin{eqnarray}
\nonumber
\Gamma^{b(3a)}_\gamma(E)&=&-{\pi\over 4}
\left\{E_{\alpha\alpha'}\myla{E_{\alpha\alpha'}}
J^{\gamma z}_{\alpha\alpha'}(\hat{J}^\perp\hat{J}^\perp)_{\alpha'\alpha}+\right.\\
\nonumber
&&\hspace{-2cm}
\left.+(E_{\alpha\alpha'}-\tilde{h})\mylc{E_{\alpha\alpha'}-\tilde{h}}
J^{\gamma\perp}_{\alpha\alpha'}(\hat{J}^z\hat{J}^\perp+\hat{J}^\perp\hat{J}^z)_{\alpha'\alpha}\right\}\\
\label{gamma_current_b_3a}
&&\hspace{2cm}
+(E\rightarrow -E)\quad.
\end{eqnarray}
As expected, the logarithmic terms for the real parts of $\Gamma^{3z}$ and $\Gamma^{b}_\gamma$ start
in order $J^3$. This means that the logarithmic terms for the
stationary magnetization (\ref{rd_stationary_kondo}) and the stationary current
(\ref{current_stationary_kondo}) start in third order in the exchange couplings.
Note that we have not calculated the terms in order $O(J^3)$ without a logarithmic contribution.

The remaining logarithmic contributions in third order are contained in the component $(3b)$, 
given by (\ref{L_3b_final_omega=0}) and (\ref{double_logarithmic_omega=0_approx}). With
$A\equiv \bar{G}^{(1)c}$ and $B\equiv \bar{I}^{\gamma(1)c}$ we have to evaluate expressions
of the form [note that $(A^{\chi s})^T=A$ and $(B^{\chi s})^T=-B$ according to Eqs.~(\ref{G_bar_leading_kondo}),
(\ref{I_leading_kondo}), and (\ref{hermiticity_exchange})]
\begin{eqnarray}
\nonumber
&&\hspace{-1cm} 
(A/B)_{12}R(\Delta_{12})A_{\bar{2}3}S(\Delta_{13})A_{\bar{3}\bar{1}}=\\
\nonumber
&&\hspace{-0.7cm}
=(A/B)^{\chi s}_{\alpha_1\alpha_2}A^{\chi' s'}_{\alpha_2\alpha_3} A^{\chi'' s''}_{\alpha_3\alpha_1} \cdot\\
\nonumber
&&\hspace{-0.5cm}
\cdot\text{Tr}_\sigma\left\{\hat{L}^\chi_s R(\Delta_{\alpha_1\alpha_2}) \hat{L}^{\chi'}_{s'}
S(\Delta_{\alpha_1\alpha_3})\hat{L}^{\chi''}_{s''}-\right.\\
\label{product_decompose_7}
&&\hspace{-0.3cm}
\left.-(\hat{L}^\chi_s)^T R(\Delta_{\alpha_2\alpha_1}) (\hat{L}^{\chi'}_{s'})^T
S(\Delta_{\alpha_3\alpha_1}) (\hat{L}^{\chi''}_{s''})^T \right\}\,,
\end{eqnarray}
where $\Delta_{\alpha_i\alpha_j}=E_{\alpha_i\alpha_j}-\tilde{L}_S$, and the two functions $R(E)$
and $S(E)$ are either both symmetric or antisymmetric: $R(-E)S(-E)=R(E)S(E)$. This identity can
be derived by summing over the two possibilities $\eta_i=\pm$ and using the 
representations (\ref{G_representation}), (\ref{I_representation}), (\ref{G_representation_T}) 
and (\ref{I_representation_T}). The functions
$(R/S)(\Delta_{\alpha_i\alpha_j})$ on the r.h.s. can be expressed in terms of the basis
superoperators by using Eq.~(\ref{delta_decomposition_RG}),
\begin{eqnarray}
\nonumber
(R/S)(\Delta_{\alpha_i\alpha_j})&\rightarrow& 
(R/S)_{\alpha_i\alpha_j}(E+i\tilde{\Gamma}_1)(1-L^c)\\
\nonumber
&&\hspace{-2.5cm}
+(R/S)_{\alpha_i\alpha_j}(E-s\tilde{h}+i\tilde{\Gamma}_2){1\over 2}(L^c+sL^h)\,,\\
\label{RS_decomposition}
\end{eqnarray}
where $R_{\alpha_i\alpha_j}(E)=R(E+\mu_{\alpha_i}-\mu_{\alpha_j})$. Inserting this result
in Eq.~(\ref{product_decompose_7}), we find after some straightforward algebra
\begin{widetext}
\begin{eqnarray}
\nonumber
\bar{G}^{(1)c}_{12}R(\Delta_{12})\bar{G}^{(1)c}_{\bar{2}3}S(\Delta_{13})
\bar{G}^{(1)c}_{\bar{3}\bar{1}}&=&
-{1\over 8}J^z_{\alpha_1\alpha_2}J^\perp_{\alpha_2\alpha_3}J^\perp_{\alpha_3\alpha_1}
R_{\alpha_1\alpha_2}(s'E-s\tilde{h})S_{\alpha_1\alpha_3}(s'E)(L^c+ss'L^h)\\
\nonumber
&&-{1\over 8}J^\perp_{\alpha_1\alpha_2}J^\perp_{\alpha_2\alpha_3}J^z_{\alpha_3\alpha_1}
R_{\alpha_1\alpha_2}(s'E)S_{\alpha_1\alpha_3}(s'E-s\tilde{h})(L^c+ss'L^h)\\
\label{L_3b_decomposition}
&&-{1\over 4}J^\perp_{\alpha_1\alpha_2}J^z_{\alpha_2\alpha_3}J^\perp_{\alpha_3\alpha_1}
R_{\alpha_1\alpha_2}(s'E-s\tilde{h})S_{\alpha_1\alpha_3}(s'E-s\tilde{h})(L^a-L^c)\quad,
\end{eqnarray}
and for $E=0$ 
\begin{eqnarray}
\nonumber
\bar{I}^{\gamma(1)c}_{12}R(\Delta_{12})\bar{G}^{(1)c}_{\bar{2}3}S(\Delta_{13})
\bar{G}^{(1)c}_{\bar{3}\bar{1}}|_{E=0}&=&
s{1\over 2}c^\gamma_{\alpha_1\alpha_2}
J^z_{\alpha_1\alpha_2}J^\perp_{\alpha_2\alpha_3}J^\perp_{\alpha_3\alpha_1}
R_{\alpha_1\alpha_2}(0)S_{\alpha_1\alpha_3}(s\tilde{h})L^{1z}\\
\label{sigma_current_3b_decomposition}
&&+s{1\over 2}c^\gamma_{\alpha_1\alpha_2}
J^\perp_{\alpha_1\alpha_2}J^z_{\alpha_2\alpha_3}J^\perp_{\alpha_3\alpha_1}
R_{\alpha_1\alpha_2}(s\tilde{h})S_{\alpha_1\alpha_3}(s\tilde{h})L^{1z}\quad.
\end{eqnarray}
\end{widetext}
Using this result in Eq.~(\ref{L_3b_final_omega=0}), we have to distinguish many different
cases in order to evaluate the logarithmic contributions according to
Eq.~(\ref{double_logarithmic_omega=0_approx}). To simplify the discussion, we will assume
in the following only two reservoirs with chemical potentials $\mu_\alpha=\alpha {V\over 2}$.
Furthermore, we take w.l.o.g. $V,\tilde{h}>0$. As we will see in 
Sec.~\ref{sec:relaxation_dephasing_rates}, we need only the two cases $E=0$ and $E=\tilde{h}$
in Eq.~(\ref{L_3b_decomposition}) for the calculation of the renormalized $g$ factor and the 
spin relaxation and dephasing rates $\tilde{\Gamma}_{1/2}$ up to the first logarithmic
correction. For the magnetization and the current, we consider the stationary case $E=0$.

We start with $E=0$. In this case, there are four possibilities for the parameters $|\bar{a}|$ and
$|\bar{b}|$ in (\ref{double_logarithmic_omega=0_approx}):
\begin{equation}
\label{E=0_ab_parameters}
|\bar{a}|,|\bar{b}|\,=\,{V\over\Lambda_c}\,,\,{\tilde{h}\over\Lambda_c}\,,\,
{|V\pm\tilde{h}|\over\Lambda_c}\quad.
\end{equation}
According to Eq.~(\ref{double_logarithmic_omega=0_approx}), logarithmic contributions can only occur
if at least one of these parameters is small compared to one. Therefore, we can distinguish
three cases
\begin{eqnarray}
\nonumber
V\ll \tilde{h}\quad&\rightarrow&\quad{V\over\Lambda_c}\ll 1\quad,\\
\nonumber
V\sim \tilde{h}\quad&\rightarrow&\quad{|V-\tilde{h}|\over\Lambda_c}\ll 1\quad,\\
\label{E=0_cases}
V\gg \tilde{h}\quad&\rightarrow&\quad{\tilde{h}\over\Lambda_c}\ll 1\quad,
\end{eqnarray}
and all the other parameters not indicated are of order one. Discussing these three cases
separately and collecting only the logarithmic terms according to 
Eq.~(\ref{double_logarithmic_omega=0_approx}), we find after some lengthy calculation
\begin{eqnarray}
\nonumber
&&\hspace{-1.5cm}
\Gamma^{a(3b)}(0)=\pi\tilde{h}\mylc{\tilde{h}}J^z_\alpha(J^\perp_{\alpha})^2\\
\nonumber
&&\hspace{0cm}
+{\pi\over 2}|V-\tilde{h}|\mylc{V-\tilde{h}}J^z_\alpha (J^\perp_{\text{nd}})^2\\
\label{gamma_a_3b}
&&\hspace{0cm}
-{\pi\over 2}(V-\tilde{h})\mylc{V-\tilde{h}}J^z_{\text{nd}}
J^\perp_{\text{nd}}J^\perp_\alpha\\
\nonumber
&&\hspace{-1.5cm}
\Gamma^{a(3b)}(0)+\Gamma^{c(3b)}(0)=
{\pi\over 2}\tilde{h}\mylc{\tilde{h}}J^z_\alpha(J^\perp_{\alpha})^2\\
\label{gamma_ac_3b}
&&\hspace{-1cm}
+{\pi\over 4}(V-\tilde{h}+|V-\tilde{h}|)\mylc{V-\tilde{h}}J^z_{\text{nd}}
J^\perp_{\text{nd}}J^\perp_\alpha\,,\\
\nonumber
&&\hspace{-1.5cm}
\Gamma^{1z(3b)}_\gamma(0)=
-\pi V\myla{V}J^{\gamma z}_{\text{nd}}J^\perp_{\text{nd}}J^\perp_{\alpha}\\
\nonumber
&&\hspace{-0cm}
-{\pi\over 2}\tilde{h}\mylc{\tilde{h}}
(J^{\gamma z}_{\text{nd}}J^\perp_{\text{nd}}+
J^{\gamma \perp}_{\text{nd}}J^z_{\text{nd}})J^\perp_{\alpha}\\
\nonumber
&&\hspace{0cm}
+{\pi\over 2}|V-\tilde{h}|\mylc{V-\tilde{h}}J^{\gamma\perp}_{\text{nd}} J^\perp_{\text{nd}}J^z_\alpha\\
\label{gamma_current_1z_3b}
&&\hspace{-1cm}
+{\pi\over 4}(V-\tilde{h})\mylc{V-\tilde{h}}
(J^{\gamma z}_{\text{nd}}J^\perp_{\text{nd}}-
J^{\gamma \perp}_{\text{nd}}J^z_{\text{nd}})J^\perp_{\alpha}\,,
\end{eqnarray}
with
\begin{eqnarray}
\label{J_d_nd_definition}
J^{z/\perp}_\alpha&=&J^{z/\perp}_{\alpha\alpha}\,,\,
J^{z/\perp}_{\text{nd}}=J^{z/\perp}_{LR}=J^{z/\perp}_{RL}\quad,\\
J^{\gamma,z/\perp}_{\text{nd}}&=&J^{\gamma,z/\perp}_{LR}=-J^{\gamma,z/\perp}_{RL}\quad.
\end{eqnarray}
All other components do not contain any logarithmic contribution in third order in $J$.

Next we consider $E=\tilde{h}$. In this case, there are six possibilities for the parameters $|\bar{a}|$ and
$|\bar{b}|$ in (\ref{double_logarithmic_omega=0_approx}),
\begin{equation}
\label{E=h_ab_parameters}
|\bar{a}|,|\bar{b}|\,=\,{V\over\Lambda_c}\,,\,{\tilde{h}\over\Lambda_c}\,,\,
{|V\pm\tilde{h}|\over\Lambda_c}\,,\,{|V\pm 2\tilde{h}|\over\Lambda_c}\quad.
\end{equation}
In addition to the three cases shown in Eq.~(\ref{E=0_cases}), we have to consider
the additional case
\begin{equation}
\label{E=h_additional_case}
V\sim 2\tilde{h}\quad\rightarrow\quad{|V-2\tilde{h}|\over\Lambda_c}\ll 1\quad.\\
\end{equation}
Discussing these four cases, we find after a lengthy calculation the following 
logarithmic terms for the effective Liouvillian at $E=\tilde{h}$:
\begin{eqnarray}
\nonumber
&&\hspace{-1.5cm}
\Gamma^{a(3b)}(\tilde{h})=\frac\pi2\tilde{h}\mylc{\tilde{h}}J^z_\alpha(J^\perp_{\alpha})^2\\
\nonumber
&&\hspace{0cm}
+{\pi\over 2}V\mylc{V}(J^z_\alpha J^\perp_{\text{nd}}+ J^z_{\text{nd}}J^\perp_\alpha)J^\perp_{\text{nd}}\\
\nonumber
&&\hspace{0cm}
+{\pi\over 4}|V-2\tilde{h}|\mylc{V-2\tilde{h}}J^z_\alpha (J^\perp_{\text{nd}})^2\\
\label{gamma_a_3b_E=h}
&&\hspace{0cm}
-{\pi\over 4}(V-2\tilde{h})\mylc{V-2\tilde{h}}J^z_{\text{nd}}J^\perp_{\text{nd}}J^\perp_\alpha,\\
\nonumber
&&\hspace{-1.5cm}
\Gamma^{a(3b)}(\tilde{h})+\Gamma^{c(3b)}(\tilde{h})=\\
\nonumber
&&\hspace{0cm}
=\frac\pi2\tilde{h}\left(\frac12\mylc{\tilde{h}}+\myla{\tilde{h}}\right)J^z_\alpha(J^\perp_{\alpha})^2\\
\label{gamma_ac_3b_E=h}
&&\hspace{-1cm}
-{\pi\over 4}\theta(\tilde{h}-V)(V-\tilde{h})\myla{V-\tilde{h}}
J^z_\alpha (J^\perp_{\text{nd}})^2,\\
\nonumber
&&\hspace{-1.5cm}
\text{Im}h^{(3b)}(\tilde{h})=
\frac\pi4\tilde h \mylc{\tilde h}J^z_\alpha(J^\perp_\alpha)^2
\\
\nonumber
&&\hspace{0cm}
-{\pi\over 4}(V-\tilde{h})\myla{V-\tilde{h}}J^z_{\text{nd}}J^\perp_{\text{nd}}J^\perp_\alpha\\
\label{h_3b_E=h}
&&\hspace{-1cm}
-{\pi\over 4}\theta(V-\tilde{h})(V-\tilde{h})\myla{V-\tilde{h}}J^z_\alpha(J^\perp_{\text{nd}})^2\,.
\end{eqnarray}

\section{Results}
\label{sec:kondo_results}

In this section, we summarize the zero temperature results for the anisotropic
nonequilibrium Kondo model in a finite magnetic field, introduced at the beginning
of Sec.~\ref{sec:kondo_model_algebra}. We will calculate quantities
characterizing the exponential time decay of the magnetization, i.e., the spin 
relaxation and dephasing rates $\tilde{\Gamma}_{1/2}$ together with the renormalized 
magnetic field $\tilde{h}$. Furthermore, we present results for the stationary current
$\langle I^\gamma \rangle^{st}$ and the stationary magnetization $M$.
All quantities are calculated one order beyond leading order, i.e., up to the first term
leading to logarithmic enhancements (suppressions) at resonance.

For simplicity, we evaluate the results for two reservoirs with voltages
\begin{equation}
\label{two_reservoirs}
\mu_\alpha=\alpha {V\over 2}\quad,\quad
\alpha\equiv \text{L/R}\equiv \pm\quad.
\end{equation}
The diagonal and nondiagonal exchange couplings are denoted by 
\begin{equation}
\label{notation_couplings}
J^{z/\perp}_{\alpha}=J^{z/\perp}_{\alpha\alpha}\quad,\quad
J^{z/\perp}_{\text{nd}}=J^{z/\perp}_{LR}=J^{z/\perp}_{RL}\quad.
\end{equation}
The various couplings are renormalized couplings evaluated at (note that the Laplace 
variable $E$ is fixed to either $E=0$ or $E=\tilde{h}$ in the following) the cutoff scale
\begin{equation}
\label{Lambda_c_results}
\Lambda_c=\max\{V,\tilde{h}\}\quad.
\end{equation}
Unrenormalized (bare) couplings are denoted by the index ``0''. Furthermore, we consider
w.l.o.g. the case $V,\tilde{h}>0$.
Although our results of the previous section include all cases for the ratios between diagonal
and nondiagonal couplings, we will treat in this section the realistic case where the
exchange couplings fulfil the relation (\ref{J_form}), i.e.,
\begin{equation}
\label{exchange_couplings_special}
J^{z/\perp}_{\alpha\alpha'}=2\sqrt{x_\alpha x_{\alpha'}}J^{z/\perp}\quad,\quad
x_L+x_R=1\quad.
\end{equation}
In this case, the two couplings $J^z$ and $J^\perp$ can be calculated analytically 
from Eq.~\eqref{RG_leading_solution} with 
\begin{equation}
\label{value_c_TK}
c^2\,=\,(J^z_0)^2-(J^\perp_0)^2\quad,\quad
T_K\,\equiv\,\Lambda_0\sqrt{J^\perp_0}\left({J^z_0-c\over J^z_0+c}\right)^{1\over 4c}\quad,
\end{equation}
according to Eqs.~(\ref{invariant_c_TK}) and (\ref{T_K_2_loop}). Furthermore, we note that some
of the third-order terms can be taken together because
\begin{equation}
\label{eq:3rd-order-Jz-Jp}
J^z_\alpha\,(J^\perp_{\text{nd}})^2\,=\,J^\perp_\alpha\,J^z_{\text{nd}}\,J^\perp_{\text{nd}}\quad.
\end{equation}

Finally, for the logarithmic terms we use the shorthand notation 
\begin{equation}
\label{log_short_hand}
\myla{x}\,=\,\ln{\Lambda_c\over \sqrt{x^2 + \tilde{\Gamma}_1^2}}\quad,\quad
\mylc{x}\,=\,\ln{\Lambda_c\over \sqrt{x^2 + \tilde{\Gamma}_2^2}}\quad,
\end{equation}
and close to resonance, we use the broadened sign function 
\begin{equation}
\label{sign_replacement}
\text{sign}_i(x)\,=\,{2\over\pi}\arctan{x\over\tilde{\Gamma}_i}\quad,
\end{equation}
where $\tilde{\Gamma}_1$ ($\tilde{\Gamma}_2$) has to be chosen when the magnetic
field does not (does) occur in $x$ (this rule refers to the case $E=0$, for $E=\tilde{h}$
also other cases can occur, see below). Furthermore, we use the broadening of 
the sign function also for expressions involving the absolute value or
the theta function
via
\begin{eqnarray}
\label{absolute_gamma_broadening}
{|x|}_i\,&=&\,x\,\text{sign}_i(x)\quad,\\
\label{theta_gamma_broadening} 
\theta_i(x)\,&=&\,{1\over 2}[1+\text{sign}_i(x)]\quad.
\end{eqnarray}

\subsection{$g$~factor, spin relaxation and dephasing rates}
\label{sec:relaxation_dephasing_rates}

In this section, we determine the renormalized magnetic field $\tilde{h}$ or the
renormalized $g$~factor 
\begin{equation}
\label{g_renormalized}
\tilde{g}\,=\,2\,{d\tilde{h}\over dh_0}\quad,
\end{equation}
together with the renormalized spin relaxation and dephasing rates $\tilde{\Gamma}_{1/2}$.
These quantities enter all formulas, especially they determine the resonance
position $V=\tilde{h}$ and how the logarithmic divergencies are cut off by
the rates. We note that previous works have only calculated $\tilde{\Gamma}_{1/2}$
up to $O(J^2)$ in bare perturbation theory,\cite{paaske_rosch_kroha_woelfle_PRB04} and 
the renormalized magnetic field has so far been calculated only up to $O(J)$, see, e.g.,
Ref.~\onlinecite{garst_etal_PRB05}. In contrast, here we perform a renormalized perturbation theory for
the rates and calculate in addition the logarithmic terms of $O(J^3\ln)$ for
$\tilde{\Gamma}_{1/2}$ and those of $O(J^2\ln)$ for $\tilde{h}$ (corresponding to
contributions in three-loop by using the conventional classification). Furthermore,
we emphasize that we do not need to combine self-energy terms with vertex corrections as
in slave-particle formalism \cite{paaske_rosch_kroha_woelfle_PRB04} since we have directly 
set up a kinetic equation from
the very beginning where the renormalized rates and the renormalized magnetic field
can be directly read off by studying the poles of the reduced density matrix of the
quantum dot in Laplace space, see the discussion at the end of 
Sec.~\ref{sec:generic_model}. 
We now calculate $\tilde{h}$ and $\tilde{\Gamma}_{1/2}$ perturbatively in the
renormalized exchange couplings by using our results (\ref{h_1})--(\ref{gamma_c_1}),
(\ref{re_h_2})--(\ref{im_gamma_ac_2}), (\ref{gamma_a_3b}), 
(\ref{gamma_ac_3b}), and (\ref{gamma_a_3b_E=h})--(\ref{h_3b_E=h}) for $h(z)$ and
$\Gamma^{a/c}(z)$ (note that we take the analytic continuation $E\rightarrow E+i\omega$
of these equations). Thereby, we consider all terms of $O(J^2,J^3\ln)$ for $\tilde{\Gamma}_{1/2}$,
and all terms of $O(J^0)$, $O(J)$, $O(J^2\ln)$ for $\tilde{h}$. Terms of $O(J^3)$ [$O(J^2)$] without
any logarithmic contribution are consistently neglected for $\tilde{\Gamma}_{1/2}$
($\tilde{h}$).

To get the perturbative solution, we insert first the lowest order results 
(\ref{h_1})--(\ref{gamma_c_1}) into Eqs.~(\ref{z_1}) and (\ref{z_+-}). This gives 
\begin{eqnarray}
\nonumber
z_1 &=& -i{1\over 1-J^z_\alpha+(J^z_\alpha)_0}\Gamma^{a(2)}(z_1)-i\Gamma^{a(3)}(z_1)\,,\\
\label{z_1_zw}\\
\nonumber
z_+ &=& {1+{1\over 2}\left(J^z_\alpha-(J^z_\alpha)_0\right)+2c^2 l_c \over 
1+J^z_\alpha-(J^z_\alpha)_0+2c^2 l_c}h_0\\
\label{z_+_zw}
&&\hspace{-0.5cm}
+\sum_{k=2,3}\left(h^{(k)}(z_+)-i(\Gamma^{a(k)}+\Gamma^{c(k)})(z_+)\right)\,.
\end{eqnarray}
Using the forms (\ref{z_1_form}) and (\ref{z_+-_form}) for $z_1$ and $z_+$, we see
that $\tilde{\Gamma}_{1/2}$ starts at $O(J^2)$. Therefore, neglecting terms of $O(J^4)$,
we can replace $z_1\rightarrow 0$ and $z_+\rightarrow\tilde{h}$ in the arguments of 
$\Gamma^{a/c(k)}$ and $h^{(k)}$ for $k=2,3$. Furthermore, we can neglect all contributions
of $O(J^3)$ for the first term on the r.h.s. of Eq.~(\ref{z_1_zw}), and all terms of $O(J^2)$
for the first term on the r.h.s. of Eq.~(\ref{z_+_zw}) (since they do not contain any 
logarithmic contribution). Thus, we finally obtain
\begin{eqnarray}
\label{tilde_gamma_1}
\tilde{\Gamma}_1 &=& \Gamma^{a(2)}(0)+\Gamma^{a(3)}(0)\,,\\
\nonumber
\tilde{\Gamma}_2 &=& \sum_{k=2,3}\left\{-\text{Im}h^{(k)}(\tilde{h})
+\text{Re}(\Gamma^{a(k)}+\Gamma^{c(k)})(\tilde{h})\right\}\,,\\
\label{tilde_gamma_2}\\
\label{tilde_h}
\tilde{h} &=& h+\text{Re}h^{(2)}(\tilde{h})+\text{Im}(\Gamma^{a(2)}+\Gamma^{c(2)})(\tilde{h})\,,
\end{eqnarray}
where
\begin{equation}
\label{renormalized_h}
h=\left(1-{1\over 2}(J^z_\alpha-(J^z_\alpha)_0)\right)h_0
\end{equation}
is the renormalized magnetic field up to first order in $J$.

Inserting the results (\ref{re_h_2})--(\ref{im_gamma_ac_2}), (\ref{gamma_a_3b}), 
(\ref{gamma_ac_3b}), and (\ref{gamma_a_3b_E=h})--(\ref{h_3b_E=h}) into 
Eqs.~(\ref{tilde_gamma_1})--(\ref{tilde_h}), and specializing to the case 
(\ref{two_reservoirs}) of two reservoirs and the special form (\ref{exchange_couplings_special})
of the exchange couplings, we obtain
\begin{widetext}
\begin{eqnarray}
\nonumber
\tilde{\Gamma}_1 &=& {\pi\over 2}\tilde{h}\left(J^\perp_\alpha\right)^2 +
{\pi\over 2}\left({|V-\tilde{h}|}_2+V+\tilde{h}\right)\left(J^\perp_{\text{nd}}\right)^2 \\
\label{tilde_gamma_1_final}
&& + \pi\tilde{h}\mylc{\tilde{h}}J^z_\alpha\left(J^\perp_\alpha\right)^2 +
{\pi\over 2}{|V-\tilde{h}|}_2\mylc{V-\tilde{h}}J^z_\alpha\left(J^\perp_{\text{nd}}\right)^2 - 
{\pi\over 2}(V-\tilde{h})\mylc{V-\tilde{h}}J^z_{\text{nd}}J^\perp_{\text{nd}}J^\perp_\alpha \quad,\\
\nonumber
\tilde{\Gamma}_2 &=& {\pi\over 2}V\left(J^z_{\text{nd}}\right)^2 +
{\pi\over 4}\tilde{h}\left(J^\perp_\alpha\right)^2 + 
{\pi\over 4}\left({|V-\tilde{h}|}_1+V+\tilde{h}\right)\left(J^\perp_{\text{nd}}\right)^2 \\
\nonumber
&& + \frac\pi2\tilde{h}\myla{\tilde{h}}J^z_\alpha\left(J^\perp_\alpha\right)^2 +
{\pi\over 4}{|V-\tilde{h}|}_1\myla{V-\tilde{h}}J^z_\alpha\left(J^\perp_{\text{nd}}\right)^2 + 
{\pi\over 4}(V-\tilde{h})\myla{V-\tilde{h}}J^z_{\text{nd}}J^\perp_{\text{nd}}J^\perp_\alpha \,,\\
\label{tilde_gamma_2_final}\\
\label{tilde_h_final}
\tilde{h} &=& h-{1\over 2}\tilde{h}\myla{\tilde{h}}\left(J^\perp_\alpha\right)^2 +
{1\over 2}(V-\tilde{h})\myla{V-\tilde{h}}\left(J^\perp_{\text{nd}}\right)^2 \quad.
\end{eqnarray}
\end{widetext}
These formulas show precisely which rate cuts off the various logarithmic terms. Since the determination
of the renormalized rates and the renormalized magnetic field involves also the value of the
Liouvillian $L_S(E)$ at $E=\tilde{h}$, we can not just use the simple rule that $\tilde{\Gamma}_1$
($\tilde{\Gamma}_2$) cuts off the logarithms where the magnetic field does not (does) occur (as it is
the case for the stationary case $E=0$). In contrast, we find that the logarithmic terms of $\tilde{h}$ 
and $\tilde{\Gamma}_2$ are cut off by $\tilde{\Gamma}_1$, and those of $\tilde{\Gamma}_1$ by
$\tilde{\Gamma}_2$.

$h$ is the well-known renormalized magnetic field up to first order in $J$ which determines the precise 
resonance positions. Note that the sign of the linear terms in $J$ in Eq.~(\ref{renormalized_h}) is different 
from the one in Eq.~(\ref{L_first_kondo}), showing that it is important to consider the linear
term in frequency of the renormalized Liouvillian to determine the correct renormalization of the
magnetic field (similiar to field-theoretical rescaling techniques
from the $Z$ factor\cite{zinn_justin}).
Note that the renormalized magnetic field is linear in the renormalized coupling, but starts only
in second order with respect to the bare coupling. Furthermore, we note that the determination of
the renormalized magnetic field up to first order in $J$ is usually associated with a two-loop calculation,
see, e.g., Ref.~\onlinecite{garst_etal_PRB05}, i.e., the first diagram of Fig.~\ref{fig:RG_Liouvillian_G} is called
a two-loop diagram. We are not adopting this notation in the present paper because this diagram 
arises from closing the one-loop diagram for the renormalization of the vertex (the third diagram
in Fig.~\ref{fig:RG_Liouvillian_G}). Therefore we call this diagram also one-loop and the second diagram
of Fig.~\ref{fig:RG_Liouvillian_G} a two-loop diagram (which usually would be classified as three-loop). We
note that the classification in one-loop and two-loop contributions is not unambiguous, e.g., if one uses 
a not-normaled ordered version of the RG formalism, the renormalization of the magnetic field in lowest 
order in $J$ would arise from closing the renormalized vertex with itself, i.e., in this case one would 
classify it as one-loop. Therefore, we prefer in this paper a classification according to the orders in $J$
and not with respect to the loop topology of the diagrams.

Some insight into the logarithmic terms in $\tilde{\Gamma}_{1/2}$ and $\tilde{h}$
can be gained by differentiating these quantities with respect to
the bare magnetic field $h_0$ [for the rates $\tilde{\Gamma}_{1/2}$ and the
second-order terms in $\tilde{h}$, it is actually equivalent to
differentiate with respect to $\tilde{h}$ because multiplying with
$\frac{\partial \tilde{h}}{\partial   h_0}\sim O(J)$ would just bring the expression
one order higher in $J$, leading to terms which we have neglected anyhow]. Furthermore,
we differentiate only the prefactor of the logarithmic terms and disregard the dependence
of the renormalized exchange couplings on $h_0$ (which are small corrections since the
logarithm varies only slowly). This gives
\begin{widetext}
\begin{eqnarray}
\label{derivative_tilde_gamma_1_final}
{d\tilde{\Gamma}_1\over d h_0} &=& {\pi\over 2}\left(J^\perp_\alpha\right)^2 +
\pi\theta_2(\tilde{h}-V)\left(J^\perp_{\text{nd}}\right)^2 +
\pi\mylc{\tilde{h}}J^z_\alpha\left(J^\perp_\alpha\right)^2 +
\pi\theta_2(\tilde{h}-V)\mylc{V-\tilde{h}}J^z_\alpha\left(J^\perp_{\text{nd}}\right)^2 \quad,\\
\label{derivative_tilde_gamma_2_final}
{d\tilde{\Gamma}_2\over d h_0} &=& 
{\pi\over 4}\left(J^\perp_\alpha\right)^2 + 
{\pi\over 2}\theta_1(\tilde{h}-V)\left(J^\perp_{\text{nd}}\right)^2 
+ \frac\pi2\myla{\tilde{h}}J^z_\alpha\left(J^\perp_\alpha\right)^2 -
{\pi\over 2}\theta_1(V-\tilde{h})\myla{V-\tilde{h}}
J^z_\alpha\left(J^\perp_{\text{nd}}\right)^2 \,,\\
\label{derivative_tilde_h_final}
\tilde{g}=2{d\tilde{h}\over d h_0} &=& 2-\left(J^z_\alpha-(J^z_\alpha)_0)\right)
-\myla{\tilde{h}}\left(J^\perp_\alpha\right)^2 
-\myla{V-\tilde{h}}\left(J^\perp_{\text{nd}}\right)^2 \quad,
\end{eqnarray}
\end{widetext}
where we have used Eq.~(\ref{eq:3rd-order-Jz-Jp}) to take some terms together.

\begin{figure}[tbp]
  \includegraphics[width=\linewidth]{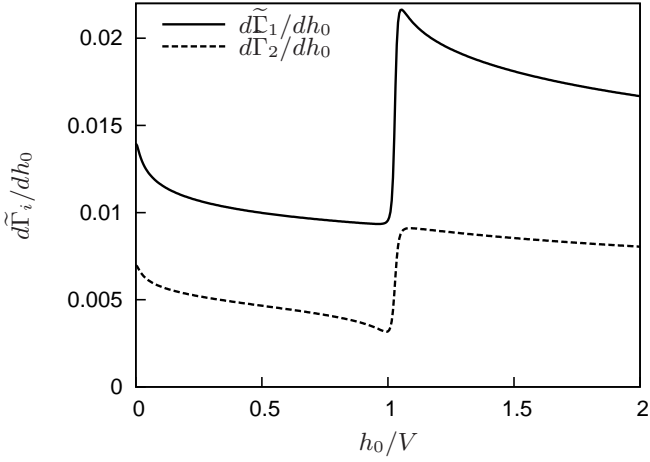}
  \caption{The relaxation and dephasing rates $\tilde\Gamma_1$ and
    $\tilde\Gamma_2$, derived with respect to the magnetic field $h_0$, for
      the isotropic Kondo model with $V=10^{-4}D$ and
      $T_K=10^{-8}D$. $\frac{\partial\tilde\Gamma_1}{\partial h_0}$ exhibits a
      logarithmic enhancement for $\tilde{h}>V$ whereas
      $\frac{\partial\tilde\Gamma_2}{\partial h_0}$ is suppressed for 
      $\tilde{h}<V$.}
    \label{fig:rates}  
\end{figure}
\begin{figure}[tbp]
  \includegraphics[width=\linewidth]{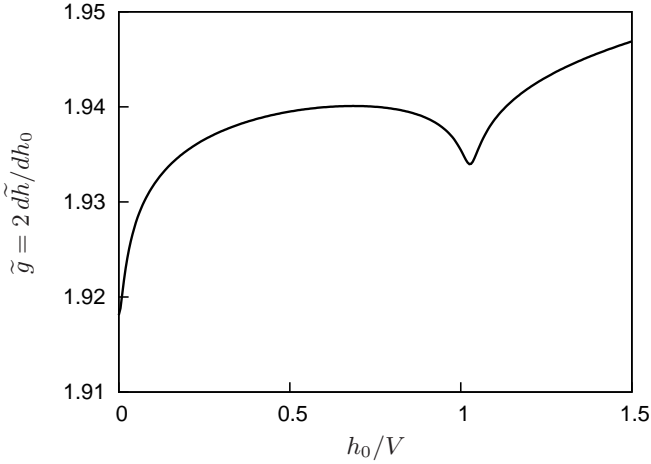}
  \caption{$g$~factor $\tilde g=2\,d\tilde h/d h_0$, derived with respect to
    the magnetic field $h_0$, for the isotropic Kondo model with
    $V=10^{-4}D$ and $T_K=10^{-8}D$.}
    \label{fig:gFactor}  
\end{figure}

For both rates $\tilde{\Gamma}_{1/2}$, we get a jump in
$\frac{\partial\tilde{\Gamma}_{1/2}}{\partial h_0}$ at $\tilde{h}=V$ in the leading 
order. In the next to leading order, there is a logarithmic
enhancement for $\tilde{h}\rightarrow0$ for both rates and a logarithmic
enhancement for $\tilde{h}>V$ in $\frac{\partial\tilde{\Gamma}_1}{\partial h_0}$ while
we find a logarithmic suppression in $\frac{\partial\tilde{\Gamma}_2}{\partial h_0}$ 
for $\tilde{h}<V$, see Fig.~\ref{fig:rates}.

\begin{figure}[htbp]
  \includegraphics[scale=0.5]{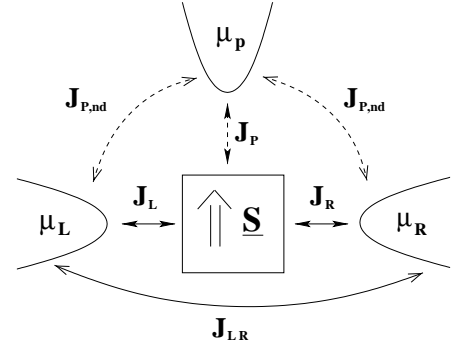}
  \caption{Three-terminal setup including a third weakly coupled probe lead,
with $\mu_P=eV_P$, in order to measure the renormalized $g$ factor as function of the
applied voltage $eV=\mu_L-\mu_R$.}
\label{fig:3_lead_configuration}
\end{figure}

The renormalized $g$~factor shows a logarithmic
suppression for $\tilde{h}\rightarrow0$ and another logarithmic suppression
for $\tilde{h}\approx V$ which is symmetric, in contrast to the effects we
observed for the rates, see Fig.~\ref{fig:gFactor}. As one can see, the second-order
logarithmic terms lead to significant changes of the first-order shift of the $g$ factor.
Especially at resonance $V=\tilde{h}$, where one usually reads off the renormalized
magnetic field from the conductance (see below), the change in the $g$ factor by the
logarithmic contributions is very important. However, fixing the voltage at $V=\tilde{h}$,
we obtain 
\begin{equation}
\label{g_resonance}
\tilde{g}|_{V=\tilde{h}}\,\approx\,2-\left[J_\alpha^z-(J_\alpha^z)_0\right]-(J^\perp_{\text{nd}})^2\,
\left(\ln{V\over\tilde{\Gamma}_1}\right)|_{V=\tilde{h}}\quad,
\end{equation}
with
\begin{equation}
\label{gamma_1_resonance}
\left(\tilde{\Gamma}_1\right)|_{V=\tilde{h}}\,\approx\,
{\pi\over 2}V\left[(J^\perp_\alpha)^2+2(J^\perp_{\text{nd}})^2\right]\quad.
\end{equation}
As a consequence, the $g$ factor at $V=\tilde{h}$ shows only a weak dependence on the bare magnetic
field $h_0$ via the renormalization of the exchange couplings. The second-order terms just lead to
an additional overall decrease $\sim J^2\ln{J}$. To see the suppression of the $g$ factor 
more clearly at resonance, it is necessary to measure it directly for various values of $h_0/V$.
In principle this can be achieved by electron spin resonance (ESR), but, similiar to previous measurements of the splitting
of the Kondo resonance in the spectral density,\cite{ensslin_kondo} we propose here a simpler
setup with a weakly coupled third lead, see Fig.~\ref{fig:3_lead_configuration}. We use the
form (\ref{J_form}) and, for simplicity, choose a symmetric coupling to the left and right leads,
\begin{eqnarray}
\label{x_probe}
&& x_L=x_R=x\quad,\quad x_P\ll 1\quad,\\
\label{J_probe}
&& J^{z/\perp}_{P,nd}=J^{z/\perp}_{PL}=J^{z/\perp}_{PR}\quad.
\end{eqnarray}

\begin{figure}[htbp]
  \includegraphics[width=\linewidth]{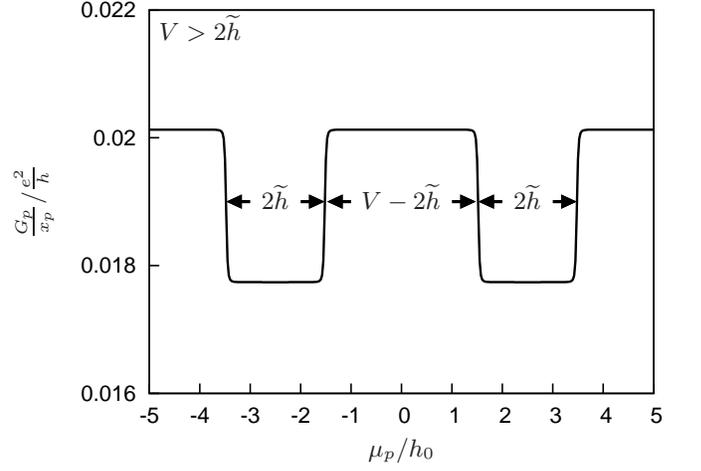}
  \caption{Probe-lead conductance in a three-terminal setup for
    $h_0=10^{-4}D$, $V=5\times10^{-4}D$, and $T_K=10^{-8}D$.}
\label{fig:3-Terminal-1}
\end{figure}

\begin{figure}[htbp]
  \includegraphics[width=\linewidth]{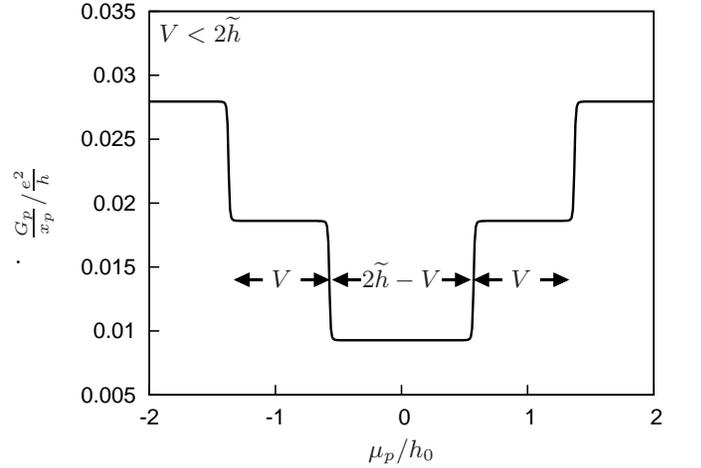}
  \caption{Probe-lead conductance in a three-terminal setup for
    $h_0=10^{-4}D$, $V=8\times10^{-5}D$, and $T_K=10^{-8}D$.}
\label{fig:3-Terminal-2}
\end{figure}

By changing the
voltage $V_P$ of the probe lead and measuring the differential conductance $G_P={dI_P\over dV_P}$,
one can probe the renormalized $g$ factor as function of the voltage $V=\mu_L-\mu_R$ between the
strongly coupled left and right leads. The reason is that the contribution of inelastic cotunneling
to the probe current changes when the probe voltage crosses the resonance points 
$V_P={V\over 2}\pm\tilde{h},-{V\over 2}\pm\tilde{h}$.
The probe current up to second order in $J_{P,nd}$ can be calculated from
Eqs.~\eqref{current_stationary_kondo},~\eqref{gamma_current_b_2}, and~\eqref{re_gamma_current_1z_2} with the
result
\begin{equation}
\label{I_probe}
\langle I^P \rangle^{st}\,=\,\Gamma^{b(2)}_P(0)\,+\,2M\,\Gamma^{1z(2)}_P(0) \quad,
\end{equation}
where
\begin{eqnarray}
\label{Gamma_b_probe}
\Gamma^{b(2)}_P(0)~&=&~{\pi\over 4}\,V_P\,\left[\left(J^z_{P,nd}\right)^2+
2\left(J^\perp_{P,nd}\right)^2\right]\quad,\\
\nonumber
\Gamma^{1z(2)}_P(0)~&=&~-{\pi\over 4}\,
\left(|V_P-{V\over 2}-\tilde{h}|_2+|V_P+{V\over 2}-\tilde{h}|_2-\right.\\
\label{Gamma_1z_probe}
&&\hspace{-2cm}
\left.-|V_P-{V\over 2}+\tilde{h}|_2-|V_P+{V\over 2}+\tilde{h}|_2\right)
\,\left(J^\perp_{P,nd}\right)^2\quad.
\end{eqnarray}
Here, $M$ is the magnetization, which in lowest order is independent of the probe lead
and is calculated in the next section. Taking the derivative with respect to $V_P$, we
find that ${d\Gamma^{b(2)}_P(0)\over dV_P}$ is independent of $V_P$,
whereas ${d\Gamma^{1z(2)}_P(0)\over dV_P}$ gives rise to a steplike structure shown in 
Figs.~\ref{fig:3-Terminal-1} and \ref{fig:3-Terminal-2} for the two regimes
$V>2\tilde{h}$ and $V<2\tilde{h}$, respectively. Sufficiently far away from resonance, we obtain
analytically from Eq.~\eqref{Gamma_1z_probe} for $V>2\tilde{h}$
\begin{equation}
\label{Gamma_1z_probe_V>2h}
{d\Gamma^{1z(2)}\over dV_P}\,=\,{\pi\over 2}\,\left(J^\perp_{P,nd}\right)^2\,
\left\{
\begin{array}{cl}
1, &\mbox{for }{V\over 2}+\tilde{h}>V_P>{V\over 2}-\tilde{h}\,\\
&\mbox{or}
-{V\over 2}+\tilde{h}>V_P>-{V\over 2}-\tilde{h} \\
0, &\mbox{otherwise,}
\end{array}
\right.
\end{equation}
and for $V<2\tilde{h}$
\begin{equation}
\label{Gamma_1z_probe_V<2h}
{d\Gamma^{1z(2)}\over dV_P}\,=\,{\pi\over 2}\,\left(J^\perp_{P,nd}\right)^2\,
\left\{
\begin{array}{cl}
2, &\mbox{for }-{V\over 2}+\tilde{h}>V_P>{V\over 2}-\tilde{h}\,\\
1, &\mbox{for }{V\over 2}+\tilde{h}>V_P>-{V\over 2}+\tilde{h} \\
&\mbox{or  }\,
{V\over 2}-\tilde{h}>V_P>-{V\over 2}-\tilde{h} \\
0, &\mbox{otherwise.}
\end{array}
\right.\quad.
\end{equation}
From the width of the steplike features one can read off the renormalized
magnetic field $\tilde{h}$ at given voltage $V$ and magnetic field $h_0$. Thus, by varying $V$
or $h_0$, one can measure the renormalized $g$ factor as function of $h_0/V$,
thereby revealing the suppression at resonance $V=\tilde{h}$ as shown
in Fig.~\ref{fig:gFactor}. We note
that the experimental situation is usually in the strong coupling
regime $J\sim O(1)$, where the qualitative features are expected to be more
pronounced.

\subsection{Magnetization and susceptibility}
\label{sec:magnetization}

The stationary magnetization (which is nonzero only if a magnetic field
$h_0\neq0$ is applied) can be calculated from the stationary density
matrix (\ref{rd_stationary_kondo}). This density matrix is the solution of the kinetic equation
and is a diagonal matrix. The occupation probabilities for the states
$|\uparrow\rangle$ and $\downarrow\rangle$ are
\begin{equation}
  \label{occupation_probabilities}
  p_\uparrow=\frac{\Gamma^a(0)-\Gamma^{3z}(0)}{2\Gamma^a(0)}\quad,\quad
  p_\downarrow=\frac{\Gamma^a(0)+\Gamma^{3z}(0)}{2\Gamma^a(0)}\quad,
\end{equation}
and the magnetization is given by
\begin{equation}
  \label{eq:Magnetization}
  M=\frac12\left(p_\uparrow-p_\downarrow\right)=-\frac{\Gamma^{3z}(0)}{2\Gamma^a(0)}\quad.
\end{equation}
$\Gamma^a(0)=\tilde{\Gamma}_1$ is given by Eq.~(\ref{tilde_gamma_1_final}). $\Gamma^{3z}(0)$ 
can be calculated from Eqs.~(\ref{gamma_3z_2}) and (\ref{gamma_3z_3a}):
\begin{eqnarray}
\nonumber
\Gamma^{3z}(0) &=& 
{\pi\over 2}\tilde{h}\left\{\left(J^\perp_\alpha\right)^2 +2\left(J^\perp_{\text{nd}}\right)^2\right\}\\
\label{gamma_3z_final}
&&\hspace{-1cm}
+ \pi\tilde{h}\mylc{\tilde{h}}\left\{J^z_\alpha\left(J^\perp_\alpha\right)^2+ 
J^\perp_\alpha J^z_{\text{nd}} J^\perp_{\text{nd}}\right\}\\
\nonumber
&&\hspace{-1cm}
-{\pi\over 2}(V-\tilde{h})\mylc{V-\tilde{h}}\left\{J^z_\alpha\left(J^\perp_{\text{nd}}\right)^2+ 
J^\perp_\alpha J^z_{\text{nd}} J^\perp_{\text{nd}}\right\}\,,
\end{eqnarray}
where we have replaced $h_0\rightarrow\tilde{h}$ in the first term, which gives only rise
to negligible terms of $O(J^3,J^4\ln)$. Because we have considered all terms
of $O(J^2,J^3\ln)$ for the rates, we can calculate the leading order $O(J^0)$ (which is 
independent of the couplings) and logarithmically enhanced terms in $O(J\ln)$ of the
magnetization.

Using Eq.~(\ref{eq:3rd-order-Jz-Jp}) and the abbreviation
\begin{eqnarray}
\nonumber
X &=& \tilde{h}(J^\perp_\alpha)^2+2\tilde{h}(J^\perp_{\text{nd}})^2
+2\tilde{h}\mylc{\tilde{h}}J^z_\alpha(J^\perp_\alpha)^2\\
\label{X_abbr}
&&-2(V-\tilde{h})\mylc{V-\tilde{h}}J^z_\alpha(J^\perp_{\text{nd}})^2\,,
\end{eqnarray}
we can write the magnetization in the compact form
\begin{eqnarray}
\label{magnetization_compact}
M &=&-{1\over 2}\cdot\\
\nonumber
&&\hspace{-1cm}
\cdot{X+2\tilde{h}\mylc{\tilde{h}}J^z_\alpha(J^\perp_{\text{nd}})^2 \over
X+({|V-\tilde{h}|}_2+V-\tilde{h})(1+\mylc{V-\tilde{h}}J^z_\alpha)
(J^\perp_{\text{nd}})^2}\,.
\end{eqnarray}
We note that, in the isotropic case, and when expanded systematically in $J$,
the terms of $O(J^0,J\ln)$ are in agreement with the perturbative
results obtained in Ref.~\onlinecite{paaske_rosch_woelfle_PRB04} and the poor man
scaling results of Ref.~\onlinecite{rosch_paaske_kroha_woelfle_PRL03}
if the different definition of the magnetization is taken into account 
(the magnetization in those references is $M=1$ in the
ground state, whereas it is $M=-1/2$ here). However, we have shown here 
how the spin dephasing rate cuts off all the logarithmic divergencies (the 
treatment in Ref.~\onlinecite{paaske_rosch_kroha_woelfle_PRB04} was only perturbative
in the bare coupling and did not treat the case of finite magnetic field).
 
If the voltage is smaller than the renormalized magnetic field $\tilde{h}$,
the logarithmic term $\sim \mylc{\tilde{h}}$ in the numerator of Eq.~(\ref{magnetization_compact})
must be consistently neglected and
the magnetization is equal to $-1/2$. This is because the energy difference
provided by the voltage is insufficient to flip the spin $1/2$ in the
dot out of its ground state $|\downarrow\rangle$.

In the limit $V\gg \tilde{h}$, the logarithmic term $\sim \mylc{V-\tilde{h}}$
in the denominator of Eq.~(\ref{magnetization_compact}) must be neglected and the magnetization 
is given by
\begin{equation}
\label{magnetization_V>>h}
M=-{1\over 2}\,\,{X+2\tilde{h}\mylc{\tilde{h}}J^z_\alpha(J^\perp_{\text{nd}})^2 \over
X+2V(J^\perp_{\text{nd}})^2}\,,
\end{equation}
which, for $J^{z/\perp}_{\text{nd}}\sim J^{z/\perp}_{\alpha}$, can be expanded as
\begin{eqnarray}
\nonumber
M&=&-{\tilde{h}\over 2V}\left\{1+{1\over 2}\left({J^\perp_\alpha\over J^\perp_{\text{nd}}}\right)^2+\right.\\
\label{magnetization_V>>h_expanded}
&&\left.+\mylc{\tilde{h}}J^z_\alpha\left(1+\left({J^\perp_\alpha\over J^\perp_{\text{nd}}}\right)^2\right)
\right\}\,.
\end{eqnarray}
As we see, there is an interesting logarithmic enhancement at $\tilde{h}=0$, which is a pure 
nonequilibrium effect [if we set $J^\perp_{\text{nd}}=0$ in Eq.~(\ref{magnetization_V>>h}), we
get $M=-{1\over 2}$].

In the limit $V>\tilde{h}$, $V-\tilde{h}\ll \tilde{h}$, (i.e., for voltages which are
slightly larger than the renormalized magnetic field), we can neglect all logarithmic
terms $\sim \mylc{\tilde{h}}$ and find after expanding in $J$
\begin{eqnarray}
\nonumber
M&\approx& -\frac12 + \frac{V-\tilde{h}}{\tilde{h}}\frac{\left(J^\perp\nd\right)^2}
{\left(J^\perp\iL+J^\perp\iR\right)^2}\cdot\\
\label{magnetization_at_resonance}
&&\cdot\left[1+(J^z\iL+J^z\iR)\mylc{V-\tilde{h}}\right]\,.
\end{eqnarray}
This result shows the logarithmic enhancement of the magnetization slightly above the
resonance.

\begin{figure}[htbp]
  \includegraphics[width=\linewidth]{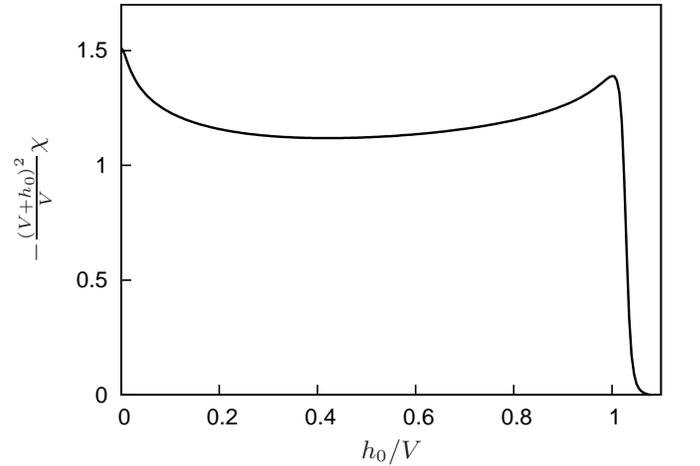}
  \caption{The susceptibility $\chi$, multiplied with
    $-\frac{(V+h_0)^2}V$, as function of the magnetic field $h_0$ for
      the isotropic Kondo model with $V=10^{-4}D$ and $T_K=10^{-8}D$.}
\label{fig:susceptibility}  
\end{figure}

From the magnetization, we can also calculate the susceptibility
\begin{equation}
  \label{susceptibility}
  \chi=\frac{\partial M}{\partial h_0}\approx\frac{\partial M}{\partial \tilde{h}}\quad,
\end{equation}
which we calculate by differentiating the
magnetization~\eqref{eq:Magnetization} with respect to the bare magnetic
field $h_0$ [or the renormalized field $\tilde{h}$ by multiplying with 
${d\tilde{h}\over d h_0}=1+O(J,J^2\ln)$]. 
In the special case of symmetric couplings
($J^\chi\iL=J^\chi\iR=J^\chi\nd$ for $\chi=z,\perp$), the
magnetization for $V>\tilde{h}$ can be expanded as
\begin{eqnarray}
\nonumber
M&=&-\frac{\tilde{h}}{V+\tilde{h}}
+\frac{V-\tilde{h}}{V+\tilde{h}}\mylc{V-\tilde{h}}J^z\\
\label{magnetization_symmetric_case}
&&-\frac{2V\tilde{h}}{(V+\tilde{h})^2}\mylc{\tilde{h}}J^z\quad,
\end{eqnarray}
and the susceptibility becomes
\begin{eqnarray}
\nonumber
\chi&=&-\frac{V}{(V+\tilde{h})^2}
-\frac{2V}{(V+\tilde{h})^2}\mylc{V-\tilde{h}}J^z\\
\label{chi_symmetric_case}
&&-\frac{2V(V-\tilde{h})}{(V+\tilde{h})^3}\mylc{\tilde{h}}J^z\quad.
\end{eqnarray}
To visualize the logarithmic effects better, we multiply $\chi$ with
$-(V+\tilde{h})^2/V$ and get
\begin{equation}
  -\frac{(V+\tilde{h})^2}{V}\chi=1
  +2J^z\mylc{V-\tilde{h}}+2\frac{V-\tilde{h}}{V+\tilde{h}}J^z\mylc{h}.
  \label{eq:SusceptibilityPlot}
\end{equation}
Plotting Eq.~\eqref{eq:SusceptibilityPlot} as function of the magnetic
field $h_0$ emphasizes the logarithmic enhancements at $\tilde{h}\ll V$ and
$\tilde{h}\approx V$, see Fig.~\ref{fig:susceptibility}.
In this plot, we multiply the susceptibility $\chi$ not with
$-\frac{(V+\tilde{h})^2}V$, but with $-\frac{(V+h_0)^2}V$ because this change
causes only subleading corrections and because the unrenormalized
magnetic field $h_0$ is easier to access experimentally. 

In summary, from the various expansions Eqs.~(\ref{magnetization_V>>h_expanded}),
(\ref{magnetization_at_resonance}), and (\ref{magnetization_symmetric_case}), we find 
that, for $V>\tilde{h}$, the logarithmic terms at $\tilde{h}=0$ and $\tilde{h}=V$ are 
essentially proportional to the longitudinal exchange coupling $J^z$. This means that even in
the limit $J^\perp\rightarrow 0$, these logarithmic contributions survive, which is
an interesting nonequilibrium effect.

\subsection{Current}
\label{sec:current}

The stationary current can be calculated from Eq.~(\ref{current_stationary_kondo}),
\begin{equation}
\label{current_stationary_kondo_final}
\langle I^\gamma\rangle^{st}\,=\,\Gamma^b_\gamma(0)\,+\,2M\,\Gamma^{1z}_\gamma(0) \quad,
\end{equation}
where the stationary magnetization $M$ follows from Eq.~\eqref{eq:Magnetization} and the components
$\Gamma_\gamma^b(0)$ and $\Gamma_\gamma^{1z}(0)$ of the current kernel 
are listed in Eqs.~\eqref{gamma_current_b_2},~\eqref{re_gamma_current_1z_2},
~\eqref{gamma_current_b_3a}, and~\eqref{gamma_current_1z_3b}. For the special case 
(\ref{two_reservoirs}) of two reservoirs and the relation (\ref{exchange_couplings_special})
between the exchange couplings, we obtain 
\begin{eqnarray}
\nonumber
&&\hspace{-1cm}
\Gamma^b_\gamma(0) = {\pi\over 4}\gamma V\left\{
\left(J^z_{\text{nd}}\right)^2+2\left(J^\perp_{\text{nd}}\right)^2\right\}\\
\nonumber
&&\hspace{0.5cm}
 +{\pi\over 2}\gamma V\myla{V}J^z_{\text{nd}}J^\perp_\alpha J^\perp_{\text{nd}}\\
\label{gamma_current_b_final}
&&\hspace{-0.5cm}
+{\pi\over 4}\gamma (V-\tilde{h})\mylc{V-\tilde{h}}J^\perp_{\text{nd}}
\left(J^z_\alpha J^\perp_{\text{nd}}+J^\perp_\alpha J^z_{\text{nd}}\right),\\
\nonumber
&&\hspace{-1cm}
\Gamma^{1z}_\gamma(0) = -{\pi\over 4}\gamma \left({|V-\tilde{h}|}_2-V-\tilde{h}\right)
\left(J^\perp_{\text{nd}}\right)^2\\
\nonumber
&&\hspace{0.5cm}
+{\pi\over 2}\gamma \left(V\myla{V}+\tilde{h}\mylc{\tilde{h}}\right)
J^z_{\text{nd}}J^\perp_\alpha J^\perp_{\text{nd}}\\
\label{gamma_current_1z_final}
&&\hspace{0.5cm}
-{\pi\over 4}\gamma {|V-\tilde{h}|}_2\mylc{V-\tilde{h}}
J^z_\alpha \left(J^\perp_{\text{nd}}\right)^2\,,
\end{eqnarray}
where we have used
$c^\gamma_{\alpha\alpha'}=-{1\over 2}\alpha\gamma\delta_{\alpha,-\alpha'}$.

For $V\ll\tilde{h}$, the logarithmic terms $\sim\mylc{\tilde{h}},\mylc{V-\tilde{h}}$ can
be neglected and the magnetization is given by $M=-{1\over 2}$. According to 
Eq.~(\ref{current_stationary_kondo_final}), this leads to 
$\langle I^\gamma\rangle^{st}=\Gamma^b_\gamma(0)-\Gamma^{1z}_\gamma(0)$, and we see
that the logarithmic terms $\sim\myla{V}$ cancel. This gives the result of elastic 
cotunneling
\begin{equation}
\label{elastic_cotunneling}
\langle I^\gamma\rangle^{st}={\pi\over 4}\gamma V\left(J^z_{\text{nd}}\right)^2\quad,
\end{equation}
but with renormalized exchange couplings.

\begin{figure}[htbp]
  \includegraphics[width=\linewidth]{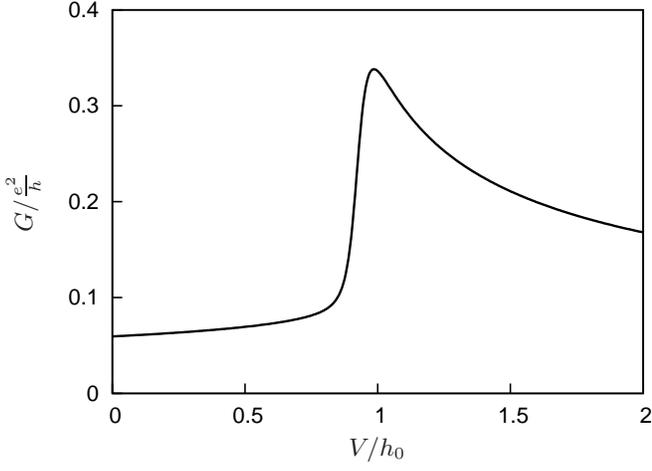}
  \caption{The differential conductance $G=\frac{dI}{dV}$ as function
    of bias voltage at $h_0=10^{-4}D$ for the isotropic Kondo model
    with $T_K=10^{-6}D$.}
\label{fig:conductance}  
\end{figure}

At $V\approx\tilde{h}$, the differential conductance $G_\gamma=\frac{dI_\gamma}{dV}$ has an
interesting feature, see Fig.~\ref{fig:conductance}. For voltages slightly below the
renormalized magnetic field ($V<\tilde{h}$, $\tilde{h}-V\ll \tilde{h}$), we get in units of 
$G_0=e^2/h$ (as before, we list the leading order and logarithmic terms in next-to-leading order)
\begin{equation}
  G_\gamma/G_0=\gamma\tfrac{\pi^2}2\left[
    \left(J^z\nd\right)^2
    +\left(J^z\iL+J^z\iR\right)
    \left(J^\perp\nd\right)^2\mylc{V-\tilde{h}}\right].
  \label{inelastic_cotunneling_1}
\end{equation}
For voltages slightly above the renormalized magnetic field ($V>\tilde{h}$,
$V-\tilde{h}\ll \tilde{h}$), we get
\begin{multline}
  G_\gamma/G_0=\gamma\tfrac{\pi^2}2\Bigg\{
  \left(J^z\nd\right)^2
  +\left(J^\perp\nd\right)^2
  \left[2+\left(\frac{2J^\perp\nd}{J^\perp\iL+J^\perp\iR}\right)^2\right]\\
  +\left(J^z\iL+J^z\iR\right)\left(J^\perp\nd\right)^2
  \left[3+\left(\frac{2J^\perp\nd}{J^\perp\iL+J^\perp\iR}\right)^2\right]
  \mylc{V-\tilde{h}}
  \Bigg\}.
  \label{inelastic_cotunneling_2}
\end{multline}
Two interesting features happen at $V\approx\tilde{h}$: There is a jump in
the leading order term in the conductance (if $J^\perp\nd\neq0$) that
is due to inelastic cotunneling which sets in at this
voltage. The jump is superposed by a logarithmic term which
becomes largest for $V=\tilde{h}$. The experimentally accessible parameters
characterizing the line shape are given by the position,
broadening and height of the resonance. The position is approximately at 
$V=\tilde{h}$ [up to unimportant terms of the $O(\tilde{\Gamma}_2)$], and the
broadening of the left side of the resonance is given by $\tilde{\Gamma}_2$. 
Using Eqs.~\eqref{tilde_h_final} and \eqref{tilde_gamma_2_final}, we obtain 
at resonance
\begin{eqnarray}
\label{h_tilde_resonance}
\hspace{-0.5cm}
\tilde{h}|_{V=\tilde{h}}\,&\approx&\,h \quad,       \\
\label{tilde_gamma_2_resonance}
\hspace{-0.5cm}
\tilde{\Gamma}_2|_{V=\tilde{h}}\,&\approx&\,
{\pi\over 2}V\left\{\left(J^z\nd\right)^2+\left(J^\perp\nd\right)^2+
{1\over 2}\left(J^\perp_\alpha\right)^2\right\}.
\end{eqnarray}
The right side of the resonance has no
characteristic broadening: asymptotically it reveals the logarithmic voltage dependence
of the exchange couplings. The value of the conductance at resonance can be
calculated in good approximation by taking the extrapolation of 
Eq.~(\ref{inelastic_cotunneling_2}) at $V=\tilde{h}$, i.e., by replacing
$\mylc{V-\tilde{h}}\rightarrow \ln{V\over{\tilde{\Gamma}_2}|_{V=\tilde{h}}}$ in
Eq.~(\ref{inelastic_cotunneling_2}).
Together with Eq.~\eqref{tilde_gamma_2_resonance}, this reveals the Kondo-induced logarithmic 
enhancement $\sim J^3\ln{J}$ of the conductance at resonance superposed on the enhancement 
from inelastic cotunneling. The result for the maximal conductance can be simplified by 
using the relation (\ref{exchange_couplings_special}), which gives
\begin{eqnarray}
\nonumber
(G_\gamma/G_0)_{max}\,&\approx&\,\gamma\, 2\pi^2\,x_L x_R \\
\nonumber
&&\hspace{-2cm}
\cdot\left\{
\left(J^z\right)^2+2(1+2x_L x_R)\left(J^\perp\right)^2+\right.\\
\label{G_max_simplified}
&&\hspace{-1.5cm}
\left.+2(3+4x_L x_R)J^z(J^\perp)^2\ln{V\over{\tilde{\Gamma}_2}|_{V=\tilde{h}}}\right\},
\end{eqnarray}
with
\begin{equation}
\label{tilde_gamma_2_resonance_simplified}
\tilde{\Gamma}_2|_{V=\tilde{h}}\,=\,\pi V\left\{2x_L x_R\left(J^z\right)^2+\left(J^\perp\right)^2\right\}.
\end{equation}
This result for the value of the maximum conductance at 
resonance has first been noted in Ref.~\onlinecite{glazman_pustilnik_05} for the case
of the isotropic Kondo model and $x_L=x_R={1\over 2}$.

Our results for the differential conductance agree with those of 
Refs.~\onlinecite{rosch_paaske_kroha_woelfle_PRL03} and~\onlinecite{glazman_pustilnik_05}.
However, note that the precise line shape at resonance can only be obtained if one
includes the cutoff scales $\tilde{\Gamma}_{1/2}$ into the logarithms and uses the correct
smearing (\ref{sign_replacement}) of the sign function. In the numerical plots, we have
always used these smeared forms and we calculated $\tilde{h}$ and $\tilde{\Gamma}_{1/2}$
self-consistenly from Eqs.~(\ref{tilde_gamma_1_final})--(\ref{tilde_h_final}) (the same
procedure has been used for the susceptibility). 

\subsection{Results for the anisotropic Kondo model}
\label{sec:kondo_results_anisotropic}
In this section, we discuss how the results for the differential
conductance, the susceptibility, the relaxation and dephasing rates,
and the renormalized $g$~factor are affected if we consider the
symmetric, but anisotropic Kondo model with
$J^z_{\alpha\alpha'}=J^z\neq J^\perp=J^\perp_{\alpha\alpha'}$. As
recently proposed,\cite{SMM_theory} such models can, e.g., be experimentally 
realized by studying quantum transport through single molecular magnets,
where the transverse exchange coupling is induced by magnetic quantum
tunneling terms due to transverse anisotropies.
To investigate this, we vary $J^z$ and $J^\perp$ such that the Kondo
temperature $T_K$ remains constant and only the other invariant of
the RG equations, i.e., $c^2=(J^z)^2-(J^\perp)^2$ is changed,
see~\eqref{invariant_c_TK}. If $c^2$ is positive, $J^z$ is larger and
$J^\perp$ is smaller than the isotropic coupling in the case $c^2=0$.

\begin{figure}[htbp]
  \includegraphics[width=\linewidth]{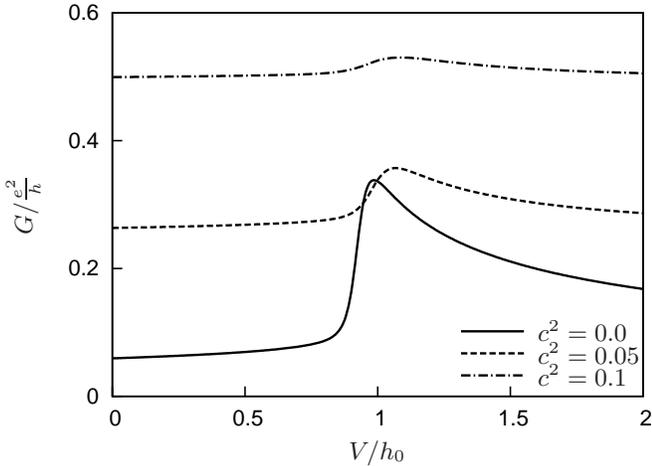}
  \caption{The differential conductance $G=\frac{dI}{dV}$ as function
    of bias voltage at $h_0=10^{-4}D$ for the isotropic Kondo model
    (solid line) and the anisotropic Kondo model with two different
    values of $c^2=(J^z)^2-(J^\perp)^2$ (dashed and dash-dotted
    lines). The Kondo temperature $T_K=10^{-6}D$ is the same in all
    cases.}
\label{fig:anisotropic-conductance}  
\end{figure}
The differential conductance is shown for three different values of
$c^2$ in Fig.~\ref{fig:anisotropic-conductance}. For growing $c^2$,
i.e., increasing $J^z$ and decreasing $J^\perp$, we can observe a
number of different effects compared to the isotropic case: First, the
conductance for $V\ll\tilde{h}$ which is proportional to
$(J^z\nd)^2$ according to Eq.~\eqref{elastic_cotunneling} is increased.
Second, the step at $V=\tilde{h}$ that is due to inelastic
cotunneling decreases its height which is proportional to
$(J^\perp\nd)^2$ in leading order according
to Eqs.~\eqref{inelastic_cotunneling_1}
and~\eqref{inelastic_cotunneling_2}. Third, also the logarithmic
enhancement at $V=\tilde{h}$ is reduced for increasing $c^2$.  The
prefactors of the logarithmic terms in Eqs.~\eqref{inelastic_cotunneling_1}
and~\eqref{inelastic_cotunneling_2} are proportional to
$J^z(J^\perp)^2$, a quantity which decreases for increasing $c^2$.
Finally, the position of the step and the logarithmic enhancement
shifts to the right, i.e., $\tilde{h}$ is increased. The reason is
that the leading contribution to $\tilde{h}-h_0$ is negative and
proportional to $J^z_\alpha-(J^z_\alpha)_0$, i.e., the renormalization
of the coupling $J^z$, according to Eq.~\eqref{renormalized_h}. This
renormalization is in leading order determined by $(J^\perp\nd)^2$,
see Eq.~\eqref{poor_man_scaling_kondo}. An increase in $c^2$ which is
connected to a decrease in $J^\perp$ therefore leads to a weaker
renormalization of $J^z$ and hence to a larger value of $\tilde
h$.

\begin{figure}[htbp]
  \includegraphics[width=\linewidth]{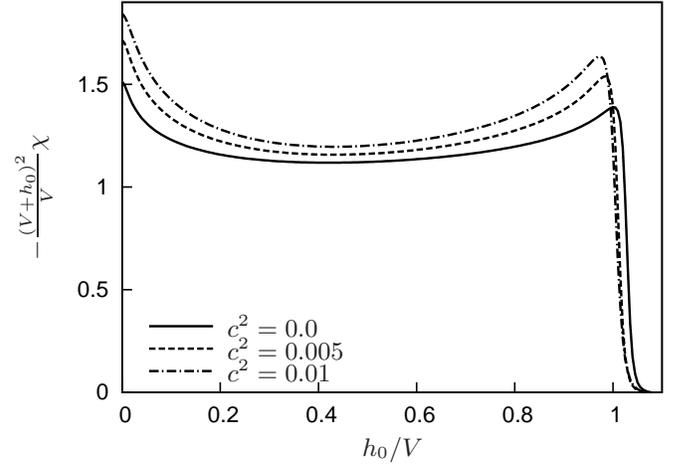}
  \caption{The susceptibility $\chi$, multiplied with $-\frac{(V+h_0)^2 }
    V$, as function of the magnetic field at $V=10^{-4}D$ for the
    isotropic Kondo model (solid line) and the anisotropic Kondo model
    with two different values of $c^2=(J^z)^2-(J^\perp)^2$ (dashed and
    dash-dotted lines). The Kondo temperature $T_K=10^{-8}D$ is the same
    in all cases.}
\label{fig:anisotropic-susceptibility}  
\end{figure}

Figure~\ref{fig:anisotropic-susceptibility} shows the magnetic
susceptibility $\chi=\frac{dM}{dh_0}$, multiplied with
$-\frac{(V+h_0)^2}V$, for different values of the anisotropy.
For larger $c^2$, i.e., increased $J^z$, the logarithmic enhancements
become larger, see Eq.~\eqref{eq:SusceptibilityPlot}. This is an effect
already mentioned at the end of Sec.~\ref{sec:magnetization}. The susceptibility
depends only on the longitudinal coupling $J^z$ and the logarithmic 
contributions survive even in the limit $J^\perp\rightarrow 0$.
Especially for molecular magnets with a very small quantum tunneling term, this
might be an interesting possibility to observe logarithmic contributions in
the susceptibility although the Kondo temperature might be quite small.
Furthermore, the resonance at the point where $\tilde{h}=V$ shifts slightly, this is
due to the dependence of $\tilde{h}$ on the anisotropy which was
discussed above.

\begin{figure}[htbp]
  \includegraphics[width=\linewidth]{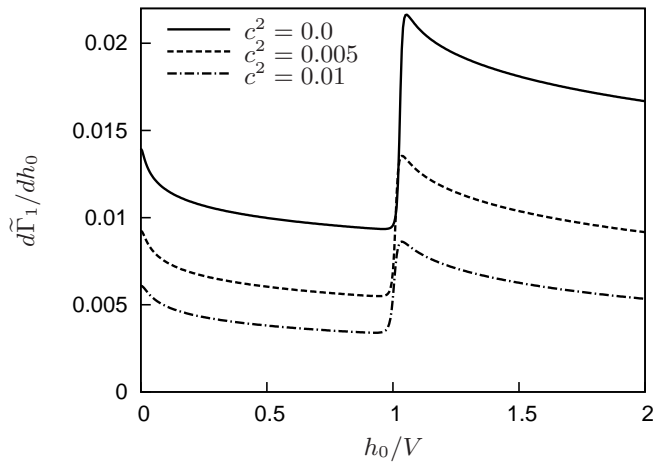}
  \caption{The rate $\tilde\Gamma_1$, derived with respect to the
    magnetic field, as function of the magnetic field at $V=10^{-4}D$
    for the isotropic Kondo model (solid line) and the anisotropic
    Kondo model with two different values of $c^2=(J^z)^2-(J^\perp)^2$
    (dashed and dash-dotted lines). The Kondo temperature $T_K=10^{-8}D$
    is the same in all cases.}
\label{fig:anisotropic-Gamma1}  
\end{figure}

\begin{figure}[htbp]
  \includegraphics[width=\linewidth]{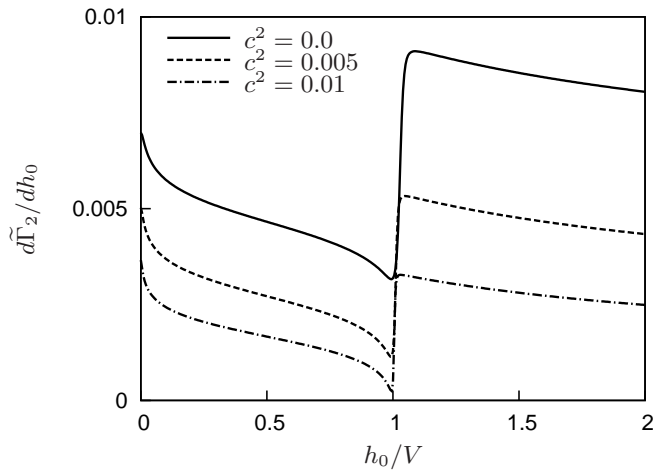}
  \caption{The rate $\tilde\Gamma_2$, derived with respect to the
    magnetic field, as function of the magnetic field at $V=10^{-4}D$
    for the isotropic Kondo model (solid line) and the anisotropic
    Kondo model with two different values of $c^2=(J^z)^2-(J^\perp)^2$
    (dashed and dash-dotted lines). The Kondo temperature $T_K=10^{-8}D$
    is the same in all cases.}
\label{fig:anisotropic-Gamma2}  
\end{figure}

The derivatives of the rates $\tilde\Gamma_1$ and
$\tilde\Gamma_2$ with respect to the magnetic field $h_0$ are
shown in Figs.~\ref{fig:anisotropic-Gamma1} and~\ref{fig:anisotropic-Gamma2}, respectively.
Increasing $c^2$, i.e., decreasing $J^\perp$, leads to a decrease in
the leading-order contributions to the derivatives of the rates which
are proportional to $(J^\perp\nd)^2$, see
Eqs.~\eqref{derivative_tilde_gamma_1_final}
and~\eqref{derivative_tilde_gamma_2_final}.
The dependence of the logarithmic effects at $\tilde{h}=0$ and
$\tilde{h}=V$ on the anisotropy is different for the derivatives of
the two different rates: The features in $d\tilde\Gamma_1/dh_0$
become less pronounced whereas the resonances in
$d\tilde\Gamma_2/dh_0$ become sharper. The reason is as follows:
all logarithmic terms in next to leading order in the rates are
proportional to $J^z(J^\perp\nd)^2$ which decreases for larger
anisotropy, but the logarithms are broadened by $\tilde\Gamma_2$
in the case of $d\tilde\Gamma_1/dh_0$ [see
Eq.~\eqref{derivative_tilde_gamma_1_final}] and by
$\tilde\Gamma_1$ in the case of $d\tilde\Gamma_2/dh_0$ [see
Eq.~\eqref{derivative_tilde_gamma_2_final}]. The fact that
$\tilde\Gamma_1$, which is proportional to $(J^\perp)^2$ in
leading order according to Eq.~\eqref{tilde_gamma_1_final}, decreases for
larger $c^2$, and $\tilde\Gamma_2$, which contains a term
$\propto(J^z)^2$, see Eq.~\eqref{tilde_gamma_2_final}, increases with
$c^2$, explains the observed behavior.

\begin{figure}[htbp]
  \includegraphics[width=\linewidth]{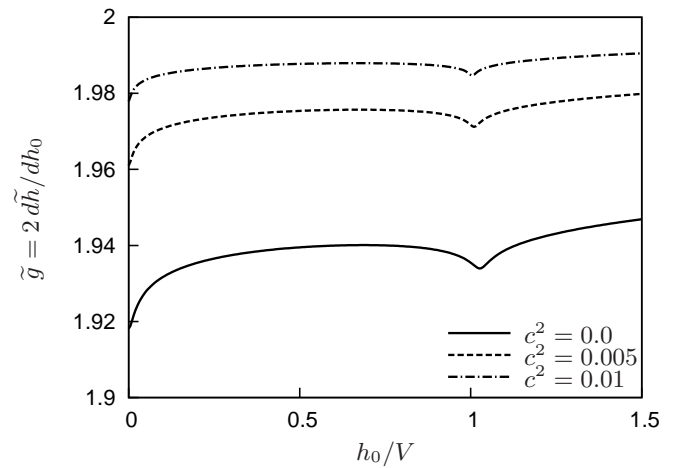}
  \caption{The renormalized $g$~factor $\tilde g=2\,d\tilde{h}/dh_0$ as
    function of the magnetic field at $V=10^{-4}D$ for the isotropic
    Kondo model (solid line) and the anisotropic Kondo model with two
    different values of $c^2=(J^z)^2-(J^\perp)^2$ (dashed and
    dash-dotted lines). The Kondo temperature $T_K=10^{-8}D$ is the same
    in all cases.}
\label{fig:anisotropic-g-factor}  
\end{figure}

Finally, we discuss how the renormalized $g$~factor is affected by the
anisotropy, see Fig.~\ref{fig:anisotropic-g-factor}.
For growing $c^2$, the deviation of $\tilde{g}$ from the unrenormalized value
2 becomes smaller. The reason has already been discussed above: Like
the renormalized field $\tilde{h}$, the
$g$~factor~\eqref{derivative_tilde_h_final} is in first order given by
the renormalization of the coupling $J^z$ which is determined by
$(J^\perp)^2$ and is therefore decreased for positive $c^2$. Also the
prefactors of the logarithmic terms depend on $(J^\perp)^2$ and
therefore become smaller for growing $c^2$.

\section{Summary}
\label{sec:conclusion}

We have presented a systematic and analytic weak-coupling analysis in two-loop
for a generic quantum dot coupled to reservoirs via spin or orbital fluctuations.
We have used a nonequilibrium RG formalism in Liouville space (or, equivalently, on the 
Keldysh contour), which is a natural and formally exact generalization of conventional 
poor man scaling methods. The essential difference between the RTRG-FS method 
and other RG formalisms in nonequilibrium 
\cite{rosch_paaske_kroha_woelfle_PRL03,glazman_pustilnik_05,kehrein_PRL05} is
the fact that RTRG-FS analyzes the dynamics of the reduced dot density matrix 
in Laplace space by calculating an effective dot Liouvillian. In each step of the
RG approach an effective kinetic equation is obtained. The
RG equations are coupled differential equations for the kernel determining this 
kinetic equation and for the vertices in Liouville space. The structure in 
Liouville space is essential for the generation of dissipation and to obtain
the {\it transport} relaxation and dephasing rates together with the {\it transport}
oscillation frequencies of the various exponentially damped modes (rather than
quantities of quantum decay theory, which describe the decay of a
discrete dot state into the reservoir continuum). Technically, it is known from the Keldysh
formalism \cite{keldysh} that dissipation (or the collision term of a quantum
Boltzmann equation) is generated by diagrams connecting the
upper with the lower part of the Keldysh contour. Therefore, to get the correct
transport rates and the precise position of resonances from the
oscillation frequencies, it is essential to include these contributions in the theory
and, at the same time, combine it with RG methods to get rid of logarithmic 
divergencies. We note that the RTRG-FS formalism is not based on the usual
Keldysh formalism as, e.g., methods based on slave particles
\cite{rosch_paaske_kroha_woelfle_PRL03} or nonequilibrium functional
RG methods.\cite{jakobs_meden_hs_PRL07}  Since the unperturbed part of the Hamiltonian
contains strong interactions, the RTRG-FS integrates out only the noninteracting reservoirs
but takes the correlation effects on the quantum dot exactly into account. This leads
to a natural formalism in terms of Liouville operators and vertices in Liouville
space, which we have combined with quantum field theoretical methods of diagrammatic
representations and RG. In this sense, we have combined the advantages of Liouville
operator techniques (providing very compact notations for the time evolution of 
the reduced density matrix in Laplace space), diagrammatic methods (to identify and
calculate irreducible blocks leading to the oscillation frequencies and the
relaxation/dephasing rates), and RG methods (reorganizing the perturbation
theory in such a way that logarithmic divergencies are absent). In particular, it is
possible to prove generically in all orders of perturbation theory and within all 
truncation schemes that the cutoff scales for the vertices are the physical 
relaxation and dephasing rates of the exponentially decaying modes of the reduced 
density matrix of the dot. This provides a reliable calculation of the line shape
at resonances. In addition, one obtains the kernel
of the kinetic equation in Laplace space from which the time evolution into the 
stationary state can be calculated. It is important to note that the possibility
to calculate the time evolution is closely related to the fact that we obtain
a systematic theory for the line shape at resonances, since the oscillation frequencies 
and the relaxation/dephasing rates governing the exponential decay of the 
{\it time evolution} are the {\it same} scales which determine the resonance positions and 
the cutoff of logarithmic divergencies at resonances for {\it stationary} quantities 
such as, e.g., the magnetization or the conductance. 

In comparism to previous formulations of RTRG 
(Refs.~\onlinecite{hs_koenig_PRL00}--\onlinecite{korb_reininghaus_hs_koenig_PRB07}) (where
the physics of cutoff scales by relaxation/dephasing modes was also included), the
recently proposed RTRG-FS method \cite{RTRG_FS} is a formulation in pure Matsubara
frequency space and avoids the need of the Keldysh indices after a first discrete
RG step where the symmetric part of the Fermi distribution is integrated out. The
main advantage of this development is the possibility to prove that the perturbation theory 
in the renormalized vertices on the r.h.s. of the RG equation is a
well defined series
in the weak-coupling regime. Essential for this proof is the fact that the imaginary
part of the denominators of all resolvents consists of a sum of strictly positive 
Matsubara frequencies (bounded from above by the cutoff), the cutoff scale $\Lambda$, 
and some positive relaxation or dephasing rate $\tilde{\Gamma}_i$. As a consequence,
it is possible to solve the RG equations analytically in the weak-coupling regime
by a systematic perturbative expansion in the renormalized couplings. In this paper,
we have described this procedure for a generic quantum dot in the Coulomb blockade 
regime where spin and/or orbital fluctuations dominate the transport properties. We
have included all two-loop contributions to calculate physical quantities up to the
first logarithmic contribution at resonances. We have shown that the latter are
well defined after renormalization and can be treated within perturbation theory
in the renormalized couplings.

We have applied the formalism to the anisotropic Kondo model at finite magnetic field
and finite bias voltage. We calculated stationary properties (magnetic susceptibility and
conductance) and the parameters characterizing the exponentially damped modes of the
time evolution (renormalized magnetic field, spin relaxation and dephasing rate). In 
comparism to the case of zero magnetic field (which was treated within flow equation 
methods already in Ref.~\onlinecite{kehrein_PRL05} for the isotropic
Kondo model), the central issue is the calculation of logarithmically enhanced 
contributions at the resonance positions $\tilde{h}\approx V$ and $\tilde{h}\ll V$.
The one-loop leading order theory of Ref.~\onlinecite{rosch_paaske_kroha_woelfle_PRL03} was able
to obtain these logarithmic contributions but it was not shown how to include 
microscopically the cutoff scales from relaxation and dephasing modes into the RG formalism.
Furthermore, the renormalization of the magnetic field was not taken into account and
the calculation was restricted to the stationary values of the magnetic susceptibility
and the conductance (the spin relaxation and dephasing rates were calculated only
perturbatively in the bare coupling and in the absence of a magnetic field in 
Ref.~\onlinecite{paaske_rosch_kroha_woelfle_PRB04}). In this paper, we have included the
renormalization of the magnetic field in $O(J)$ and $O(J^2\ln)$ together with the
spin relaxation and dephasing rate in $O(J^2)$ and $O(J^3\ln)$. This provides a
consistent theory for the resonance position and the line shape at resonance, and answers
the question which rate occurs in the various logarithmic terms. For stationary 
quantities, the obvious physical result that the spin dephasing rate cuts
off the logarithmic contributions is obtained. For the rates and the renormalized field itself, it
turns out that the spin relaxation rate determines the resonant line shape for the spin
dephasing rate and the $g$ factor, whereas the spin dephasing rate controls the
resonant line shape for the spin relaxation rate. In addition, it turns out that
the logarithmic enhancements or suppressions at the resonance $\tilde{h}=V$ are quite 
different for the
various physical quantities regarding the symmetry for $\tilde{h}>V$ and $\tilde{h}<V$.
The $g$ factor shows a symmetric suppression, the conductance an asymmetric enhancement,
the suspectibility shows only an enhancement for $\tilde{h}<V$, the derivative 
${d\tilde{\Gamma}_1\over dh_0}$ of the spin relaxation rate only an enhancement for $V>\tilde{h}$,
and the corresponding quantity for the spin dephasing rate only a suppression for $V<\tilde{h}$.
We proposed an experimental setup with a weakly coupled probe lead to measure the
logarithmic suppression of the renormalized $g$ factor as function of
voltage $V$ or
magnetic field $h_0$. This has to be contrasted to the usual measurement of the $g$ factor
at resonance $V=\tilde{h}$ from the position of the cotunneling- and Kondo-enhanced
conductance peak, which provides the $g$ factor only for one special value of the voltage.
In the anisotropic case, we find the obvious result that all logarithmic contributions
involving the transverse coupling are suppressed for increasing longitudinal coupling 
since the renormalization of the transverse coupling is reduced.
However, for the magnetic susceptibility this does not hold since the transverse coupling
cancels out in the prefactor of the logarithmic terms. Therefore, the logarithmic enhancements
become larger for the susceptibility if one increases the longitudinal coupling. This result
might be helpful to find Kondo physics in transport through single molecular magnets, where the 
transverse coupling can be smaller than the longitudinal one. Concerning the line shape at resonance
when increasing the longitudinal coupling, it turns out that 
the logarithmic features become less pronounced if the spin dephasing rate determines the
line shape, whereas it becomes sharper when the spin relaxation rate is involved. The reason
is that the spin dephasing rate involves a voltage-induced term $\sim V(J^z)^2$, which
increases with increasing longitudinal coupling, whereas the spin relaxation rate involves
only the transverse coupling in order $O(J^2)$.

\section{List of symbols}
\label{sec:list_of_symbols}

In this section we provide a complete list of all symbols used in this paper. We present two tables, one
for the symbols used in Sec.~\ref{sec:generic} of the generic case, and one for Sec.~\ref{sec:kondo}
where the method is applied to the Kondo model.

The table for the symbols used in Sec.~\ref{sec:generic} is given by

  \begin{longtable}{l|l|l}
    Symbol & What the symbol means & Ref. \\
    \hline
    $H$ & Hamiltonian & (\ref{H_total})  \\
    $H_{res}$ & reservoir Hamiltonian & (\ref{H_res})  \\
    $H_S$ & dot Hamiltonian & (\ref{H_S})  \\
    $V$ & coupling dot $\leftrightarrow$ reservoir & (\ref{coupling})  \\
    $H_0$ & $H_{res}+H_S$ & (\ref{H_total})  \\
    $\omega$ & frequency & (\ref{coupling})  \\
    $\eta$ & creation/annihilation index & (\ref{coupling})  \\
    $\nu$ & index for $\sigma,\alpha,\dots$ & (\ref{coupling})  \\
    $\sigma$ & spin index & (\ref{coupling})  \\
    $\alpha$ & reservoir index & (\ref{coupling})  \\
    $1$ & index for $\eta\nu\omega$ & (\ref{spin_orbital})  \\
    $\bar{1}$ & index for $-\eta\nu\omega$ & (\ref{contraction})  \\
    $\delta_{11'}$ & $\delta_{\eta\eta'}\delta_{\nu\nu'}\delta(\omega-\omega')$ & (\ref{contraction})  \\
    $a_1$ & field operator & (\ref{coupling})  \\
    $g_{11'}$ & coupling vertex & (\ref{coupling})  \\
    $\rho_{res}$ & reservoir density matrix & (\ref{contraction})  \\
    $T_\alpha$ & temperature of reservoir $\alpha$ & (\ref{contraction})  \\
    $\mu_\alpha$ & chemical potential of reservoir $\alpha$ & (\ref{H_res})  \\
    $f_\alpha(\omega)$ & Fermi function & (\ref{contraction})  \\
    $D$ & reservoir band width & (\ref{band_cutoff})  \\
    $\rho(\omega)$ & $D^2/(D^2+\omega^2)$ & (\ref{band_cutoff})  \\
    $I^\gamma$ & current operator for reservoir $\gamma$ & (\ref{current_operator})  \\
    $i^\gamma_{11'}$ & current vertex & (\ref{current_vertex})  \\
    $c^\gamma_{11'}$ & $-{1\over 2}(\eta\delta_{\alpha\gamma}+\eta'\delta_{\alpha'\gamma})$ & (\ref{c_symbol})  \\
    $\rho_S(t)$ & reduced dot density matrix & (\ref{rho_S_formal})  \\
    $[A,B]_\pm$ & $AB\pm BA$ & (\ref{L_LI})  \\
    $L$ & $[H,\cdot]_-$ & (\ref{L_LI})  \\
    $L_{I^\gamma}$ & ${i\over 2}[I^\gamma,\cdot]_+$ &  (\ref{L_LI}) \\
    $\tilde{\rho}_S(E)$ & Laplace transform of $\rho_S(t)$ & (\ref{laplace_rho})  \\
    $\langle \tilde{I}^\gamma \rangle(E)$ & Laplace transform of $\langle I^\gamma\rangle(t)$ 
    &(\ref{laplace_current}) \\
    $\mbox{Tr}_{res}$ & trace over the reservoir &(\ref{laplace_rho}) \\
    $\mbox{Tr}_S$ & trace over the dot &(\ref{laplace_current}) \\
    $L_S^{(0)}$ & $[H_S,\cdot]_-$ &(\ref{L_S_(0)}) \\
    $L_V$ & $[V,\cdot]_-$ &(\ref{coupling_product}) \\
    $p$ & Keldysh index &(\ref{liouville_field_operators}) \\
    $G^{pp'}_{11'}$ & vertex in Liouville space &(\ref{G_vertex_liouville}) \\
    $J^p_1$ & field operator in Liouville space &(\ref{liouville_field_operators}) \\
    $S$ & symmetry factor for equivalent diagrams &(\ref{value_diagram}) \\
    $N_p$ & number of permutations of field operators &(\ref{value_diagram}) \\
    $X_i$  &  energy variables crossed by vertical line  & (\ref{value_diagram})  \\
    $\gamma^{pp'}_{11'}$ & contraction ${J_1^p\,J_{1'}^{p'}
      \begin{picture}(-20,11) 
        \put(-22,8){\line(0,1){3}} 
        \put(-22,11){\line(1,0){12}} 
        \put(-10,8){\line(0,1){3}}
      \end{picture}
      \begin{picture}(20,11) 
      \end{picture}}$ &(\ref{liouville_contraction}) \\
    $\Pi_{1\dots n}$ & $(E_{1\dots n}+\bar{\omega}_{1\dots n}-L_S^{(0)})^{-1}$ &(\ref{pi_bare}) \\
    & $(E_{1\dots n}+\bar{\omega}_{1\dots n}-L_S(E_{1\dots n}+\bar{\omega}_{1\dots n}))^{-1}$
    & (\ref{pi_energy}) \\
    $E_{1\dots n}$ & $E+\sum_{i=1}^n \bar{\mu}_i$ &(\ref{E_omega_definition}) \\
    $\bar{\omega}_{1\dots n}$ & $\sum_{i=1}^n \bar{\omega}_i$ &(\ref{E_omega_definition}) \\
    $\bar{\mu}_i$ & $\eta_i\mu_{\alpha_i}$ &(\ref{mu_omega_bar}) \\
    $\bar{\omega}_i$ & $\eta_i\omega_i$ &(\ref{mu_omega_bar}) \\
    $L_S^{eff}$ & effective Liouville operator &(\ref{reduced_dm}) \\
    $\Sigma(E)$ & $L_S^{eff}(E)-L_S^{(0)}$ &(\ref{L_eff}) \\
    $\Sigma_\gamma(E)$ & current kernel &(\ref{current}) \\
    $(I^\gamma)^{pp'}_{11'}$ & current vertex in Liouville space &(\ref{current_liouvillian_vertex}) \\
    $\rho^{st}_S$ & stationary dot density matrix &(\ref{rd_stationary}) \\
    $\Pi(z)$ & $(z-L_S^{eff}(z))^{-1}$ &(\ref{pi_spectral_representation}) \\
    $\lambda_i(z)$ & eigenvalues of $L_S^{eff}(z)$ &(\ref{def_right_eigenvectors}) \\
    $|x_i(z)\rangle$ & right eigenvector of $L_S^{eff}(z)$  &(\ref{def_right_eigenvectors}) \\
    $|x_0(z)\rangle$ & eigenvec. with zero eigenval. of $L_S^{eff}(z)$ &(\ref{right_zero_eigenvector}) \\
    $\langle \bar{x}_i(z)|$ & left eigenvector of $L_S^{eff}(z)$ &(\ref{def_left_eigenvectors}) \\
    $z_i$ & poles of $\Pi(z)$: $z_i=\lambda_i(z_i)$ &(\ref{pole_equation}) \\
    $a_i$ & residua of $\Pi(z)$: $(1-{d\lambda_i\over dz}(z_i))^{-1}$ &(\ref{residuum}) \\
    $\tilde{Z}$,$\tilde{L}_S$ & $\Pi(z)\approx \tilde{Z}/(z-\tilde{L}_S)$ &(\ref{Pi_general_approximation}) \\
    $\tilde{h}_i$,$\tilde{\Gamma}_i$ & $z_i=\tilde{h}_i-i\tilde{\Gamma}_i$ &(\ref{effective_h_Gamma}) \\
    $\Gamma$ & short hand notation for $\tilde{\Gamma}_i\equiv \Gamma$ &  \\
    $A^c$ & $(A^c)_{ss',\bar{s}\bar{s}'}=A^*_{s's,\bar{s}'\bar{s}}$ &(\ref{c_transformation}) \\
    $\bar{G}_{11'}$ & $\sum_p G^{pp}_{11'}$ &(\ref{vertex_bar}) \\
    $\tilde{G}_{11'}$ & $\sum_p pG^{pp}_{11'}$ &(\ref{vertex_decomposition}) \\
    $\bar{I}^\gamma_{11'}$ & $\sum_p (I^\gamma)^{pp}_{11'}$ &(\ref{vertex_bar}) \\
    $\gamma_1^{s/a}$ & symmetric/antisymmetric part of $\gamma^{pp'}_{11'}$ &(\ref{sym_antisym_contraction}) \\
    $L_S^a$ & effective Liouvillian including $\gamma_1^s$ &(\ref{L_initial}) \\
    $\Sigma^a$ & effective kernel including $\gamma_1^s$ & (\ref{Sigma_initial}) \\
    $\bar{G}^a_{11'}$ & effective vertex including $\gamma_1^s$ &(\ref{G_initial}) \\
    $\Lambda$ & RG cutoff parameter &(\ref{Fermi_cutoff}) \\
    $\Lambda_0$ & initial cutoff  &(\ref{Lambda_initial}) \\
    $\Lambda_{T_\alpha}$ & Matsubara freq. $\omega_n^\alpha$ lying closest to $T_\alpha$ &
    (\ref{gamma_lambda_derivative}) \\
    $\Lambda_c$ & $\text{max}\{|E|,|\mu_\alpha|,|\tilde{h}_i|\}$  &(\ref{Lambda_c}) \\
    $l$ & $\ln{\Lambda_0\over\Lambda}$ &(\ref{poor_man_scaling_kondo}) \\
    $f_\alpha^\Lambda(\omega)$ & cutoff-dependent Fermi function &(\ref{Fermi_cutoff}) \\
    $\omega_n^\alpha$ & Matsubara frequencies for reservoir $\alpha$ &(\ref{Fermi_cutoff}) \\
    $\theta_T(\omega)$ & smeared $\theta$-function &(\ref{theta_T}) \\
    $\gamma_1^\Lambda$ & cutoff-dependent antisymmetric part  &(\ref{gamma_cutoff}) \\
    $L_S(E,\omega)$ & $L_S(E+i\omega)$ &(\ref{L_matsubara}) \\
    $\bar{G}_{12}(\dots)$ & 
    $\bar{G}_{12}(E,\omega,\omega_1,\omega_2)=
    \bar{G}_{12}(E+i\omega)|_{\bar{\omega}_i\rightarrow i\omega_i}$ &(\ref{G_matsubara}) \\
    $\Pi(E,\omega)$ & $(E+i\omega-L_S(E,\omega))^{-1}$ &(\ref{pi_matsubara}) \\
    $J_\Lambda$ & order of $\bar{G}_{11'}\sim O(J_\Lambda)$ &  \\
    $J_c$ & $J_{\Lambda=\Lambda_c}$ & \\
    $J_0$  &  $J_{\Lambda=\Lambda_0}$  &    \\
    $K_\Lambda(z)$ & $\ln{2\Lambda-iz\over \Lambda-iz}$ &(\ref{K_function}) \\
    $\tilde{K}_\Lambda(z)$ & $K_\Lambda(z)-{iz\over 2\Lambda}$ &(\ref{K_tilde_function}) \\
    $F_\Lambda(z)$ & $K_\Lambda(z)={d\over d\Lambda}F_\Lambda(z)$ &(\ref{F_function}) \\
    $\tilde{F}_\Lambda(z)$ & $\tilde{K}_\Lambda(z)={d\over d\Lambda}\tilde{F}_\Lambda(z)$ &
    (\ref{F_tilde_function}) \\
    $\tilde{F}^\prime_\Lambda(z)$ & $\tilde{F}_\Lambda(z)-{iz\over 2}\ln{2}$ &(\ref{tilde_F_prime}) \\
    $L_S^{(n)}$  &  $O(J^n)$ of $L_S(E,\omega)$  & (\ref{L_expansion}) \\
    $\bar{G}_{12}^{(n)}$  &  $O(J^n)$ of $\bar{G}_{12}(E,\omega,\omega_1,\omega_2)$  & (\ref{G_expansion}) \\
    $L_S^{(1)}$,$Z^{(1)}$  &  $L_S^{(1)}(E,\omega)=L_S^{(1)}-(E+i\omega)Z^{(1)}$  & 
    (\ref{L_first_decomposition}) \\
    $L_S^{(n)c}$  &  $L_S^{(n)}(E,\omega)|_{\Lambda=\Lambda_c}$  &  \\
    $\bar{G}_{12}^{(n)c}$  &  $\bar{G}_{12}^{(n)}(E,\omega,\omega_1,\omega_2)|_{\Lambda=\Lambda_c}$  &  \\
    $\tilde{L}_S^{(1)}$  &  $L_S^{(1)}-{1\over 2}(Z^{(1)}L_S^{(0)}+L_S^{(0)}Z^{(1)})$  & (\ref{tilde_L}) \\
    $\bar{G}_{11'}^{(2a/b)}$  &  $\bar{G}_{11'}^{(2)}=\bar{G}_{11'}^{(2a)}+\bar{G}_{11'}^{(2b)}$ & 
    (\ref{G_second_decomposition}) \\
    $\bar{G}_{11'}^{(2a_{1/2})}$  &  $\bar{G}_{11'}^{(2a)}=i\bar{G}_{11'}^{(2a_1)}+\bar{G}_{11'}^{(2a_2)}$ & 
    (\ref{G_2a_decomposition}) \\
    $L_S^{(2a/b/c)}$  &  decomposition of $L_S^{(2)}(E,\omega)$  & (\ref{L_2_decomposition}) \\
    $Z^{(2b/c)}$  &  decomposition of $L_S^{(2)}(E,\omega)$  & (\ref{L_2_decomposition}) \\
    $L_S^{(3a/b)}$  &  decomposition of $L_S^{(3)}(E,\omega)$  & (\ref{L_expansion_below_c}) \\
    $\Delta_{1\dots n}$  &  $E_{1\dots n}+i\omega-\tilde{L}_S$  & (\ref{low_energy_cutoff}) \\
    $\Delta$  &  short hand notation for $\Delta_{1\dots n}\equiv \Delta$  
  \end{longtable}
The table for the symbols used in Sec.~\ref{sec:kondo} is given by
  \begin{longtable}{l|l|l}
    Symbol & What the symbol means & Ref. \\
    \hline
    $h_0$  &  bare magnetic field  &  (\ref{H_S_kondo}) \\
    $h$  &  $(1-{1\over 2}(J_\alpha^z-(J_\alpha^z)_0))h_0$  &  (\ref{renormalized_h}) \\
    $\tilde{h}$  &  renormalized magnetic field  &  (\ref{z_+-_cc}) \\
    $\tilde{\Gamma}_1$  &  spin relaxation rate  &  (\ref{z_1_cc}) \\
    $\tilde{\Gamma}_2$  &  spin dephasing rate  &  (\ref{z_+-_cc}) \\
    $\underline{S}$  &  spin operator  &  (\ref{g_kondo}) \\
    $(J^i_{\alpha\alpha'})_0$  &  bare exchange couplings ($i=x,y,z,\perp$)  &  (\ref{g_kondo}) \\
    $J^i_{\alpha\alpha'}$  &  renormalized exchange couplings   &  (\ref{poor_man_scaling_kondo}) \\
    $J^{z/\perp}$  &  $J^{z/\perp}_{\alpha\alpha'}=2\sqrt{x_\alpha x_{\alpha'}}J^{z/\perp}$  &  
    (\ref{J_form_renormalized}) \\
    $J^{z/\perp}_\alpha$  &  $J^{z/\perp}_{\alpha\alpha}$  &  (\ref{notation_couplings})  \\
    $J^{z/\perp}\nd$  &  $J^{z/\perp}_{LR}=J^{z/\perp}_{RL}$  &  (\ref{notation_couplings})  \\
    $x_\alpha$  &  factor for coupling to reservoir $\alpha$  &  (\ref{J_form}) \\
    $J^{\gamma,z/\perp}_{\alpha\alpha'}$  &  $c^\gamma_{\alpha\alpha'}J^{z/\perp}_{\alpha\alpha'}$  &  
    (\ref{current_exchange_renormalized}) \\
    $c^\gamma_{\alpha\alpha'}$  &  $-{1\over 2}(\delta_{\gamma\alpha}-\delta_{\gamma\alpha'})$  &  
    (\ref{current_exchange_renormalized}) \\
    $T_K$  &  Kondo temperature  &  (\ref{invariant_c_TK}) \\
    $T_K^\prime$  &  $\sqrt{J^\perp_0}T_K$  &  (\ref{T_K_2_loop}) \\
    $c^2$  &  ${1\over 4}\text{Tr}((\hat{J}^z)^2-(\hat{J}^\perp)^2)$  &  (\ref{invariant_c_TK}) \\
    $V$  &  voltage  &  (\ref{voltage}) \\
    $\underline{L}^\pm$  &  $\underline{L}^+A=\underline{S}A$,$\underline{L}^-A=-A\underline{S}$  &  
    (\ref{L_basis_operators}) \\
    $L^a$  &  $\tfrac34\cdot\Id+\underline{L}^+\cdot\underline{L}^-$  &  (\ref{eq:ScalarSuperoperators}) \\
    $L^b$  &  $\tfrac14\cdot\Id-\underline{L}^+\cdot\underline{L}^-$  &  (\ref{eq:ScalarSuperoperators}) \\
    $L^c$  &  $\tfrac12\cdot\Id+2L^{+z}L^{-z}$  &  (\ref{eq:ScalarSuperoperators}) \\
    $L^h$  &  $L^{+z}+L^{-z}$  &  (\ref{eq:ScalarSuperoperators}) \\
    $\underline{L}_1$  &  $\tfrac12\left(\underline{L}^+-\underline{L}^-
      -2i\underline{L}^+\times\underline{L}^-\right)$  &  (\ref{eq:VectorSuperoperators}) \\
    $\underline{L}_2$  &  $-\tfrac12\left(\underline{L}^++\underline{L}^-\right)$  &  
    (\ref{eq:VectorSuperoperators}) \\
    $\underline{L}_3$  &  $\tfrac12\left(\underline{L}^+-\underline{L}^-
      +2i\underline{L}^+\times\underline{L}^-\right)$  &  (\ref{eq:VectorSuperoperators}) \\
    $L^i_\pm$  & $L^{ix}\pm iL^{iy}$; $i=1,2,3,\pm$   &  (\ref{eq:L_i_pm}) \\
    $L^4_\pm$  &  $L^2_\pm\pm\left(L^+_\pm L^{-z}+L^{+z} L^-_\pm\right)$  &  (\ref{L_4}) \\
    $L^5_\pm$  &  $L^2_\pm\mp\left(L^+_\pm L^{-z}+L^{+z} L^-_\pm\right)$  &  (\ref{L_5}) \\
    $h(E,\omega)$  &  term $h(E,\omega)L^h$ in $L_S(E,\omega)$  &  (\ref{L_representation}) \\
    $\Gamma^{i}(E,\omega)$  &  term $-i\Gamma^{i}(E,\omega)L^{i}$ in $L_S(E,\omega)$; $i=a,c,3z$  &  
    (\ref{L_representation}) \\
    $\Gamma_\gamma^{i}(E,\omega)$  &  term $i\Gamma_\gamma^{i}(E,\omega)L^{i}$ 
    in $\Sigma_\gamma(E,\omega)$; $i=b,1z$  &  (\ref{sigma_current_representation}) \\
    $\lambda_i(E,\omega)$  &  eigenvalues of $L_S(E,\omega)$; $i=0,1,\pm$  &  (\ref{eigenvalues}) \\
    $P_i(E,\omega)$  &  projectors on eigenvectors of $L_S(E,\omega)$  &  (\ref{projectors}) \\
    $|x_i(E,\omega)\rangle$ & right eigenvector of $L_S(E,\omega)$  &(\ref{eigenvectors}) \\
    $\langle \bar{x}_i(E,\omega)|$ & left eigenvector of $L_S(E,\omega)$ &(\ref{eigenvectors}) \\
    $z_0$  &  $z_0=0$  &  (\ref{z_0}) \\
    $z_1$  &  $z_1=-i\tilde{\Gamma}_1$  &  (\ref{z_1}) \\
    $z_\pm$  &  $z_\pm=\pm\tilde{h}-i\tilde{\Gamma}_2$  &  (\ref{z_+-}) \\
    $\sigma^z_\pm$  &  ${1\over 2}(1\pm\sigma^z)$  &  (\ref{spin_matrices}) \\
    $\sigma_\pm$  &  ${1\over 2}(\sigma_x\pm i\sigma_y)$  &  (\ref{spin_matrices}) \\
    $\hat{L}_\pm^\chi$  &  $L^\chi\sigma^z_\pm$; $\chi=a,b,c,h,1z,3z$  &  (\ref{L_spin_1}) \\
    $\hat{L}_\pm^\chi$  &  $L^\chi_\pm\sigma_\mp$; $\chi=1,3,4,5$  &  (\ref{L_spin_2}) \\
    $\bar{G}^{\chi s}(\dots)$  &  comp. of 
    $\bar{G}_{12}(E,\omega,\omega_1,\omega_2)|_{\eta_1=-\eta_2=+}$  &  (\ref{G_representation}) \\
    $\bar{I}^{\gamma\chi s}(\dots)$  &  comp. of 
    $\bar{I}^\gamma_{12}(E,\omega,\omega_1,\omega_2)|_{\eta_1=-\eta_2=+}$  &  (\ref{I_representation}) \\
    $\bar{G}^{\chi s}(\dots)^T$  &  $(\bar{G}^{\chi s})^T_{\alpha\alpha'}=\bar{G}^{\chi s}_{\alpha'\alpha}$  
    &  (\ref{transpose}) \\
    $\bar{I}^{\gamma\chi s}(\dots)^T$  &  $(\bar{I}^{\gamma\chi s})^T_{\alpha\alpha'}=
    \bar{I}^{\gamma\chi s}_{\alpha'\alpha}$    &  (\ref{transpose}) \\
    $(\hat{L}^\chi_s)^T$  &  $(\hat{L}^\chi_s)^T_{\sigma\sigma'}=(\hat{L}^\chi_s)_{\sigma'\sigma}$  
    &  (\ref{transpose}) \\
    $\hat{K}^{z/\perp}$  &  ${1\over 2}\hat{K}^z s\hat{L}^h_s
    -{1\over 2}\hat{K}^\perp(\hat{L}^4_s+\hat{L}^5_s)$  &  (\ref{ansatz_G_2a_2_kondo}) \\
    $K^{z/\perp}$  &  $K^{z/\perp}_{\alpha\alpha'}=2\sqrt{x_\alpha x_{\alpha'}}\,K^{z/\perp}$  &  
    (\ref{K_form}) \\
    $\hat{K}^{\gamma,z/\perp}$  &  ${1\over 2}\hat{K}^{\gamma z} s\hat{L}^{1z}_s
    +{1\over 2}\hat{K}^{\gamma\perp} \hat{L}^{1}_s$  &  (\ref{ansatz_I_2a_2_kondo}) \\
    $\hat{R}^{\gamma,z/\perp}$  &  $K^{\gamma,z/\perp}_{\alpha\alpha'}=
    c^\gamma_{\alpha\alpha'}K^{z/\perp}_{\alpha\alpha'}+R^{\gamma,z/\perp}_{\alpha\alpha'}$  &  
    (\ref{K_current_K_2a_2}) \\
    $\hat{R}^{z/\perp}$  &  $R^{\gamma,z/\perp}_{\alpha\alpha'}=
    c^\gamma_{\alpha\alpha'}R^{z/\perp}_{\alpha\alpha'}$  &  (\ref{R_gamma_form}) \\
    $R^{z/\perp}$  &  $R^{z/\perp}_{\alpha\alpha'}=2\sqrt{x_\alpha x_{\alpha'}}\,R^{z/\perp}$  &  
    (\ref{R_form}) \\
    ${\hat{\cal{J}}}^{z/\perp}$  &  $\hat{J}^{z/\perp}+\hat{K}^{z/\perp}$  &  (\ref{total_vertex}) \\
    $\mathcal{L}_i(x)$  &  $\ln\frac{\Lambda_c}{\sqrt{x^2+(\tilde{\Gamma}_i)^2}}$  &  (\ref{eq:log}) \\
    $E_{\alpha\alpha'}$  &  $E+\mu_\alpha-\mu_{\alpha'}$  &  (\ref{product_decompose_2}) \\
    $\text{sign}_i(x)$  &  ${2\over\pi}\arctan{x\over\tilde{\Gamma}_i}$  &  (\ref{sign_replacement}) \\
    ${|x|}_i$  &  $x\,\text{sign}_i(x)$  &  (\ref{absolute_gamma_broadening}) \\
    $\theta_i(x)$  &  ${1\over 2}[1+\text{sign}_i(x)]$  &  (\ref{theta_gamma_broadening}) \\
    $p_{\uparrow/\downarrow}$  &  occupation probabilities  &  (\ref{occupation_probabilities}) \\
    $M$  &  magnetization  &  (\ref{rd_stationary_kondo}) \\
    $X$  &  auxiliary symbol to express $M$  & (\ref{X_abbr}) \\
    $\tilde{g}$  & renormalized $g$ factor $\tilde{g}=2{d\tilde{h}\over dh_0}$  & (\ref{g_renormalized}) \\
    $\chi$  &  magnetic susceptibility $\chi={d M\over d h_0}$  &  (\ref{susceptibility}) \\
    $G_\gamma$  &  conductance $G_\gamma={d I_\gamma\over dV}$  &  (\ref{inelastic_cotunneling_1}) \\
    $G_0$  &  conductance quantum $G_0={e^2\over h}$  &  (\ref{inelastic_cotunneling_1}) 
  \end{longtable}

\section*{Acknowledgments}

We thank N. Andrei, L. Glazman, S. Jakobs, S. Kehrein, H. Kroha, J. K\"onig, T. Korb, V. Koerting,
M. Kurz, V. Meden, J. Paaske, M. Pletyukhov, A. Rosch, D. Schuricht, M. Wegewijs, 
P. W\"olfle, and A. Zawadowski for valuable discussions. This work was supported by the 
DFG-Forschergruppe 723, the VW Foundation, and the Forschungszentrum
J\"ulich via the virtual institute IFMIT.

\begin{appendix}
  \section{\label{sec:appendix_A}
    \bf{second-order RG}}

In this appendix, we prove that the effect of the RG terms (\ref{L_expansion_term_3}) and
(\ref{L_expansion_term_4}) is the replacement 
$F_\Lambda(z)\rightarrow F_\Lambda^\prime(z)$ 
in Eq.~(\ref{L_2_term_a}), with $F_\Lambda^\prime(z)$ defined by Eq.~(\ref{tilde_F_prime}).

First we consider the term (\ref{L_expansion_term_4}). The terms of $O({\Delta\over\Lambda}J^3)$
can be extracted by using the expansion (\ref{integral_resolvent_expansion}) for the integral
over the resolvent and considering only the terms of $O(1)$ and $O({\Delta\over\Lambda})$ in this
expansion,
\begin{equation}
\label{integral_resolvent_lowest}
i\,\int_0^\Lambda\,d\omega_2\,\Pi(E_{12},\Lambda+\omega+\omega_2)\approx
\ln(2)+{i\over 2\Lambda}\,z_{12}\,,
\end{equation}
where we have defined
\begin{equation}
\label{def_z}
z_{12}\,\equiv\,(E_{12}+i\omega-L_S^{(0)})\quad.
\end{equation}
Inserting this in Eq.~(\ref{L_expansion_term_4}) for the two independent integrals over the
resolvents and collecting the terms in $O({\Delta\over\Lambda}J^3)$ leads to the result
\begin{eqnarray}
\nonumber
\hspace{-1cm}
(\ref{L_expansion_term_4})\,&\rightarrow &\,\\
\label{L_expansion_term_4_new}
&&\hspace{-1cm}
-{\ln(2)\over 2\Lambda}\left\{
\bar{G}_{12}^{(1)}z_{12}\bar{G}_{\bar{2}3}^{(1)}\bar{G}_{\bar{3}\bar{1}}^{(1)}
+\,\bar{G}_{12}^{(1)}\bar{G}_{\bar{2}3}^{(1)}
z_{13}\bar{G}_{\bar{3}\bar{1}}^{(1)}\right\}\,.
\end{eqnarray}

Next we consider the term (\ref{L_expansion_term_3}) and use for $\bar{G}^{(2b)}$ 
the result (\ref{G_second_frequency}), which gives
\begin{eqnarray}
\nonumber
\hspace{-1cm}
\bar{G}_{\bar{2}\bar{1}}^{(2b)}(E_{12},\Lambda+\omega+\omega_2,-\omega_2,-\Lambda)\,&=&\,\\
\label{G_2b_term_1}
&&\hspace{-4cm}
=\,\bar{G}_{\bar{2}3}^{(1)}
\ln{2\Lambda-iz_{13}\over \Lambda}
\bar{G}_{\bar{3}\bar{1}}^{(1)}\\
\label{G_2b_term_2}
&&\hspace{-3.5cm}
-\,\bar{G}_{\bar{1}3}^{(1)}
\ln{\Lambda+\omega_2-iz_{23}\over \Lambda}
\bar{G}_{\bar{3}\bar{2}}^{(1)}\,,
\end{eqnarray}
and
\begin{eqnarray}
\label{G_2b_term_3}
\hspace{-0.5cm}
\bar{G}_{12}^{(2b)}(E,\omega,\Lambda,\omega_2)\,&=&\,
\bar{G}_{13}^{(1)}\ln{2\Lambda-iz_{13}\over \Lambda}
\bar{G}_{\bar{3}2}^{(1)}\\
\label{G_2b_term_4}
&&\hspace{-0.5cm}
-\,\bar{G}_{23}^{(1)}
\ln{\Lambda+\omega_2-iz_{23}\over \Lambda}
\bar{G}_{\bar{3}1}^{(1)}\,.
\end{eqnarray}

Equations~(\ref{G_2b_term_1}) and (\ref{G_2b_term_3}) are independent of the
integration variable $\omega_2$, and we can replace the logarithm by
\begin{equation}
\ln{2\Lambda-iz_{13}\over \Lambda}\approx
\ln(2)-{i\over 2\Lambda}z_{13}\,.
\end{equation}
Inserting Eqs.~(\ref{G_2b_term_1}) and (\ref{G_2b_term_3}) in 
(\ref{L_expansion_term_3}), using Eq.~(\ref{integral_resolvent_lowest}), and
collecting all terms of $O({\Delta\over\Lambda}J^3)$, we find
\begin{eqnarray}
\nonumber
\hspace{-0.5cm}
(\ref{L_expansion_term_3})\,\,\text{from}\,\,
(\ref{G_2b_term_1})\,\,\text{and}\,\,(\ref{G_2b_term_3})  
\,&\rightarrow &\,\\
\nonumber
&&\hspace{-4.5cm}
{\ln(2)\over 2\Lambda}\left\{
\bar{G}_{12}^{(1)}z_{12}\bar{G}_{\bar{2}3}^{(1)}\bar{G}_{\bar{3}\bar{1}}^{(1)}
-\,\bar{G}_{12}^{(1)}\bar{G}_{\bar{2}3}^{(1)}
z_{13}\bar{G}_{\bar{3}\bar{1}}^{(1)}\right.\\
\label{L_expansion_term_3_new_1}
&&\hspace{-4cm}
\left. -\,\bar{G}_{13}^{(1)}z_{13}
\bar{G}_{\bar{3}2}^{(1)}\bar{G}_{\bar{2}\bar{1}}^{(1)}
+\bar{G}_{13}^{(1)}\bar{G}_{\bar{3}2}^{(1)}
z_{12}\bar{G}_{\bar{2}\bar{1}}^{(1)}\right\}\,=\,0\,,
\end{eqnarray}
i.e., the terms cancel each other.

In contrast, Eqs.~(\ref{G_2b_term_2}) and (\ref{G_2b_term_4}) are not independent 
of the integration variable $\omega_2$. When inserted in Eq.~(\ref{L_expansion_term_3}),
we need the integral ($z_{ij}$ is replaced by its eigenvalue in this equation)
\begin{eqnarray}
\nonumber
\int_0^\Lambda\,d\omega_2\,{1\over \Lambda+\omega_2-iz_{12}}
\,\ln{\Lambda+\omega_2-iz_{23}\over\Lambda}\,&\approx &\,\\
\nonumber
&&\hspace{-6cm}\approx\,
{1\over 2}\ln^2(2)-i{\ln(2)\over 2\Lambda}z_{12}+{i\over 2\Lambda}(z_{12}-z_{23})\,,
\end{eqnarray}
where we have expanded up to linear order in ${z_{ij}\over\Lambda}$. This gives
\begin{eqnarray}
\nonumber
\hspace{-1cm}
(\ref{L_expansion_term_3})\,\,\text{from}\,\,
(\ref{G_2b_term_2})\,\,\text{and}\,\,(\ref{G_2b_term_4})  
\,&\rightarrow &\,\\
\nonumber
&&\hspace{-4.5cm}\rightarrow\,
{\ln(2)\over 2\Lambda}\left\{
\bar{G}_{12}^{(1)}z_{12}\bar{G}_{\bar{1}3}^{(1)}\bar{G}_{\bar{3}\bar{2}}^{(1)}+
\bar{G}_{23}^{(1)}\bar{G}_{\bar{3}1}^{(1)}z_{12}\bar{G}_{\bar{2}\bar{1}}^{(1)}\right\}\\
\nonumber
&&\hspace{-4cm}
-{1\over 2\Lambda}\left\{
\bar{G}_{12}^{(1)}z_{12}\bar{G}_{\bar{1}3}^{(1)}\bar{G}_{\bar{3}\bar{2}}^{(1)}+
\bar{G}_{23}^{(1)}\bar{G}_{\bar{3}1}^{(1)}z_{12}\bar{G}_{\bar{2}\bar{1}}^{(1)}\right.\\
\label{L_expansion_term_3_new_2}
&&\hspace{-3cm}
\left.-\bar{G}_{12}^{(1)}\bar{G}_{\bar{1}3}^{(1)}z_{23}\bar{G}_{\bar{3}\bar{2}}^{(1)}-
\bar{G}_{23}^{(1)}z_{23}\bar{G}_{\bar{3}1}^{(1)}\bar{G}_{\bar{2}\bar{1}}^{(1)}\right\}
\end{eqnarray}
Using the antisymmetry property (\ref{rg_G_symmetry}), we see that the second term is 
zero and the first term agrees with Eq.~(\ref{L_expansion_term_4_new}), leading to the final
result
\begin{eqnarray}
\nonumber
\hspace{-0.5cm}
(\ref{L_expansion_term_3})+(\ref{L_expansion_term_4})\,&\rightarrow &\,\\
\nonumber
&&\hspace{-2cm}
-{\ln(2)\over \Lambda}\left\{
\bar{G}_{12}^{(1)}(E_{12}+i\omega-L_S^{(0)})
\bar{G}_{\bar{2}3}^{(1)}\bar{G}_{\bar{3}\bar{1}}^{(1)}\,+\right.\\
\label{L_expansion_term_34_new}
&&\hspace{-1.5cm}
\left. +\,\bar{G}_{12}^{(1)}\bar{G}_{\bar{2}3}^{(1)}
(E_{13}+i\omega-L_S^{(0)})\bar{G}_{\bar{3}\bar{1}}^{(1)}\right\}\quad.
\end{eqnarray}
Finally, using the leading-order RG equations (\ref{G_reference_solution})
or (\ref{I_reference_solution}), we see that Eq.~(\ref{L_expansion_term_34_new})
can be written in the form
\begin{eqnarray}
\nonumber
\hspace{-1cm}
(\ref{L_expansion_term_3})+(\ref{L_expansion_term_4})\,&\rightarrow &\,\\
\label{L_expansion_term_34_derivative}
&&\hspace{-2cm}
-{\ln(2)\over 2}{d\over d\Lambda}\left\{
\bar{G}_{12}^{(1)}(E_{12}+i\omega-L_S^{(0)})
\bar{G}_{\bar{2}\bar{1}}^{(1)}\right\}\,,
\end{eqnarray}
i.e., a similiar term to Eq.~(\ref{L_2_term_a}) results, which proves the
replacement (\ref{tilde_F_prime}).

\end{appendix}


\end{document}